\documentclass[useAMS,usedcolumn,usegraphicx,usenatbib]{mn2e}

\usepackage{times}
\usepackage{amsmath}                  
\usepackage{amssymb}                  
\usepackage{graphicx}
\usepackage{units}
\usepackage{dcolumn}
\usepackage{longtable}
\usepackage{lscape}
\usepackage{booktabs}
\usepackage{subfigure}
\usepackage{natbib}
\usepackage{aas_macros}
\usepackage[draft,pdftex,pdfpagemode={UseOutlines},bookmarks,bookmarksopen,colorlinks,linkcolor={blue},citecolor={green},urlcolor={red}]{hyperref}

\let\orgautoref\autoref
\providecommand{\Autoref}
        {\def\equationautorefname{Equation}%
         \def\figureautorefname{Fig.}%
         \def\subfigureautorefname{Fig.}%
         \def\partautorefname{Part}%
         \def\chapterautorefname{Chapter}%
         \def\sectionautorefname{Section}%
         \def\subsectionautorefname{Section}%
         \def\subsubsectionautorefname{Section}%
         \def\Itemautorefname{Item}%
         \def\tableautorefname{Table}%
         \def\lstlistingautorefname{Listing}%
         \orgautoref}
\providecommand{\Autorefs}
        {\def\equationautorefname{Equations}%
         \def\figureautorefname{Figs.}%
         \def\subfigureautorefname{Figs.}%
         \def\partautorefname{Parts}%
         \def\chapterautorefname{Chapters}%
         \def\sectionautorefname{Sections}%
         \def\subsectionautorefname{Sections}%
         \def\subsubsectionautorefname{Sections}%
         \def\Itemautorefname{Items}%
         \def\tableautorefname{Tables}%
         \def\lstlistingautorefname{Listings}%
         \orgautoref}
\renewcommand{\autoref}
        {\def\equationautorefname{equation}%
         \def\figureautorefname{Fig.}%
         \def\subfigureautorefname{Fig.}%
         \def\partautorefname{part}%
         \def\chapterautorefname{chapter}%
         \def\sectionautorefname{section}%
         \def\subsectionautorefname{section}%
         \def\subsubsectionautorefname{section}%
         \def\appendixautorefname{appendix}%
         \def\Itemautorefname{item}%
         \def\tableautorefname{Table}%
         \def\lstlistingautorefname{Listing}%
         \orgautoref}
\providecommand{\autorefs}
        {\def\equationautorefname{equations}%
         \def\figureautorefname{Figs.}%
         \def\subfigureautorefname{Figs.}%
         \def\partautorefname{parts}%
         \def\chapterautorefname{chapters}%
         \def\sectionautorefname{sections}%
         \def\subsectionautorefname{sections}%
         \def\subsubsectionautorefname{sections}%
         \def\Itemautorefname{items}%
         \def\tableautorefname{Tables}%
         \def\lstlistingautorefname{Listings}%
         \orgautoref}
         
\makeatletter
\newcolumntype{d}[1]{>{\DC@{,}{{.}}{#1}}c<{\DC@end}}
\newcolumntype{o}[1]{>{\DC@{+}{\pm}{#1}}c<{\DC@end}}
\makeatother
\makeatletter
\newcolumntype{f}[1]{>{\DC@{p}{\ldots}{#1}}c<{\DC@end}}
\makeatother
         
\title[Identifying birth places of young isolated neutron stars]{Identifying birth places of young isolated neutron stars}
\author[N. Tetzlaff et al.]{N. Tetzlaff$^{1}$\thanks{E-mail:
nina@astro.uni-jena.de}, R. Neuh\"auser$^{1}$, M. M. Hohle$^{1,2}$ and G. Maciejewski$^{1}$\\
$^{1}$Astrophysikalisches Institut und Universit\"ats-Sternwarte Jena, Schillerg\"asschen 2-3, 07745 Jena, Germany\\
$^{2}$Max-Planck-Institut f\"ur extraterrestrische Physik, Giessenbachstra{\ss}e, 85741 Garching, Germany}

\newcommand{\rxja}{RX~J1856.5-3754}
\newcommand{\rxjn}{RX~J0720.4-3125}
\newcommand{\rxjs}{RX~J1605.3+3249}

\begin{document}

\date{Accepted 23 November 2009. Received 04 November 2009; in original form 15 September 2009}

\pagerange{\pageref{firstpage}--\pageref{lastpage}} \pubyear{2009}

\maketitle

\label{firstpage}

\begin{abstract}

Young isolated radio-quiet neutron stars are still hot enough to be detectable at X-ray and optical wavelengths due to their thermal emission and can hence probe cooling curves. An identification of their birth sites can constrain their age.\\
For that reason we try to identify the parent associations for four of the so-called Magnificent Seven neutron stars for which proper motion and distance estimates are available. We are tracing back in time each neutron star and possible birth association centre to find close encounters. The associated time of the encounter expresses the kinematic age of the neutron star which can be compared to its characteristic spin-down age. Owing to observational uncertainties in the input data, we use Monte-Carlo simulations and evaluate the outcome of our calculations statistically. \\
\rxja{} most probably originated from the Upper Scorpius association about $\unit[0{.}3]{Myr}$ ago. \rxjn{} was either born in the young local association TW Hydrae about $\unit[0{.}4]{Myr}$ ago or in Trumpler 10 $\unit[0{.}5]{Myr}$ in the past. Also \rxjs{} and RBS 1223 seem to come from a nearby young association such as the Scorpius-Centraurus complex or the extended Corona-Australis association. For RBS 1223 also a birth in Scutum OB2 is possible.\\
We also give constraints on the observables as well as on the radial velocity of the neutron star. Given the birth association, its age and the flight time of the neutron star, we estimate the mass of the progenitor star.\\
Some of the potential supernovae were located very nearby ($<\unit[100]{pc}$) and thus should have contributed to the $^{10}$Be and $^{60}$Fe material found in the Earth's crust.\\ 
In addition we reinvestigate the previously suggested neutron star/ runaway pair PSR B1929+10/ $\zeta$ Ophiuchi and conclude that it is very likely that both objects were ejected during the same supernova event.

\end{abstract}

\begin{keywords}
stars: early-type -- stars: kinematics -- stars: neutron -- pulsars: individual: PSR B1929+10, \rxja{}, \rxjn{}, \rxjs{}, RBS 1223.
\end{keywords}


\section{Introduction}\label{sec:intro}

Neutron stars have very large proper motions which, with known distances, indicate high space velocities \citep[e.g.][thereafter Ho05]{1994Natur.369..127L,1997MNRAS.289..592L,1997MNRAS.291..569H,1998ApJ...505..315C,2002ApJ...568..289A,2005MNRAS.360..974H}. Those high velocities that are usually larger than those of the progenitor stars are the result of asymmetric supernova explosions assigning the new-born neutron star a kick velocity for that a number of mechanisms have been suggested \citep[e.g.][]{1996PhRvL..76..352B,1996A&A...306..167J,2005ASPC..332..363J,2006ApJ...639.1007W,2009arXiv0906.2802K}.\\
Another fact is that about $\unit[46]{\%}$ of O and $\unit[10]{\%}$ of B stars also show high space velocities (see \citealt{1991AJ....102..333S} for a discussion). Two scenarios \citep[thereafter H01]{2001A&A...365...49H} are accepted to produce those so-called ``runaway stars'' \citep{1961BAN....15..265B}. The binary-supernova scenario is related to the formation of the high velocity neutron stars: The runaway and neutron star are the products of a supernova within a binary system. The velocity of the former secondary is comparable to its original orbital velocity. The second scenario, which will not be further discussed within this paper, is the dynamical ejection due to gravitational interactions between massive stars in dense clusters. H01 proposed examples for each of the scenarios. In \autoref{sec:BSS} of this paper we will review and further investigate the binary supernova scenario for PSR B1929+10 and the runaway O9.5V star $\zeta$ Oph. Instead of restricting the unknown pulsar radial velocity to a certain interval, we will use a velocity distribution to cover the whole spectrum of possible radial velocities. We then use more recent data for the pulsar to strengthen the hypothesis.\\
Beforehand, we introduce our sample of OB associations and clusters in \autoref{sec:assocsample} and the procedure utilised in \autoref{sec:procedure}.\\
In \autoref{sec:idparassoc} we investigate the origin of four isolated radio-quiet X-ray emitting neutron stars. So far, seven such sources have been confirmed. They are not only bright X-ray sources, but are also detected in the optical due to a hot cooling surface. In two cases a parallax is obtained. Brightness, parallax and temperature yield their radii. From spectra, one can in principle determine their mass and composition, which eventually may lead to constraints on the equation of state. However, only seven such sources have been identified up to now for which they were named ``The Magnificent Seven'' (M7) (\citealt{2001ASPC..234..225T}; for a recent review see \citealt{2008AIPC..968..129K}; two new candidates were recently published by \citealt{2009A&A...498..233P} and \citealt{2008ApJ...672.1137R}). With known luminosity, one only needs to estimate the age, in order to probe cooling curves. For young neutron stars, the characteristic spin-down age only gives an upper limit. We will therefore try to estimate their age from kinematics.\\
Our aim in this paper is to identify the parent associations of four members of the M7 and find constraints on their observables and radial velocity.\\
We give a summary of our results and draw our conclusions in \autoref{sec:conclusions}.


\section{The sample of associations and clusters}\label{sec:assocsample}

Given a distance of $\unit[1]{kpc}$ and typical neutron star velocities of $100$ to $\unit[500]{km/s}$ (\citealt{2002ApJ...568..289A}; Ho05) and maximum ages of $\unit[5]{Myr}$ for neutron stars to be detectable in the optical (see cooling curves in \citealt{2005MNRAS.363..555G} and \citealt{2006PhRvC..74b5803P}), we restricted our search for birth associations and clusters of young nearby neutron stars to within $\unit[3]{kpc}$. We chose a sample of OB associations and young clusters (we use the term ``association'' for both in the following) within $\unit[3]{kpc}$ from the sun with available kinematic data and distance. We collected those from \citet{2001AstL...27...58D} and H01 and associations to which stars from the galactic O-star catalogue from \citet{2004ApJS..151..103M} are associated with. Furthermore, we added young local associations (YLA) from \citet{2008A&A...480..735F} (thereafter F08) since they are possible hosts of a few supernovae in the near past. We also included the Hercules-Lyrae association (Her-Lyr) and the Pleiades and massive star forming regions \citep{2008hsf1.book.....R,2008hsf2.book.....R}. We set the lower limit of the association age to $\approx\unit[2]{Myr}$ to account for the minimum lifetime of a progenitor star that can produce a neutron star (progenitor mass smaller than $\approx\unit[30]{M_{\odot}}$; see e.g. \citealt{2003ApJ...591..288H}).\footnote{The minimum lifetime is actually $\approx\unit[6]{Myr}$. We relax that condition to account for possible uncertainties in the association age as well as the fact that also more massive stars may produce neutron stars \citep{2008ApJ...685..400B}.} The list of all explored associations and their properties can be found in \autoref{appsec:Assoclist}. Coordinates as well as heliocentric velocity components are given for a right-handed coordinate system with the $x$ axis pointing towards the galactic centre and $y$ is positive in the direction of galactic rotation.


\section{Procedure}\label{sec:procedure}

To identify potential parent associations of neutron stars, we calculate the trajectories for both, the neutron star and the centre of the association, into the past. We account for solar motion adopting a local standard of rest of $\left(U\ V\ W\right)_{\odot} = \left(10{.}00\ 5{.}25\ 7{.}17\right)\,\mathrm{km/s}$ \citep{1998MNRAS.298..387D}. To include the effect of a potential on the vertical motion we use the vertical acceleration from \citet{2003A&A...404..519P} (and references therein) and work with an Euler-Cauchy numerical method with a fixed time step of $10^4$ years. Utilising this simple technique is fully sufficient for the treatment of some million years as done here and is consistent with results obtained by applying a fifth-order Runge-Kutta integration method for a more complicated potential as e.g. given in \citet{2001AJ....122.1397H} and references therein \citep[cf. comparison of methods in][]{2009DiplA...Nina}.\\
Simultaneously to the trajectories, we derived the separation between the neutron star and the association centre for each time step and found the minimum of those as well as the associated time in the past. Since the input consists of observables with errors, we calculate a large set of simulations varying the starting parameters normally (proper motion and parallax\footnote{For some cases we obtain the parallax from other distance estimates.}; radial velocity see below) within their confidence levels.\\
For neutron stars the radial velocity cannot easily be derived from spectra due to the large gravitational redshift and hence is unknown. To be able to investigate their three dimensional motion nonetheless, we varied the radial velocity within a probability distribution derived from the one for three-dimensional pulsar velocities obtained by Ho05. They deducted a distribution with one Maxwellian component from transverse velocities of 233 pulsars assuming the radial component is of the same order. \citet{2002ApJ...568..289A} derived a distribution that includes two Maxwellian components suggesting that there might be two different populations: those which gained their speed only due to a kick during the supernova and those which have an additional component from their former orbital velocity. However, as stated by Ho05 a one-component model better fits the observations. Infact, which radial velocity distribution is used is of minor importance since its usage shall just assure that the whole spectra of possible radial velocities is covered and the probability of occurence of excessively high values is low which is the case for both distributions. Adopting the two-component model does not affect the final results significantly \citep{2009DiplA...Nina}. As the sign of the radial velocity cannot be derived from statistics it was chosen randomly, thus covering a total range of $-1,500$ to $\unit[+1,500]{km/s}$.


\section{A potential binary supernova in Upper Scorpius}\label{sec:BSS}

The runaway O-star $\zeta$ Ophiuchi ($\zeta$ Oph = HIP 81377) is an isolated main sequence star with a space velocity of $\approx\unit[20]{km/s}$. \citet{1952BAN....11..414B} suggested its origin in the Scorpius OB2 association due to its proper motion vector which points away from that association. \\
H01 investigated the origin of $\zeta$ Oph in more detail and proposed that it gained its high velocity in a binary supernova in Upper Scorpius (US) about $\unit[1]{Myr}$ ago which is also supported by its large rotational velocity of $\geq\unit[340]{km/s}$ \citep{1992A&A...261..209H,1996ApJ...463..737P,2001MNRAS.327..353H} and high helium abundance \citep{1992A&A...261..209H,1996A&A...305..825V,2005A&A...442..263V}. They also identified a neutron star which might have been the former primary: PSR B1929+10 (PSR J1932+1059).\\

Initially, we repeat the experiment of H01 applying the same starting parameters for the pulsar (\autoref{tab:1929ZOph_par},  $\mu_{\alpha}^*$ is the proper motion in right ascension corrected for declination). Note that the pulsar radial velocity of $\unit[200\pm50]{km/s}$ is not measured, but an assumption in H01. The runaway proper motion and parallax were adopted from the Hipparcos Catalogue \citep{1997A&A...323L..49P} and its radial velocity from the Hipparcos Input Catalogue \citep{1992ESASP1136.....T}. The results are in very good agreement with those of H01. $29{,}700$ of 3 million Monte-Carlo runs yield a minimum separation between $\zeta$ Oph and PSR B1929+10 of less than $\unit[10]{pc}$ compared to $30{,}822$ runs as published by H01. The smallest separation we found was $\unit[0{.}20]{pc}$ ($\unit[0{.}35]{pc}$ found by H01). In $4{,}631$ runs both objects were additionally not farther from the centre of US than $\unit[10]{pc}$. The latter number differs slightly from that of H01 ($4{,}210$ runs) which is owing to somewhat different input parameters for the association.\\
Similar calculations were also done by \citet{2008AstL...34..686B} with the same results using the epicycle approximation \citep{1959HDP....53...21L,1982lbg6.conf..208W} for tracing back the objects. Afterwards, he used more recent parameter values for the pulsar from \citet{2004ApJ...604..339C} but increased the error intervals (factor of ten for the parallax, factor of 30 for the proper motion components), to have them of the same order as H01 before, and drew the conclusion that the scenario of a binary supernova in US involving the two objects is very likely. \\

\begin{table*}
\centering
\caption{Properties of PSR B1929+10 and $\zeta$ Oph ($\mu_{\alpha}^*$ is the proper motion in right ascension corrected for declination). Note that the radial velocity of the pulsar is an assumption from H01, not a measurement.}\label{tab:1929ZOph_par}
\begin{tabular}{l d{2} d{2} o{2.2} o{3.2} o{3.2} o{3.2} c}
\toprule
	&		\multicolumn{1}{c}{$\unit[\alpha]{[^\circ]}$}		&	\multicolumn{1}{c}{$\unit[\delta]{[^\circ]}$}	&	\multicolumn{1}{c}{$\unit[\pi]{[mas]}$} & \multicolumn{1}{c}{$\unit[\mu_{\alpha}^*]{[mas/yr]}$} & \multicolumn{1}{c}{$\unit[\mu_{\delta}]{[mas/yr]}$} & \multicolumn{1}{c}{$\unit[v_r]{[km/s]}$} & Ref.\\\midrule
	
PSR B1929+10 (H01) & 293,06 &		10,99 & 4+2	&	99+12	& 39+8	& 200+50 & 1\\
PSR B1929+10 (recent) & 293,06 &		10,99 & 2{.}76+0{.}14	&	94{.}03+0{.}14	& 43{.}37+0{.}29	& \multicolumn{1}{c}{--} & 2\\
$\zeta$ Oph	 & 249,29	&		-10,57 &	7{.}12+0{.}71	&	13{.}07+0{.}85	&	25{.}44+0{.}72 &	-9{.}0+5{.}5 & 3, 4\\\bottomrule
\multicolumn{8}{l}{1 -- H01 and references therein, 2 -- \citet{2004ApJ...604..339C}, 3 -- Hipparcos Catalogue, 4 -- Hipparcos Input Catalogue}
\end{tabular}
\end{table*}

\begin{figure}
\subfigure[]{\includegraphics[width=0.25\textwidth, viewport= 85 265 480 590]{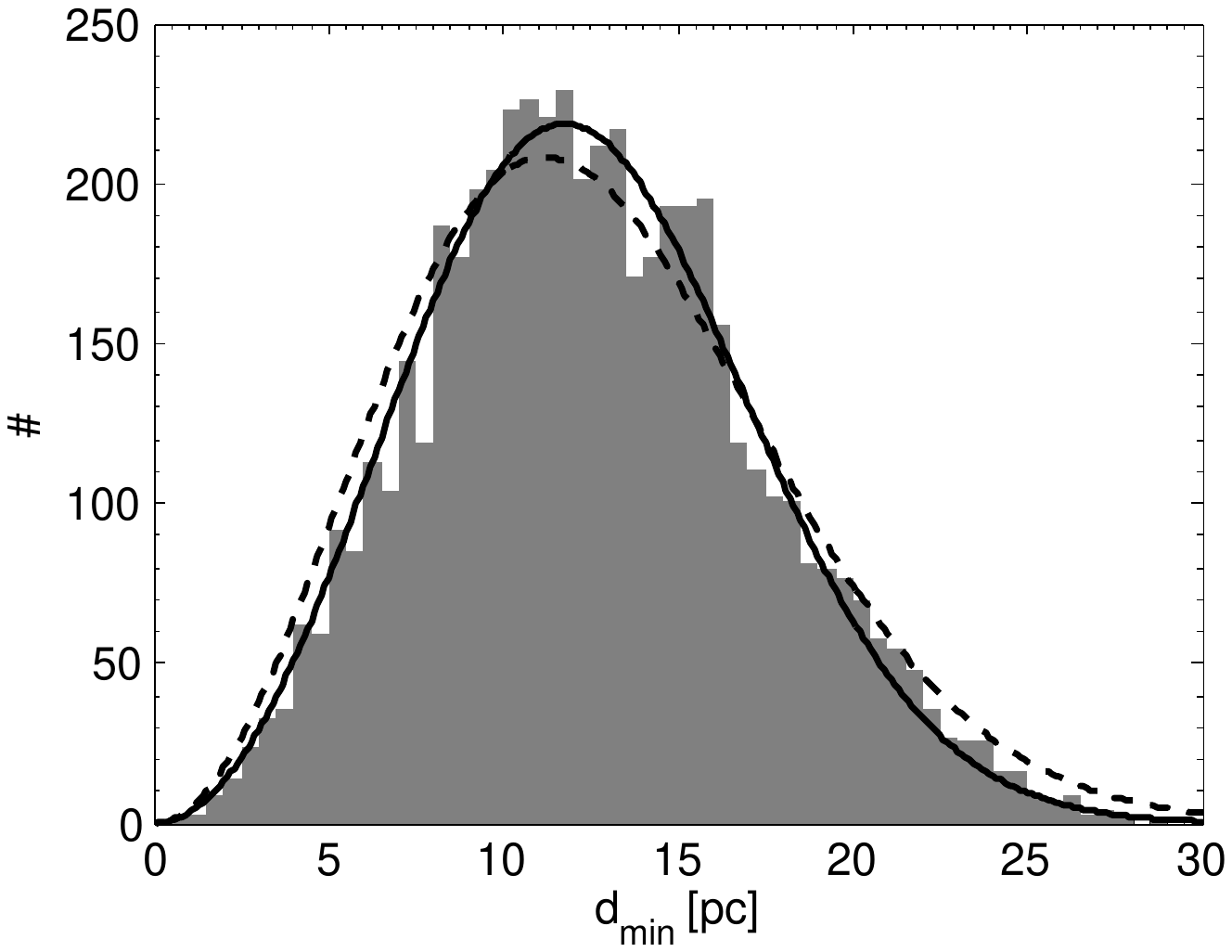}\label{subfig:B1929ZOph_hists_dist}}\nolinebreak
\subfigure[]{\includegraphics[width=0.25\textwidth, viewport= 85 265 480 590]{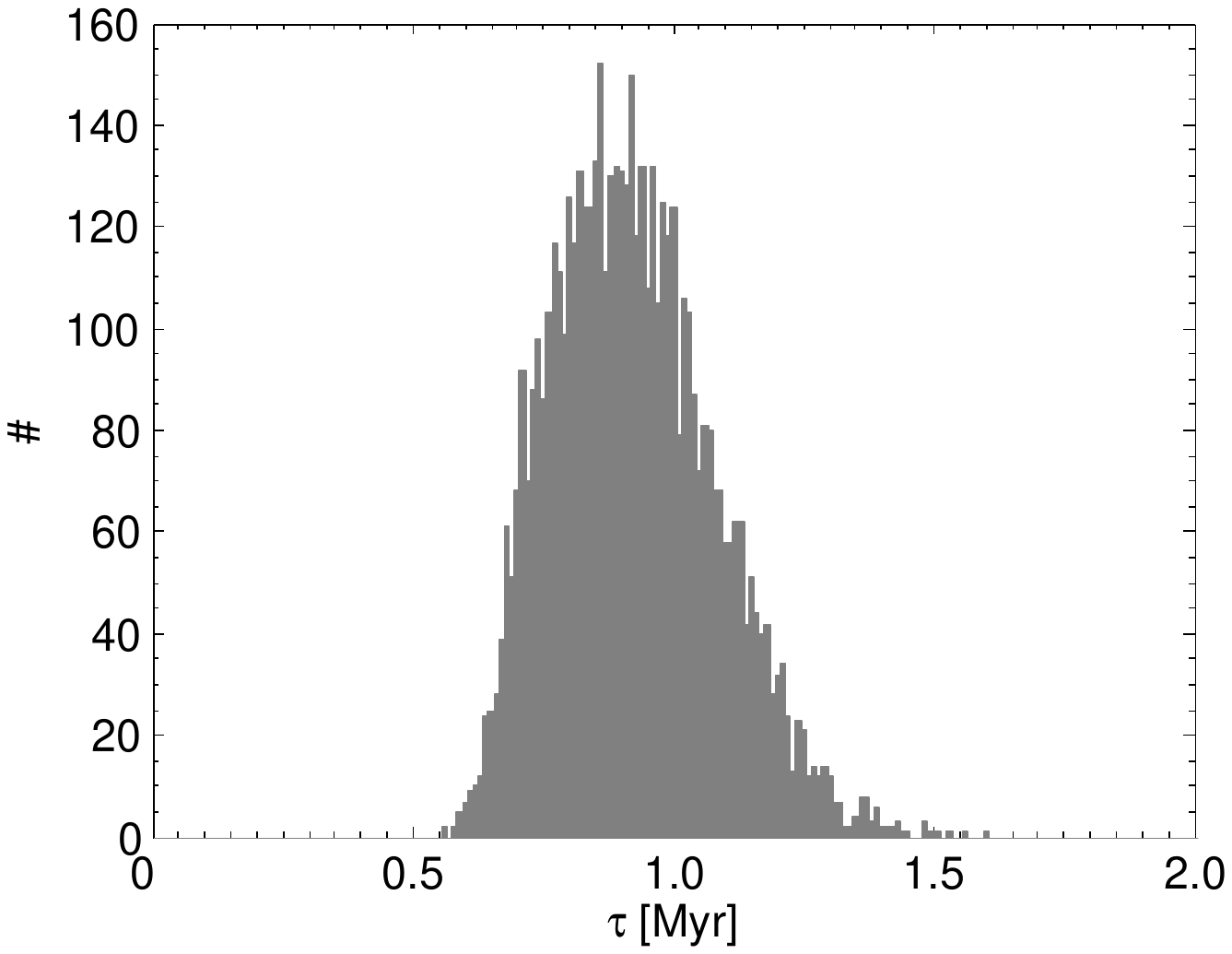}\label{subfig:B1929ZOph_hists_Zeit}}
\caption{(a) Distribution of minimum separations $d_{min}$ of the $5,367$ runs for which both objects were not farther than $\unit[10]{pc}$ from the US centre with updated pulsar data. Drawn as well are theoretical curves for three-dimensional Gaussian distributions (\autorefs{eq:3DGauss_diff} and \ref{eq:3DGaussmunull_diff}) with $\mu~=\unit[9{.}0]{pc}$ and $\sigma~=\unit[4{.}0]{pc}$ (solid) and $\mu~=0$ and $\sigma~=\unit[5{.}6]{pc}$ (dashed), respectively. Note that these are not fitted to the data. (b) Distribution of corresponding flight times $\tau$ in the past since the supernova.}
\label{fig:B1929ZOph_hists}
\end{figure}

\begin{figure}
\includegraphics[width=0.45\textwidth, viewport= 85 265 480 590]{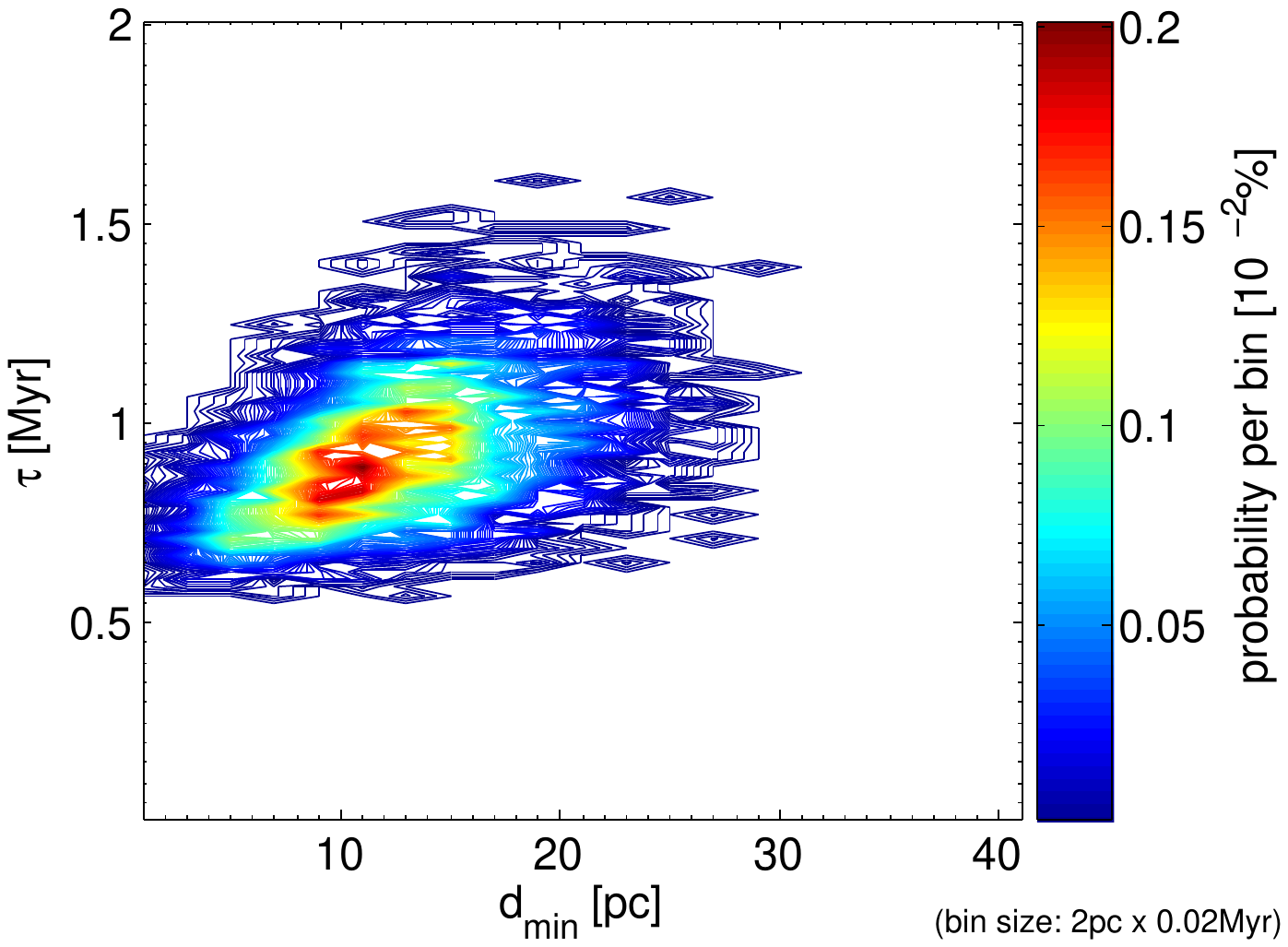}
\caption{$\tau$-$d_{min}$ contour plot to \autoref{fig:B1929ZOph_hists}.}
\label{fig:B1929ZOph_contour}
\end{figure}

We reinvestigate the issue as well using a radial velocity distribution (\autoref{sec:procedure}) instead of a specified interval. In 3 million Monte-Carlo runs the smallest separation found was $\unit[4{.}0]{pc}$. This is too large to support the hypothesis that $\zeta$ Oph and PSR B1929+10 once were at the same position at the same time in the past. However, the errors of the parallax and the proper motion components of the pulsar (\autoref{tab:1929ZOph_par}) are very small and thus might be underestimated due to unknown systematical effects. For that reason we as well increase them by a factor of ten. The smallest difference in position then was $\unit[0{.}2]{pc}$ which is consistent with the binary supernova scenario. We select those minimum separations for which both, the pulsar and the runaway, were not more than $\unit[15]{pc}$ away from the centre of US. Of 3 million runs $5{,}367$ fulfilled this requirement. The slope of the resulting histogram in \autoref{subfig:B1929ZOph_hists_dist} can be explained with a three-dimensional Gaussian distribution of the separation (see H01 for details),

\begin{equation}
W_{3D}\left(\Delta\right) = \frac{\Delta}{2\sqrt{\pi}\sigma\mu}\left\{exp\left[-\frac{1}{2}\frac{\left(\Delta-\mu\right)^2}{2\sigma^2}\right]-exp\left[-\frac{1}{2}\frac{\left(\Delta+\mu\right)^2}{2\sigma^2}\right]\right\}
\label{eq:3DGauss_diff}
\end{equation}

\noindent and in the limit $\mu \rightarrow 0$ 

\begin{equation}
W_{3D,\mu\to0}\left(\Delta\right) = \frac{\Delta^2}{2\sqrt{\pi}\sigma^3}\exp\left[-\frac{\Delta^2}{4\sigma^2}\right],
\label{eq:3DGaussmunull_diff}
\end{equation}

\noindent respectively, either with $\mu=\unit[0\pm5{.}6]{pc}$ or $\mu=\unit[9{.}0\pm4{.}0]{pc}$. $\Delta$ denotes the three-dimensional separation $|{x_{pulsar}}-{x_{\zeta Oph}}|$ between the pulsar and the runaway, $\mu$ and $\sigma$ are the expectation value and the standard deviation, respectively, and $\pi=3{.}1459\ldots$. The agreement to the curves is remarkable. \autoref{fig:B1929ZOph_contour} shows the correlation between minimum separations and corresponding times. It can be seen that there is a small spread in time for larger separations which is owing to the deviations from a Gaussian distribution and the actual four dimensional problem since the time is involved as well.\\

Adopting that PSR B1929+10 and $\zeta$ Oph were ejected in the same supernova about $\unit[1]{Myr}$ ago in US, we find that the pulsar would now have a radial velocity of $\approx\unit[250]{km/s}$ which is close to the value of $\unit[200\pm50]{km/s}$ H01 adopted previously. 
The distance to the Sun of this supernova event $\unit[1]{Myr}$ ago was $\unit[154\pm7]{pc}$ (note that this error quotes a 68\% confidence interval but is not a $1\sigma$ error, though).\\
Given an age of $\unit[5]{Myr}$ for US \citep{1989A&A...216...44D,2002AJ....124..404P} we can estimate the mass of the progenitor star of PSR B1929+10 assuming that star formation happened contemporaneously, i.e that the progenitor exploded $\unit[(5-1)]{Myr}$ after birth. Evolutionary models from \citet{1980FCPh....5..287T} (thereafter T80), \citet{1989A&A...210..155M} (thereafter MM89) and \citet{1997PhDT........31K} (thereafter K97) (see also \citet{2005A&A...430..491R} for a review) yield a progenitor mass of $\unit[44]{M_{\odot}}$, $\unit[38]{M_{\odot}}$ and $\unit[32]{M_{\odot}}$, respectively, for a star with a lifetime of $\unit[4]{Myr}$ assuming solar metallicity, corresponding to a spectral type of O6 to O7 on the main sequence \citep{Schmidt-Kaler1982}. According to \citet{2003A&A...404..913S} US is somewhat older, namely $8$ to $\unit[10]{Myr}$. This association age range would imply a smaller progenitor mass of $18$ (B0V) to $\unit[37]{M_\odot}$ (O6V) (depending on the evolutionary model) which is in better agreement to the upper star mass limit of $\approx\unit[30]{M_\odot}$ \citep{2003ApJ...591..288H} for formation of a neutron star. However, owing to mass transfer during binary evolution also more massive stars may produce neutron stars \citep{2008ApJ...685..400B}. \\
The O9.5V star $\zeta$ Oph has a mass of $\unit[20]{M_{\odot}}$ (derived from isochrones by \citet{1992A&AS...96..269S}, \citet{1994A&AS..106..275B} and \citet{2004A&A...424..919C} for solar metallicity; table in \citealt{Schmidt-Kaler1982}).\\
If $\zeta$ Oph was ejected in a supernova, the progenitor star should have an earlier spectral type, consistent with our results. With the additional constraint that the supernova produced a neutron star, i.e. the progenitor was later than $\approx$O6 to O7, the progenitor star was O6/O7 to O9.\\

It may not be justified to enlarge the errors of the input parameters of the pulsar. With the published small error bars we cannot confirm that $\zeta$ Oph was at the same position in space at the same time as PSR B1929+10, but may have been ejected by a different supernova event in US.


\section{Identifying parent associations and clusters for young isolated neutron stars}\label{sec:idparassoc}

\begin{table*}
\centering
\caption{Parameters of four of the M7. Note that only for \rxja{} and RX J0720.4-3125 $\pi$ denotes trigonometric parallax measurements whereas for RX J1605.3+3249 and RBS 1223 distance limits are model estimates (from brightness and extinction compared to \rxja{} and \rxjn{}).\newline
References: 1 -- \citet{1999A&A...351..177M}, 2 -- \citet{2002ApJ...571..447K}, 3 -- \citet{2003conf...Kaplan}, 4 -- \citet{2005A&A...441..597S}, 5 -- \citet{2006A&A...457..619Z},  6 -- \citet{2007Ap&SS.308..619S} (their model 3), V. Hambaryan (priv. comm.), 7 -- \citet{2007Ap&SS.308..171P}, 8 -- \citet{2007ApJ...660.1428K}, 9 -- \citet{2007astph.712..342M}, 10 -- \citet{2009A&A...498..811H}.
}\label{tab:vierM7_Eigensch}
\begin{tabular}{l d{2} d{2} c o{3.2} o{3.2} o{3.2} c}
\toprule
	&		\multicolumn{1}{c}{$\unit[\alpha]{[^\circ]}$}		&	\multicolumn{1}{c}{$\unit[\delta]{[^\circ]}$}	&	\multicolumn{1}{c}{$\unit[\pi]{[mas]}$} & \multicolumn{1}{c}{$\unit[\mu_{\alpha}^*]{[mas/yr]}$} & \multicolumn{1}{c}{$\unit[\mu_{\delta}]{[mas/yr]}$} & \multicolumn{1}{c}{$\unit[n_H]{[10^{20}\,cm^{-2}]}$} & Ref.\\\midrule
	
\rxja{} & 284,15 &		-37,91 &  5{.}6$\pm$0{.}6		&	326{.}7+0{.}8	& -59{.}1+0{.}7 & 0{.}74+0{.}10 & 2, 3, 7\\
RX J0720.4-3125 & 110,10	&		-31,43 &	2{.}77$\pm$1{.}29	&	-93{.}9+2{.}2	&	52{.}8+2{.}3 & 1{.}04+0{.}02 & 8, 10\\
RX J1605.3+3249  & 231,33	&		32,82  &	$>$2{.}4						&	-43{.}7+1{.}7	&	148{.}7+2{.}6 & 1{.}1+0{.}4 & 1, 5, 9\\
RBS 1223			  & 197,20	&		21,45  &	1{.}4 \ldots 13		&	-207+20				&	84+20				 & 1{.}8+0{.}2 & 4, 6, 9\\\bottomrule
\end{tabular}
\end{table*}

For four of the M7 distance estimates and proper motion values are available and thus their birth sites can in principle be found if assumptions on the radial velocity are made. \Autoref{tab:vierM7_Eigensch} gives their properties.\\

We trace back the objects for $\unit[5]{Myr}$. As a first step we perform $100{,}000$ runs for each neutron star and all 140 associations in the sample. Those associations for which the smallest separation found was less than three times the association radius or less than $\unit[100]{pc}$ were chosen for a more detailed investigation with 2 million runs. We give the results of the latter in the following sections.


\subsection{\rxja{}}\label{subsec:RXJ1856}

Investigating probable birth places of \rxja{}, nine of the 140 associations yield a smallest separation between the neutron star and the association centre found after another 2 million runs which was consistent with the respective association radius\footnote{Considering the present radius of an association ensures that we do not miss potential supernovae that might have occured near the edge. The former radius was surely smaller than the present one.} (see \autoref{tab:1856_smallestsep}). However, 
only five of the other associations show a peak in the $\tau$-$d_{min}$ contour plot which also lies within the association boundaries or at least intersects. \Autoref{tab:detailed_scan_1856} lists the respective associations along with the results of the investigation including 2 million runs with varying input parameters. Columns 2 and 3 of that table specify a region in the $\tau$-$d_{min}$ contour diagram with higher probability than its surroundings (see \autoref{appsec:colval}). 
Columns 4 to 8 give values for the present-day properties which \rxja{} would have if it originated from the particular association. The space velocity $v_{space}$ which corresponds to the ejection speed of the neutron star in the supernova is given in column 7. Assuming that the ejection speed is of the same amount as the kick velocity, this value gives an upper limit of the latter.\footnote{The kick velocity strongly depends on whether the progenitor star was single or not and in the case of binarity also on the separation between the stars. However, supernova kick theory predicts kick velocity distributions which are consistent with observed neutron star velocities \citep[e.g.][]{2006A&A...457..963S,2007ApJ...660.1357N,2009MNRAS.395.2087K}.} \\
The last three columns indicate the position of the potential supernova at the corresponding time with the distance to the Sun and equatorial coordinates. For how those values are derived, please see \autoref{appsec:colval}.\\

\begin{table}
\centering
\caption{Associations for which the smallest separation between \rxja{} and the association centre found (min. $d_{min}$) after 2 million runs was within the radius of the association. Column 2 gives the smallest separation, column 3 the radius of the association (see Appendix A for references).}\label{tab:1856_smallestsep}
\begin{tabular}{l d{2.2} d{2.2}}
\toprule
Association	&		\multicolumn{1}{c}{min. $d_{min}$ [pc]} & \multicolumn{1}{c}{$R_{Assoc}$ [pc]}\\\midrule
US					&		0,4																			& 15\\
Tuc-Hor			&		46,1																		& 50\\
$\beta$ Pic-Cap & 26,3																	& 57\\
Ext. R CrA  &   4,4																			& 31\\
AB Dor			&		32,0																		& 45\\
Her-Lyr			&		10,4																		& 13\\
Sgr OB5			&		64,1																		&	111\\
Sco OB4			&   19,0																		& 33\\
Pismis 24		&   1,6																			& 2\\
Tr 27				&   1,8																			& 6\\
\bottomrule
\end{tabular}
\end{table}

From \autoref{tab:detailed_scan_1856} we conclude that the most likely parent association for \rxja{} is either Upper Scorpius (US), the extended Corona-Australis association (Ext. R CrA) or Scorpius OB4 (Sco OB4). For the two remaining -- the $\beta$ Pictoris group ($\beta$ Pic-Cap) and AB Doradus (AB Dor) -- the radial needed velocities would be excessively high ($>\unit[1000]{km/s}$, space velocities $v_{space}>\unit[1000]{km/s}$), thus making them unlikely (see Ho05) to host the birth place of \rxja{}. The probability of those two associations of being the birth associations is more than $4,000$ times smaller than for the other three (for deviation of the probability, please see \autoref{appsec:colval}).\\

\begin{table*}
\centering
\caption{Potential parent associations of \rxja{}.\newline
Columns 2 and 3 mark the boundaries of a 68\% area in the $\tau$-$d_{min}$ contour plot for which the current neutron star parameters (columns 4 to 7, radial velocity $v_r$, proper motion $\mu^*_{\alpha}$ and $\mu_{\delta}$ and parallax $\pi$) were obtained and columns 8 to 10 indicate the distance to the Sun $d_{\odot}$ and equatorial coordinates (J2000.0) of the potential supernova. Column 7 gives the space velocity (ejection speed) $v_{space}$ derived from proper motion and radial velocity. For the deduction of the values given in columns 4 to 11, please see \autoref{appsec:colval}.}
\label{tab:detailed_scan_1856}
\setlength\extrarowheight{3pt}
\small
\begin{tabular}{c f{2.2} f{4.4} >{$}r<{$} o{4.2} o{4.2} >{$}r<{$} >{$}r<{$} >{$}c<{$} >{$}r<{$} >{$}r<{$}}
\toprule
Association	&		\multicolumn{1}{c}{$d_{min}$}	&	\multicolumn{1}{c}{$\tau$}		&	\multicolumn{1}{c}{$v_r$}			& \multicolumn{1}{c}{$\mu_{\alpha}^*$} & \multicolumn{1}{c}{$\mu_{\delta}$} & \multicolumn{1}{c}{$v_{space}$} &	\multicolumn{1}{c}{$\pi$}						&	d_{\odot} 	& \multicolumn{1}{c}{$\alpha$}	& \multicolumn{1}{c}{$\delta$}\\ 
						&	\multicolumn{1}{c}{[pc]} & \multicolumn{1}{c}{[Myr]} & \multicolumn{1}{c}{[km/s]} & \multicolumn{1}{c}{[mas/yr]} & \multicolumn{1}{c}{[mas/yr]} & \multicolumn{1}{c}{[km/s]} & \multicolumn{1}{c}{[mas]} & \multicolumn{1}{c}{[pc]} & \multicolumn{1}{c}{[$^\circ$]} & \multicolumn{1}{c}{[$^\circ$]}\\\midrule
US														&  5p13				&  0{.}28p0{.}34				& 193^{+45}_{-32} & 326{.}7+0{.}8 & -59{.}1+0{.}7 & 349^{+40}_{-32} & 5{.}41^{+0{.}33}_{-0{.}28}	& 140\ldots163			& 243{.}21^{+0{.}56}_{-0{.}52}	& -23{.}66^{+0{.}31}_{-0{.}29}\\
$\beta$ Pic-Cap 					&  35p53			& 0{.}10p0{.}15	& 1190^{+167}_{-172}& 326{.}7+0{.}8	& -59{.}1+0{.}7	& 1222^{+165}_{-173} & 5{.}71^{+0{.}49}_{-0{.}24}& 40^{+5}_{-5}			& 221\ldots233	& -18\ldots-8\\
Ext. R CrA 								&  26p54			&  0{.}10p0{.}15					& 406^{+121}_{-88}	& 326{.}7+0{.}8	& -59{.}1+0{.}7	& 492^{+113}_{-81} & 5{.}68^{+0{.}35}_{-0{.}37}& 127^{+6}_{-6}	& 256{.}76^{+2{.}06}_{-2{.}02}	& -33{.}72^{+0{.}63}_{-0{.}61}\\
AB Dor											& 44p60			& 0{.}12p0{.}16					& 1146^{+138}_{-152}	& 326{.}7+0{.}8	& -59{.}1+0{.}7	& 1177^{+137}_{-151} &
5{.}88^{+0{.}30}_{-0{.}27}& 32\ldots46		& 210{.}19^{+6{.}41}_{-4{.}13}	& 0{.}90^{+2{.}80}_{-5{.}72}\\
Sco OB4										&  25p33			&  1{.}26p1{.}51			&	-604^{+31}_{-20}	& 326{.}7+0{.}8	& -59{.}1+0{.}7	& 664^{+25}_{-35} &
5{.}72^{+0{.}32}_{-0{.}30}& 1092^{+2}_{-2}& 259{.}84^{+0{.}69}_{-0{.}69}	& -31{.}66^{+0{.}85}_{-0{.}34}\\
\bottomrule
\end{tabular}
\end{table*}

\begin{figure}
\includegraphics[width=0.45\textwidth, viewport= 85 265 480 590]{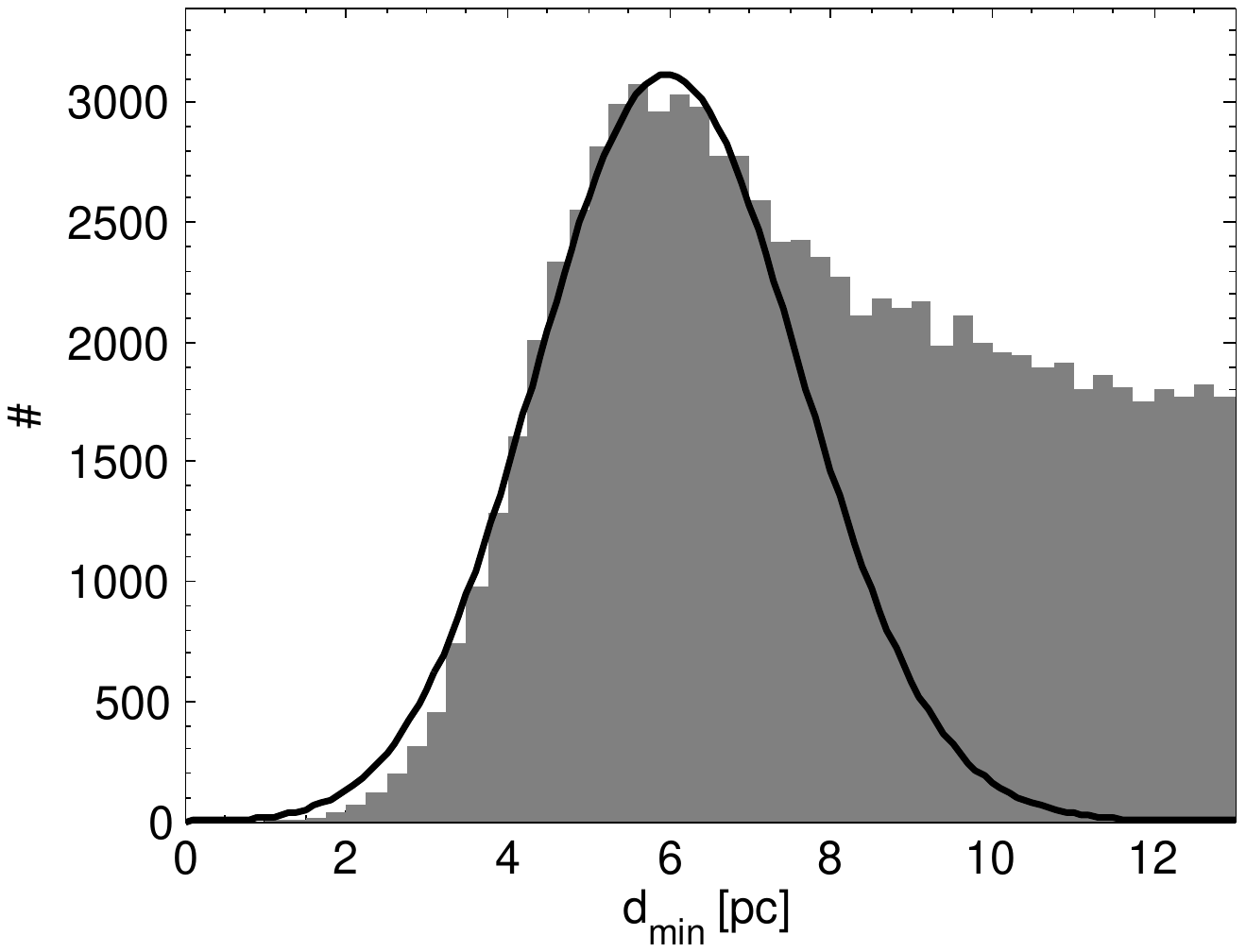}
\caption{Distribution of minimum separations $d_{min}$ between \rxja{} and the centre of US within the defined time range $\tau$ (see \autoref{tab:detailed_scan_1856}, column 3). Shown as well is a theoretical curve for a three-dimensional Gaussian distribution (\autoref{eq:3DGauss_diff}) with $\mu~=\unit[5{.}5]{pc}$ and $\sigma~=\unit[1{.}2]{pc}$. Note that the curve is not a fit to the data but shall just give an explanation of the slope of the histogram.\newline The disagreement on the right side of the diagram is due to the actual four dimensional problem since the time is involved. Furthermore, natural discrepancies to a Gaussian distribution occur since parallaxes are taken to obtain positions and a Maxwellian distribution is utilised for the velocities. Unlike in the previous section we cannot restrict to a certain number of runs owing to a third component (in the case of PSR B1929/ $\zeta$ Oph the distance to US) which gives constraints on the time.}
\label{fig:hist_1856_US}
\end{figure}

\begin{figure}
\centering
\includegraphics[clip, width=0.3\textwidth, viewport= 137 7 623 539]{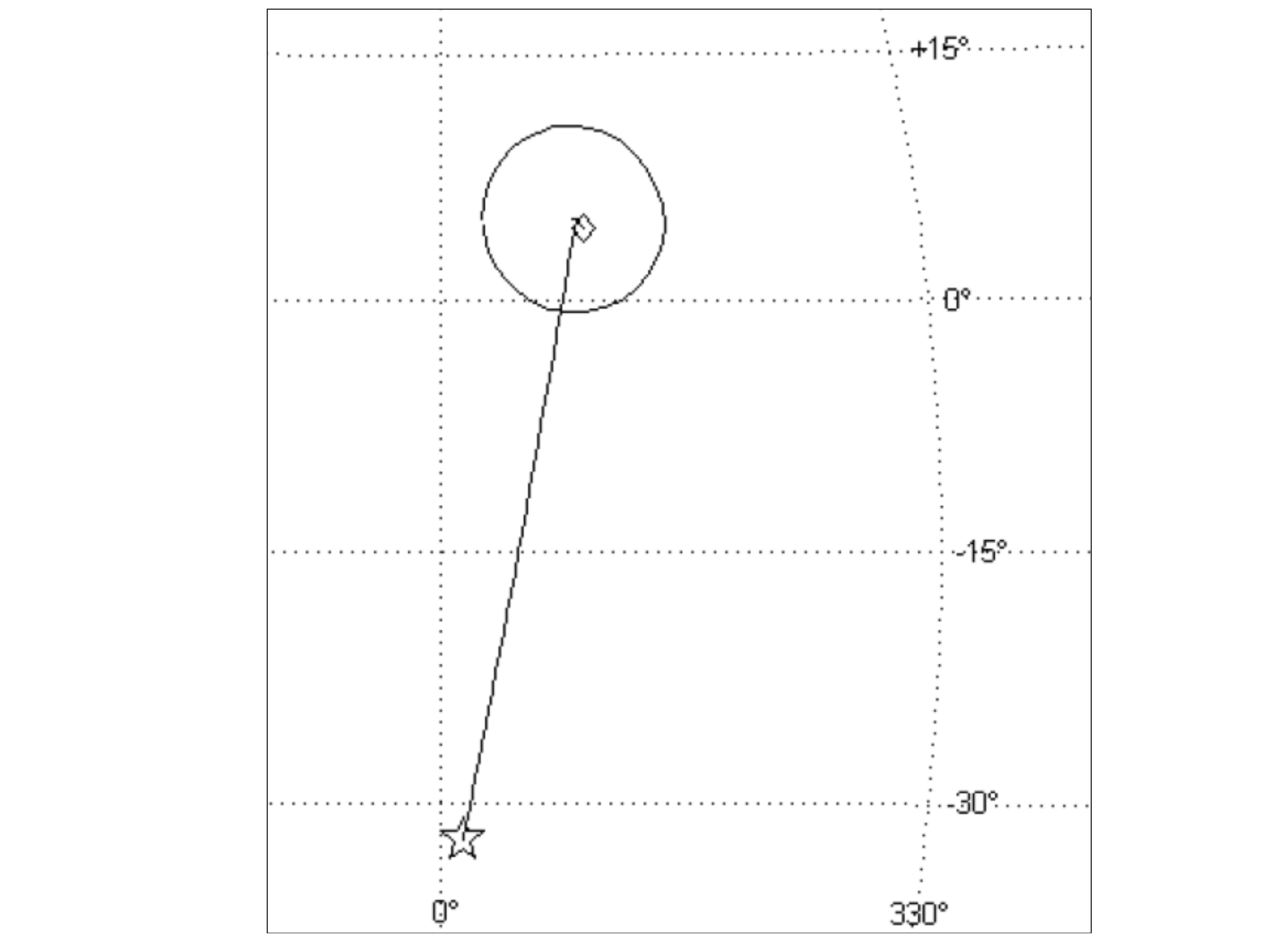}
\caption{Past trajectories for \rxja{} and US projected on a Galactic coordinate system (for a particular set of input parameters consistent with \autoref{tab:detailed_scan_1856}). Present positions are marked with a star for the neutron star and a diamond for the association. The large circle reflects an association radius of $\unit[15]{pc}$.}
\label{fig:1856US_traj}
\end{figure}

Since the smallest separation found for Sco OB4 ($\unit[19]{pc}$) is relatively high compared to those of US ($<\unit[1]{pc}$) and Ext. R CrA ($<\unit[5]{pc}$), and the modulus of the radial velocity needed of $\approx\unit[600]{km/s}$ (column 4 of \autoref{tab:detailed_scan_1856}) is rather high (space velocity $\approx\unit[665]{km/s}$), Sco OB4 may be less likely the birth place of \rxja{}. For Ext. R CrA the potential place of the supernova lies between 26 to $\unit[44]{pc}$ away from the association centre. The radius of Ext. R CrA is $\unit[31]{pc}$, thus the supernova should have occurred near the edge of the association. In comparison to that, a supernova in US which produced \rxja{} would have taken place at $\unit[5{.}5\pm1{.}2]{pc}$ from the centre of US (\autoref{fig:hist_1856_US}) which has a radius of $\unit[15]{pc}$. As supernovae are supposed to occur in denser regions of an association, the latter scenario thus seems most likely. Furthermore, the probability of US being the parent association of \rxja{} is $10$ times larger than that for Ext. R CrA and $8$ times larger than for Sco OB4.\\
The origin of \rxja{} has also been previously suggested by \citet{2002ApJ...576L.145W} to lie within the US association $\approx\unit[0{.}5]{Myr}$ ago as the projected trajectory of \rxja{} crosses the projection of US. Here, we find a kinematic age of $\approx\unit[0{.}3]{Myr}$.\\
In this case (\autoref{fig:1856US_traj}) the neutron star would now have a radial velocity of $\approx\unit[190]{km/s}$. The space velocity of $\approx\unit[340]{km/s}$ in this case is close to the peak of the velocity distribution for pulsars from Ho05.\\

\citet{2000IAUS..195..437W} suggested that $\zeta$ Oph was the former secondary of \rxja{} that exploded a few Myr ago in US. Therefore, we re-investigate the separation between \rxja{} and the runaway star as well as their separation to the centre of US. Out of 3 million runs the smallest separation found between the neutron star and $\zeta$ Oph was $\unit[22]{pc}$ which is far too large to support this former binary hypothesis.\\

Given the age of US of $\unit[5]{Myr}$ \citep{1989A&A...216...44D} we estimate the mass of the progenitor star (lifetime $\unit[4{.}7]{Myr}$) to be, depending on the model, $\unit[45]{M_{\odot}}$ (T80), $\unit[41]{M_{\odot}}$ (MM89) and $\unit[35]{M_{\odot}}$ (K97) (spectral type $\approx$O6.5, \citealt{Schmidt-Kaler1982}). 
Using the more recent age determination for the Sco-Cen associations by \citet{2003A&A...404..913S} of $8$ to $\unit[10]{Myr}$ for US, we estimate the mass of the progenitor star of \rxja{} to be between $17$ and $\unit[35]{M_{\odot}}$ (B0 to O7 on the main sequence). Currently, the earliest spectral type in US is B0. Hence, our assumption of contemporary star formation and our result on the progenitor spectral type are consistent with the progenitor star of \rxja{} being earlier than the earliest present member star.\\

Since \citet{2002ApJ...576L.145W} published a significant different parallax for \rxja{} of $\pi = \unit[8{.}5\pm0{.}9]{mas}$, we repeat our simulations adopting this value. The overall picture of potential parent associations remains (cf. \autoref{tab:1856_smallestsep}\footnote{We find additional small separations for Bochum 13 and NGC 6383; however, the fraction of such runs is very small.}). 
Still, US is the most probable association to have hosted the supernova which formed \rxja{}. While the position of this supernova using Walter's parallax does not significantly change, the neutron star's current parallax would be $\pi = \unit[7{.}44^{+1{.}27}_{-0{.}48}]{mas}$ and the radial velocity needed $v_r = \unit[-25^{+58}_{-24}]{km/s}$. Furthermore, the kinematic age would be slightly larger, $\tau\approx\unit[0{.}5]{Myr}$ (confirming the result of \citealt{2002ApJ...576L.145W}), implying a progenitor mass of $37$ to $\unit[45]{M_\odot}$ ($\approx$O6.5 on the main sequence) adopting an age of US of $\unit[5]{Myr}$ or $17$ to $\unit[29]{M_\odot}$ (B0 to O7 on the main sequence) for an association age of $\unit[10]{Myr}$ (T80, MM89, K97).\\
We again investigated the \rxja{}/ $\zeta$ Oph binary scenario using the larger parallax as an input. Out of 2 million runs we found a smallest separation of $\unit[20]{pc}$ ruling out this hypothesis.

\subsection{RX J0720.4-3125}

For RX J0720.4-3125, 12 associations were identified for which the smallest separation between the neutron star and the association centre found in the past after 2 million runs lied within the respective association boundaries (see \autoref{tab:0720_smallestsep}). Five of those show a peak in the $\tau$-$d_{min}$ contour plot at values of $d_{min}$ that at least intersect the association. For the Chamaeleon-T association (Cha T) only a very small fraction of runs yield small distances, thus we exclude it as a probable parent association for \rxjn{}. Analogues to \autoref{tab:detailed_scan_1856} for \rxja{}, they are given in \autoref{tab:detailed_scan_0720} along with the separation and corresponding time ranges in the past defined by the peak in the $\tau$-$d_{min}$ contour diagram. Also given are the neutron star parameters needed as well as the position of the potential supernova.\\ 

\begin{table}
\centering
\caption{Associations for which the smallest separation between RX J0720.4-3125 and the association centre found (min. $d_{min}$) after 2 million runs was within the radius of the association, as \autoref{tab:1856_smallestsep}.}\label{tab:0720_smallestsep}
\begin{tabular}{l d{2.2} d{2.2}}
\toprule
Association	&		\multicolumn{1}{c}{min. $d_{min}$ [pc]} & \multicolumn{1}{c}{$R_{Assoc}$ [pc]}\\\midrule
US					&		11,8																		& 15\\
UCL					&   27,5																		& 33\\
LCC					&		18,2																		& 23\\
TWA					&		0,1																			& 33\\
Tuc-Hor			&		37,6																		& 50\\
$\beta$ Pic-Cap & 18,2																	& 57\\
HD 141569   &   4,7																			& 16\\
Pup OB1			&		6,0																			& 65\\
Pup OB3			&		0,5																			& 15\\
NGC 2546		&   6,6																			& 10\\
Tr 10  			&		17,3																		& 23\\
Cha T				&		0,8																			&	21\\\bottomrule
\end{tabular}
\end{table}

\begin{figure}
\subfigure[TWA]{\includegraphics[width=0.25\textwidth, viewport= 85 265 480 590]{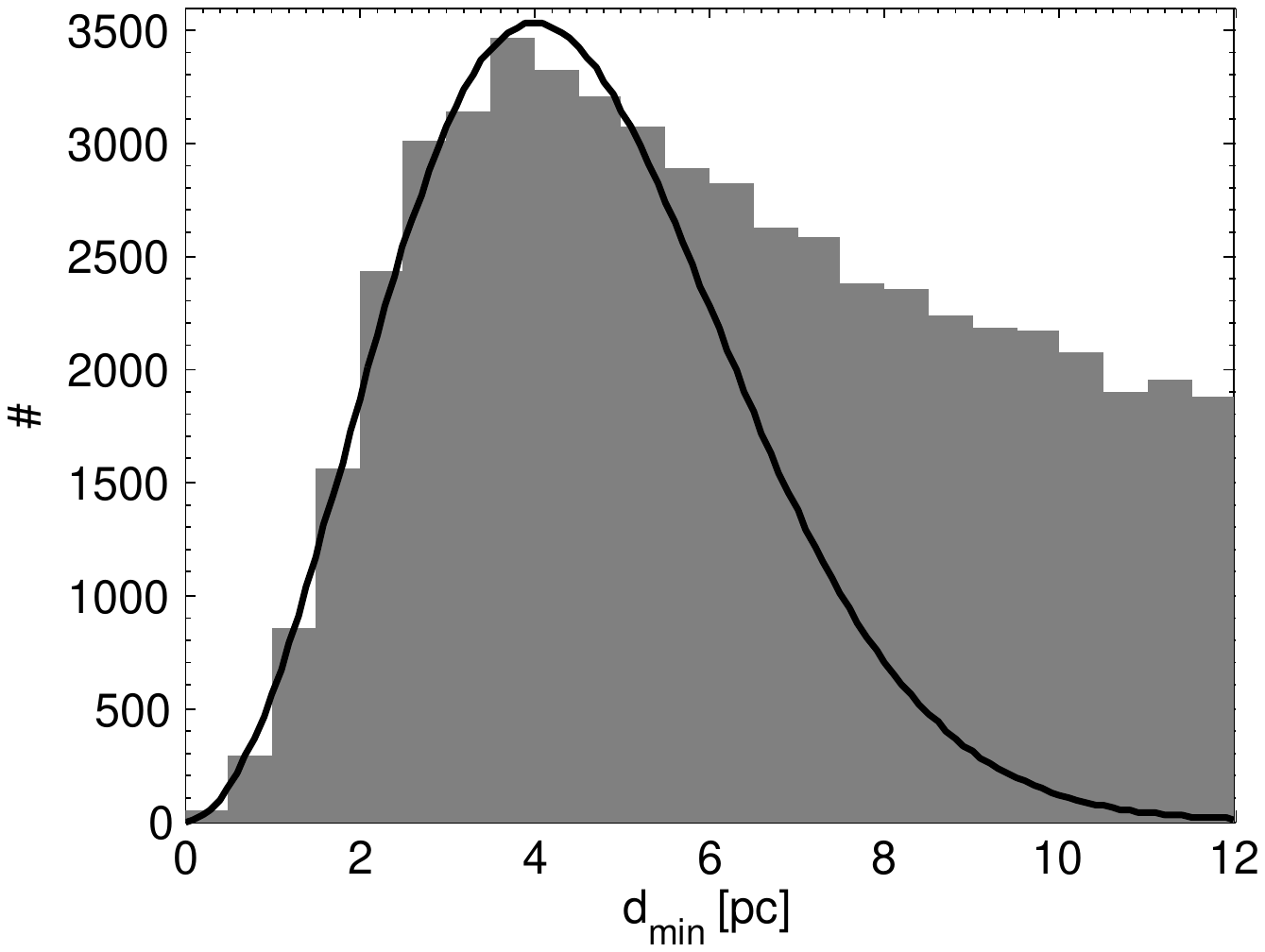}\label{subfig:0720_TWHya_hist}}\nolinebreak
\subfigure[HD 141569]{\includegraphics[width=0.25\textwidth, viewport= 85 265 480 590]{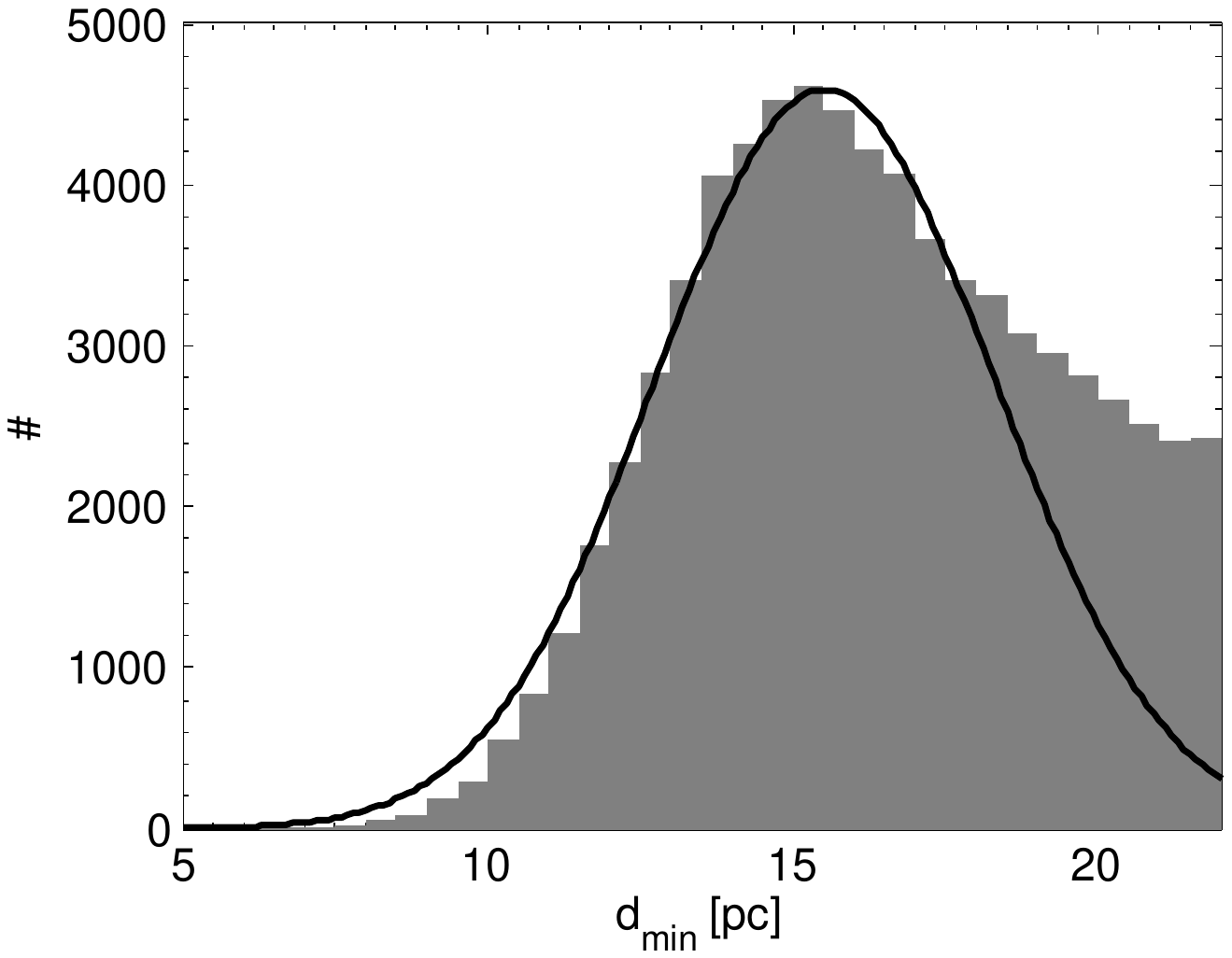}\label{subfig:0720_HD141569_hist}}\\
\subfigure[Tr 10]{\includegraphics[width=0.25\textwidth, viewport= 85 265 480 590]{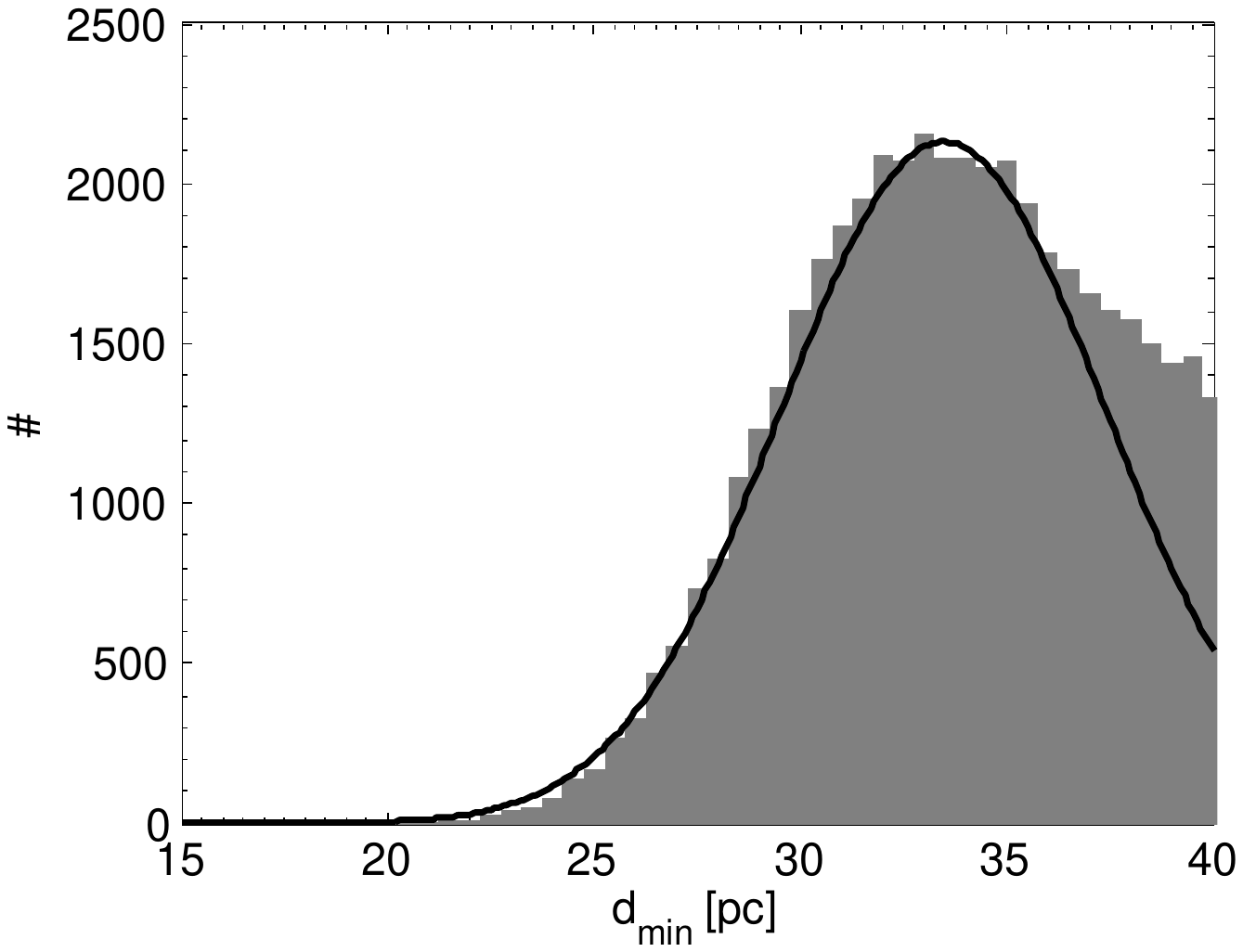}\label{subfig:0720_Tr10_hist}}\\
\caption{(a) Distribution of minimum separations $d_{min}$ between RX J0720.4-3125 and the centres of TWA (a), HD 141569 (b) and Tr 10 (c), respectively, within the defined time range $\tau$ (see \autoref{tab:detailed_scan_0720}, column 3). Shown as well are theoretical curves for three-dimensional Gaussian distributions (\autoref{eq:3DGauss_diff}) with $\mu~=\unit[0]{pc}$ and $\sigma~=\unit[2]{pc}$ for TWA, $\mu~=\unit[15]{pc}$ and $\sigma~=\unit[2]{pc}$ for HD 141569 and $\mu~=\unit[33]{pc}$ and $\sigma~=\unit[2{.}8]{pc}$ for Tr 10. Note that the curves are not fitted to the data (see also \autoref{fig:hist_1856_US}).}
\label{fig:0720_hists}
\end{figure}

\begin{figure}
\centering
\includegraphics[clip, width=0.4\textwidth, viewport= 7 18 720 522]{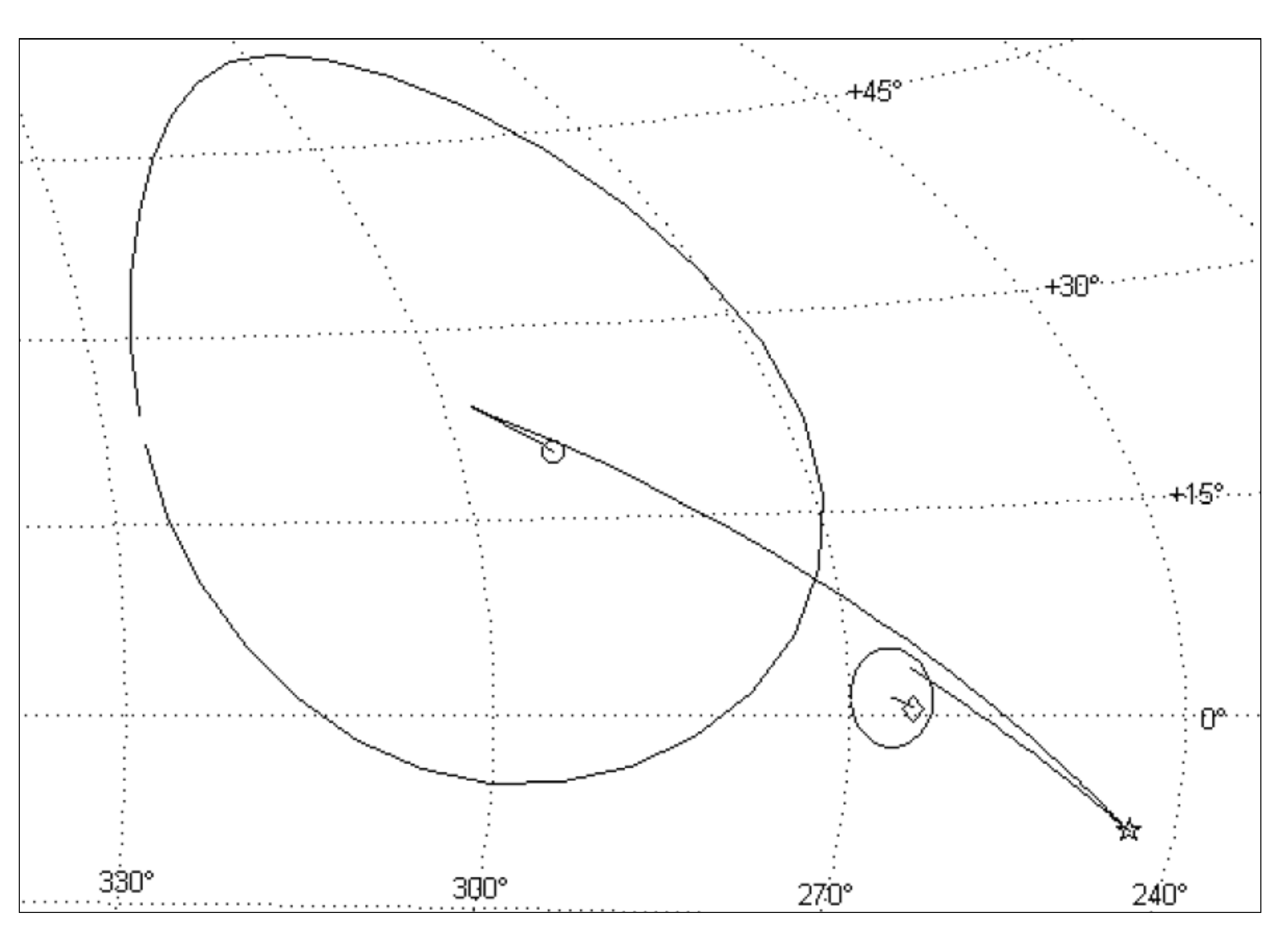}
\caption{Past trajectories for RX J0720.4-3125 and Tr 10 and TWA, respectively, projected on a Galactic coordinate system (for particular sets of input parameters consistent with \autoref{tab:detailed_scan_0720}). Present positions are marked with a star for the neutron star and a diamond for Tr 10 and an open circle for TWA. Large circles reflect association extensions (radii of $\unit[23]{pc}$ for Tr 10 and $\unit[33]{pc}$ for TWA).}
\label{fig:0720Tr10TWA_traj}
\end{figure}

It has been previously suggested that RX J0720.4-3125 originated from the Trumpler 10 (Tr 10) association by \citet{2003A&A...408..323M} who considered the general direction of the neutron star's motion and \citet{2007ApJ...660.1428K} who investigated the probability of close approaches of the neutron star to any of the OB associations given in \citet{1999AJ....117..354D} giving the proper motion randomly directions and sign as well as varying the radial velocity in a certain range and adopting specified distances. They concluded that it is very likely that Tr 10 is the birth place. Since early B-type stars are present in Tr 10, it is plausible that the association already experienced a supernova. Varying the parallax in the range $\pi\pm\sigma_{\pi}$ and the radial velocity in the range $v_r\pm0{.}935v_t$ ($0{.}935$ corresponds to $1\sigma$ in $v_r$ for random orientation, $v_t$ is the transverse velocity), \citet{2007ApJ...660.1428K} found a separation between the neutron star and the centre of Tr 10 of $\unit[17]{pc}$ $\unit[0{.}7]{Myr}$ ago for a radial velocity of $-20$ to $\unit[+50]{km/s}$. Infact, we find the smallest separation between RX J0720.4-3125 and the centre of Tr 10 to be $\unit[17]{pc}$ $\approx\unit[0{.}6]{Myr}$ in the past. We find radial velocities in the range of $-50$ to $\unit[500]{km/s}$ for which RX J0720.4-3125 would have come as close as $\unit[30]{pc}$ to the centre of Tr 10. In the $\tau$-$d_{min}$ contour plot a maximum lies at $d_{min}$ between $23$ and $\unit[40]{pc}$ and corresponding flight times $\tau$ of $0{.}44$ and $\unit[0{.}65]{Myr}$. The slope of the histogram of minimum separations $d_{min}$ within this time range (\autoref{subfig:0720_Tr10_hist}) is in good agreement with a three-dimensional Gaussian implying $d_{min}=\unit[33{.}0\pm2{.}8]{pc}$. Given a radius of Tr 10 of $\unit[23]{pc}$, this means that the potential supernova would have occurred outside the association or, if we relax the boundary condition, at least at its outer edge making the scenario less likely (the probability is two times smaller than for TWA). If we nevertheless adopt Tr 10 as the birth place of RX J0720.4-3125 (\autoref{fig:0720Tr10TWA_traj}) and consider the association age ($15$ to $\unit[35]{Myr}$), we estimate a progenitor mass of $\approx7$ to $\unit[13]{M_{\odot}}$ (models from T80, MM89 and K97), corresponding to a spectral type between B4 and B1 on the main sequence \citep{Schmidt-Kaler1982}.
\citet{2003A&A...408..323M} and \citet{2007ApJ...660.1428K} also suggest Vela OB2 (Vel OB2) as a possible birth place of \rxjn{} for which we find a minimum separation of $\unit[81{.}6]{pc}$ \citep[cf. 70\,pc stated by][]{2007ApJ...660.1428K}  which is outside the association with a radius of $\unit[35]{pc}$.\\
We exclude Pup OB3 as birth association owing to a very small parallax needed to reach the association implying a current distance to the Sun of $>\unit[1]{kpc}$. For that reason, the probability of Pup OB3 being the birth site of \rxjn{} is more than five times smaller than for TWA (and more than two times for most of the other associations listed in \autoref{tab:detailed_scan_0720}).\\

\begin{table*}
\centering
\caption{Potential parent associations of RX J0720.4-3125, columns as in \autoref{tab:detailed_scan_1856}.}
\label{tab:detailed_scan_0720}
\setlength\extrarowheight{3pt}
\small
\begin{tabular}{c f{2.2} f{4.4} >{$}r<{$} o{4.2} o{4.2} >{$}r<{$} >{$}r<{$} >{$}c<{$} >{$}r<{$} >{$}r<{$}}
\toprule
Association	&		\multicolumn{1}{c}{$d_{min}$}	&	\multicolumn{1}{c}{$\tau$}		&	\multicolumn{1}{c}{$v_r$}			& \multicolumn{1}{c}{$\mu_{\alpha}^*$} & \multicolumn{1}{c}{$\mu_{\delta}$} & \multicolumn{1}{c}{$v_{space}$} &	\multicolumn{1}{c}{$\pi$}						&	d_{\odot} 	& \multicolumn{1}{c}{$\alpha$}	& \multicolumn{1}{c}{$\delta$}\\ 
						&	\multicolumn{1}{c}{[pc]} & \multicolumn{1}{c}{[Myr]} & \multicolumn{1}{c}{[km/s]} & \multicolumn{1}{c}{[mas/yr]} & \multicolumn{1}{c}{[mas/yr]} & \multicolumn{1}{c}{[km/s]} & \multicolumn{1}{c}{[mas]} & \multicolumn{1}{c}{[pc]} & \multicolumn{1}{c}{[$^\circ$]} & \multicolumn{1}{c}{[$^\circ$]}\\\midrule
TWA 												&  0p12			&  0{.}34p0{.}46					& 502^{+111}_{-88}	& -93{.}9+2{.}2 & 52{.}8+2{.}3 & 518^{+112}_{-90} & 4{.}06^{+0{.}47}_{-0{.}49}& 48\ldots67			& 187{.}10^{+3{.}44}_{-2{.}91} & -37{.}98^{+1{.}53}_{-1{.}38}\\
Tuc-Hor										&  44p80			&  0{.}26p0{.}52				& 476^{+128}_{-73}& -93{.}8+2{.}2 & 52{.}8+2{.}3 & 492^{+129}_{-75} & 4{.}12^{+0{.}55}_{-0{.}47}& 35\ldots70		& 190{.}19^{+4{.}00}_{-4{.}97} & -37{.}29^{+1{.}30}_{-1{.}44}\\
$\beta$ Pic-Cap							&  35p84			&  0{.}31p0{.}61					& 468^{+106}_{-87}& -93{.}9+2{.}2 & 52{.}8+2{.}3	& 486^{+107}_{-89} & 3{.}94^{+0{.}53}_{-0{.}39}& 38\ldots78		& 209{.}87^{+3{.}13}_{-3{.}20}	& -28{.}52^{+2{.}04}_{-2{.}15}\\
HD 141569					&  12p22			&  0{.}52p0{.}75					& 469^{+108}_{-70}		&	-93{.}8+2{.}2	& 53{.}0+2{.}3	& 488^{+111}_{-73} & 3{.}80^{+0{.}56}_{-0{.}55}& 96\ldots116		& 240\ldots250	& -6\ldots3\\
Pup OB3			&  14p42		& 0{.}47p0{.}61			& -375^{+144}_{-150}  &  -94{.}0+2{.}1		&  52{.}9+2{.}3		& 715^{+143}_{-110} &  0{.}84^{+0.05}_{-0.05} & 1440^{+12}_{-23} & 124{.}11^{+1{.}27}_{-0{.}60}  &  -37{.}16^{+0{.}43}_{-0{.}35}\\
Tr 10						& 23p40			& 0{.}44p0{.}65					& 290^{+143}_{-110}	& -93{.}8+2{.}2	& 53{.}1+2{.}2	& 393^{+136}_{-102} & 1{.}93^{+0{.}24}_{-0{.}20}& 335\ldots375	& 135\ldots140	& -40{.}19^{+0{.}58}_{-0{.}59}\\
\bottomrule
\end{tabular}
\end{table*}

The four other associations for which we find a maximum in the $\tau$-$d_{min}$ contour plot which is located well within the association boundaries are young local associations -- TW Hydrae (TWA), Tucana/Horologium (Tuc-Hor), the $\beta$ Pictoris moving group ($\beta$ Pic-Cap) and HD 141569. For two of them, TWA and the HD 141569 group, the smallest separation between RX J0720.4-3125 and the centre of the particular association is smaller than $\unit[5]{pc}$. For that reason it seems plausible that one of those two may be the birth place of RX J0720.4-3125. However, since the radii of Tuc-Hor ($\unit[50]{pc}$) and $\beta$ Pic-Cap ($\unit[56]{pc}$) are rather large, they could also have hosted the supernova. Compared to TWA the probabilities that one of those two associations hosted the supernova which formed \rxjn{} is two to three times smaller.\\

\begin{figure}
\centering
\includegraphics[width=0.4\textwidth, viewport= 85 265 480 590]{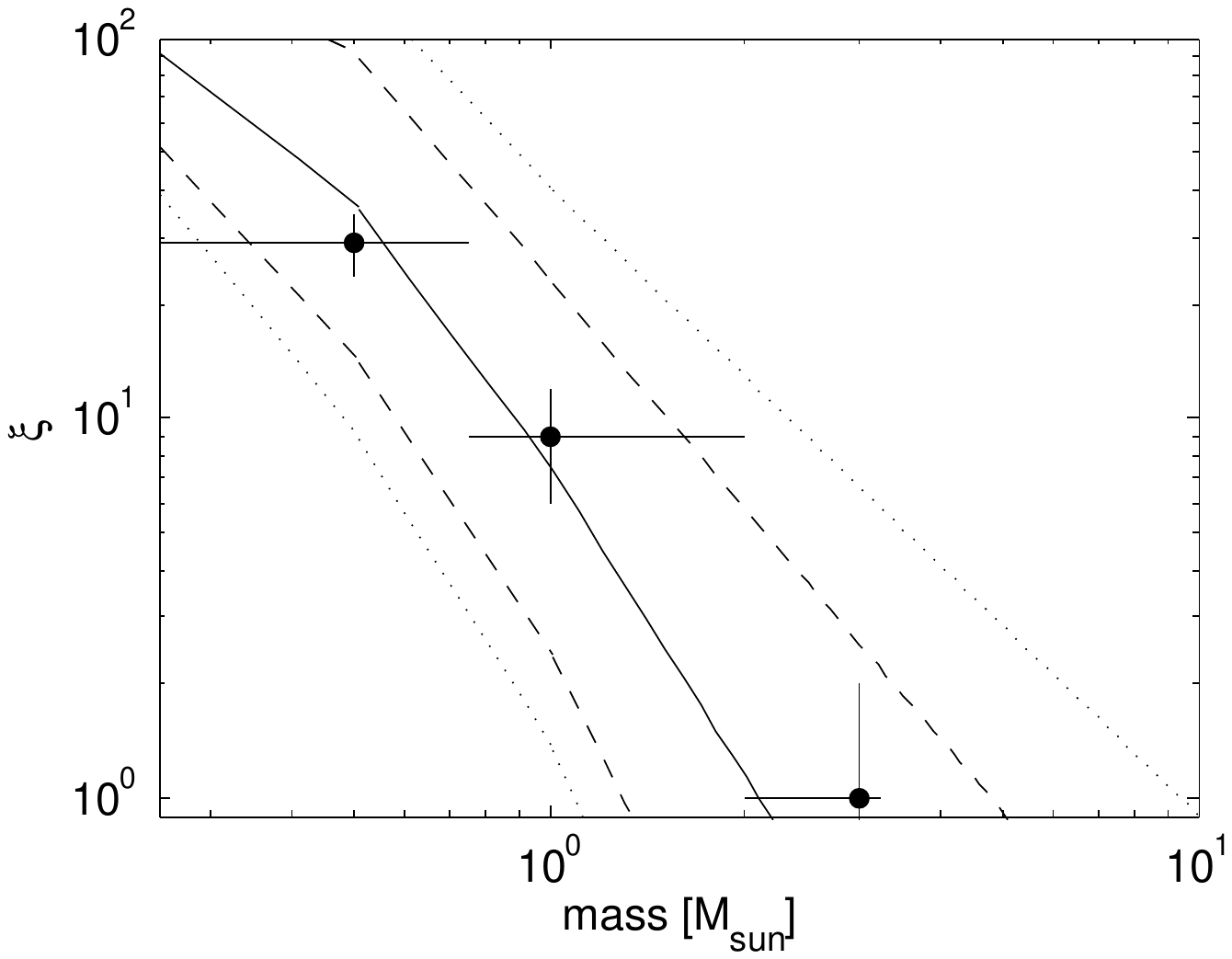}
\caption{Present TWA mass function (black circles, see footnote 5) in comparison with the initial mass function $\xi$ (IMF, solid line) as given by \citet{2005ASSL..327..175K}, their equation 2. Dashed and dotted lines represent the $1\sigma$ and $1{.}5\sigma$ IMF boundaries. Within $1{.}5\sigma$ we expect up to one star with $\unit[10]{M_\odot}$; the probability for higher mass stars is low.}
\label{fig:TWAmassfun}
\end{figure}

\Autorefs{subfig:0720_TWHya_hist} and \ref{subfig:0720_HD141569_hist} show the distributions of minimum separations $d_{min}$ to the centres of TWA and HD 141569. From the theoretical curves we can predict that a potential supernova which formed RX J0720.4-3125 in HD 141569 would have been situated $\unit[15\pm2]{pc}$ from the centre, i.e. at the edge of the association with a radius of about $\unit[15]{pc}$. Moreover, since HD 141569 only consists of five members (F08), it is rather unlikely to have hosted a supernova far from its core. The probability of such a scenario is seven times smaller than for TWA.\\
In the case of TWA we can predict the supernova very close to the association centre at a separation of $\unit[0\pm2]{pc}$. Then, RX J0720.4-3125 would have been formed $\approx\unit[0{.}4]{Myr}$ ago and its current radial velocity would be $\approx\unit[500]{km/s}$ (space velocity $\approx\unit[520]{km/s}$).
The current distance to the Sun which the neutron star would have in this case is $\approx\unit[250]{pc}$ which is in very good agreement with $235$ to $\unit[265]{pc}$ as derived by \citet{2007Ap&SS.308..171P} (thereafter P07) from the spectrum and $n_H$ (note that they obtained a slightly higher value of $\unit[1{.}2\cdot10^{20}]{cm^{-2}}$ compared to the most recent value from \citealt{2009A&A...498..811H}; \autoref{tab:vierM7_Eigensch}) as well as with the parallax of $\unit[2{.}77\pm1{.}29]{mas}$ \citep{2007ApJ...660.1428K}. \\
The potential supernova would have occurred between $48$ and $\unit[67]{pc}$ from the Sun. For such a small distance we expect to find supernova-produced radionuclides in the terrestrial crust \citep{1996ApJ...470.1227E} such as $^{10}$Be and $^{60}$Fe, with half-lives of $\unit[1{.}5]{Myr}$ \citep{1993PhLB..312..501A} and $\unit[2{.}6]{Myr}$ \citep{2009PhRvL.103g2502R}, respectively. A small but insignificant signal of $^{60}$Fe was found at a time below $\unit[1]{Myr}$ \citep{2004PhRvL..93q1103K,2008PhRvL.101l1101F}. For $^{10}$Be anomalous abundances in Antarctic ice cores corresponding to times of $\approx0{.}35$ and $\approx\unit[0{.}6]{Myr}$ in the past were identified \citep{1987Natur.326..273R}. The first signal has also been reported for deep-sea sediments in the Gulf of California \citep{1995GeoRL..22..659M} and in the Mediterranean \citep{1995ICRC....4.1204C}. More recently, \citet{2006ApJ...652.1763C} analysed ice cores from the Qinghai-Tibetan plateau in China but could not confirm previous findings. However, it seems possible that a recent supernova in TWA which produced RX J0720.4-3125 (\autoref{fig:0720Tr10TWA_traj}) contributed to one of the $^{10}$Be signals. \\
Moreover, F08 state that it is likely that one or more of the YLA hosted a supernova in the near past. Despite the small number of TWA members, we can expect one $\unit[10]{M_\odot}$-star to have been born in TWA due to its present mass function\footnote{We estimated the masses for the TWA members listed in F08, taking also into account that HD 98800 \citep{1998ApJ...498..385S} and TWA-5 \citep{2000A&A...360L..39N,2003AJ....125.3237T} are both quadruple, using the \citet{1997MmSAI..68..807D} model, and $\unit[2{.}9]{M_{\odot}}$ for the A0-type main sequence star HR 4796 A.} (\autoref{fig:TWAmassfun}). Given the association age of $3$ to $\unit[20]{Myr}$ \citep[e.g.][]{2006A&A...459..511B}, we estimate a progenitor mass of $10$ (B2V) up to $\unit[100]{M_{\odot}}$ (O4V) ($10$ to $>\unit[50]{M_{\odot}}$ for T80, $11$ to $\unit[100]{M_{\odot}}$ for MM89 and $10$ to $\unit[100]{M_{\odot}}$ for K97); spectral types from \citet{Schmidt-Kaler1982}).\\
If \rxjn{} was born in TWA, the supernova would have contributed only a small amount to $^{10}$Be and $^{60}$Fe found in the Earth's crust due to the relatively low mass progenitor ($\approx\unit[10]{M_\odot}$). This would be consistent with the findings. Furthermore, we propose a larger age ($\approx\unit[20]{Myr}$) for TWA.

\subsection{RX J1605.3+3249}

\begin{table}
\centering
\caption{Associations for which the smallest separation between RX J1605.3+3249 and the association centre found (min. $d_{min}$) after 2 million runs was within the radius of the association, as \autoref{tab:1856_smallestsep}.}\label{tab:1605_smallestsep}
\begin{tabular}{l d{2.2} d{2.2}}
\toprule
Association	&		\multicolumn{1}{c}{min. $d_{min}$ [pc]} & \multicolumn{1}{c}{$R_{Assoc}$ [pc]}\\\midrule
US					&		4,6																			& 15\\
UCL					&   13,0																		& 33\\
Tuc-Hor			&		5,2					  													& 50\\
$\beta$ Pic-Cap & 6,3 																	& 57\\
$\epsilon$ Cha & 21,4																		& 28\\
HD 141569   &   9,1																			& 16\\
Ext. R CrA	&		25,3																		& 31\\
AB Dor			&		12,7																		& 43\\
NGC 6322		&   2,2																			& 11\\
Bochum 13		&		2,1																			& 2\\
Sco OB4			&		0,9																			& 33\\\bottomrule
\end{tabular}
\end{table}

\begin{table*}
\centering
\caption{Potential parent associations of RX J1605.3+3349, columns as in \autoref{tab:detailed_scan_1856}.}
\label{tab:detailed_scan_1605}
\setlength\extrarowheight{3pt}
\small
\begin{tabular}{c f{2.2} f{4.4} >{$}r<{$} o{4.2} o{4.2} >{$}r<{$} >{$}r<{$} >{$}c<{$} >{$}r<{$} >{$}r<{$}}
\toprule
Association	&		\multicolumn{1}{c}{$d_{min}$}	&	\multicolumn{1}{c}{$\tau$}		&	\multicolumn{1}{c}{$v_r$}			& \multicolumn{1}{c}{$\mu_{\alpha}^*$} & \multicolumn{1}{c}{$\mu_{\delta}$} & \multicolumn{1}{c}{$v_{space}$} &	\multicolumn{1}{c}{$\pi$}						&	d_{\odot} 	& \multicolumn{1}{c}{$\alpha$}	& \multicolumn{1}{c}{$\delta$}\\ 
						&	\multicolumn{1}{c}{[pc]} & \multicolumn{1}{c}{[Myr]} & \multicolumn{1}{c}{[km/s]} & \multicolumn{1}{c}{[mas/yr]} & \multicolumn{1}{c}{[mas/yr]} & \multicolumn{1}{c}{[km/s]} & \multicolumn{1}{c}{[mas]} & \multicolumn{1}{c}{[pc]} & \multicolumn{1}{c}{[$^\circ$]} & \multicolumn{1}{c}{[$^\circ$]}\\\midrule
Tuc-Hor						&  7p15			&  0{.}24p0{.}39				& 452^{+115}_{-76}& -43{.}7+1{.}7 & 148{.}7+2{.}6 & 462^{+118}_{-78} & 7{.}54^{+1{.}53}_{-1{.}33}& 39^{+5}_{-4}		& 32{.}48^{+10{.}60}_{-34{.}55} & -75{.}69^{+4{.}70}_{-1{.}55}\\
$\beta$ Pic-Cap								&  8p10			&  0{.}16p0{.}24				& 537^{+103}_{-93}& -43{.}7+1{.}7 & 148{.}6+2{.}6 & 543^{+104}_{-94}	& 8{.}86^{+1{.}60}_{-1{.}09}& 18^{+2}_{-7}		& 292{.}49^{+13{.}51}_{-22{.}49}	& -75{.}21^{+8{.}25}_{-1{.}41}\\
Ext. R CrA			&  30p42			&  0{.}43p0{.}64						& 321^{+109}_{-58}	& -43{.}7+1{.}7	& 148{.}8+2{.}6 & 362^{+117}_{-66}	& 4{.}41^{+0{.}90}_{-0{.}85}& 70\ldots115	& 261{.}99^{+1{.}12}_{-0{.}99}	& -39{.}84^{+2{.}10}_{-2{.}00}\\
AB Dor				&  21p43			&  0{.}17p0{.}32						& 452^{+109}_{-84}	& -43{.}7+1{.}7	& 148{.}7+2{.}6 & 462^{+110}_{-85}	& 7{.}56^{+1{.}22}_{-0{.}78}& 21^{+7}_{-6}	& 264{.}30^{+4{.}76}_{-4{.}93}	& -44{.}11^{+6{.}98}_{-8{.}82}\\
Sco OB4					& 12p74			&  1{.}28p1{.}68						& 311^{+75}_{-69}	& -43{.}7+1{.}7	& 148{.}7+2{.}6	& 743^{+98}_{-77} & 1{.}09^{+0{.}08}_{-0{.}08}& 1042\ldots1140	& 259{.}57^{+0{.}97}_{-0{.}99}	& -35{.}0\ldots-30{.}5\\\midrule
US									& 26p36				& 0{.}46p0{.}63					& 343^{+86}_{-80} & -43{.}7+1{.}6 & 148{.}8+2{.}6 & 406^{+92}_{-85} & 3{.}38^{+0{.}47}_{-0{.}41}	& 	129\ldots171		& 255{.}09^{+0{.}57}_{-0{.}66}	& -18{.}44^{+1{.}00}_{-1{.}06}\\
UCL							& 50p63			& 0{.}48p0{.}73					&	318^{+116}_{-63} & -43{.}7+1{.}7 & 148{.}7+2{.}6 & 375^{+122}_{-72} & 3{.}70^{+0{.}71}_{-0{.}59}	& 96\ldots164			& 260{.}48^{+0{.}81}_{-0{.}86}	& -35{.}68^{+1{.}83}_{-1{.}60}\\\bottomrule
\end{tabular}
\end{table*}

Since there is currently only an upper limit of the distance ($<\unit[410]{pc}$, \citealt{2007astph.712..342M}) available for \rxjs{}, thus a lower limit for the parallax ($\pi > \unit[2{.}4]{mas}$, see \autoref{tab:vierM7_Eigensch}) we had to adopt a value for our investigations. The hydrogen column density of $n_H = \unit[1{.}1\pm0{.}4\cdot10^{20}]{cm^{-2}}$ \citep{1999A&A...351..177M} predicts a relatively small distance of RX J1605.3+3249 to the Sun. For that reason we chose a parallax value of $\pi = \unit[5\pm3]{mas}$ corresponding to a distance of $\unit[200^{+300}_{-75}]{pc}$. Infact, this is a source of large uncertainty. After the 2 million run investigation we identify 11 associations for which the smallest separation $d_{min}$ between the association centre and \rxjs{} was consistent with the association radius (\autoref{tab:1605_smallestsep}). Five of them show a maximum in the $\tau$-$d_{min}$ contour plot also being consistent with the respective association boundaries (\autoref{tab:detailed_scan_1605}, top).\\

\begin{figure}
\subfigure[Tuc-Hor]{\includegraphics[width=0.25\textwidth, viewport= 85 265 480 590]{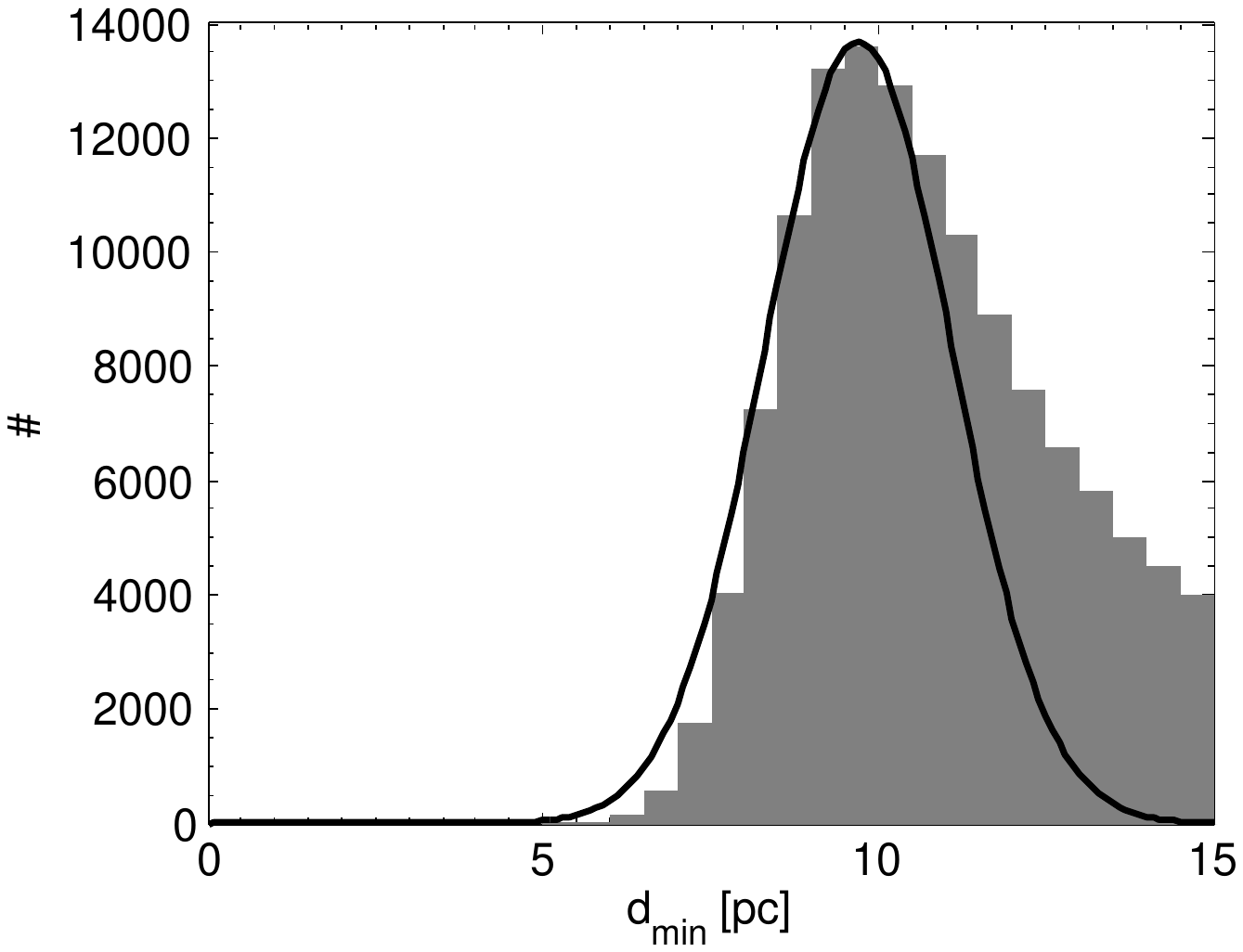}\label{subfig:1605_Tuc-Hor_dmin}}\nolinebreak
\subfigure[$\beta$ Pic-Cap]{\includegraphics[width=0.25\textwidth, viewport= 85 265 480 590]{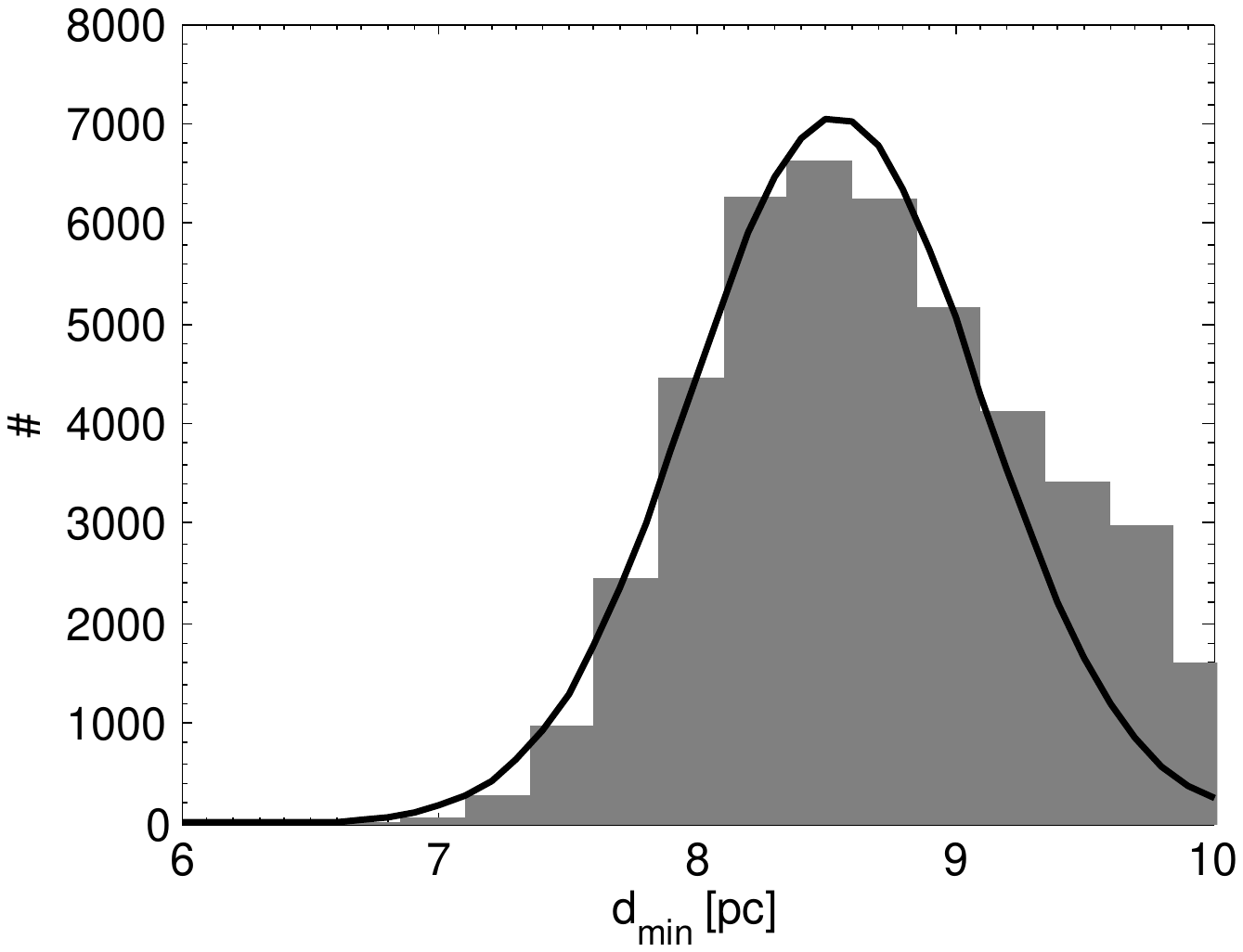}\label{subfig:1605_betPic-Cap_dmin}}\\
\subfigure[Ext. R CrA]{\includegraphics[width=0.25\textwidth, viewport= 85 265 480 590]{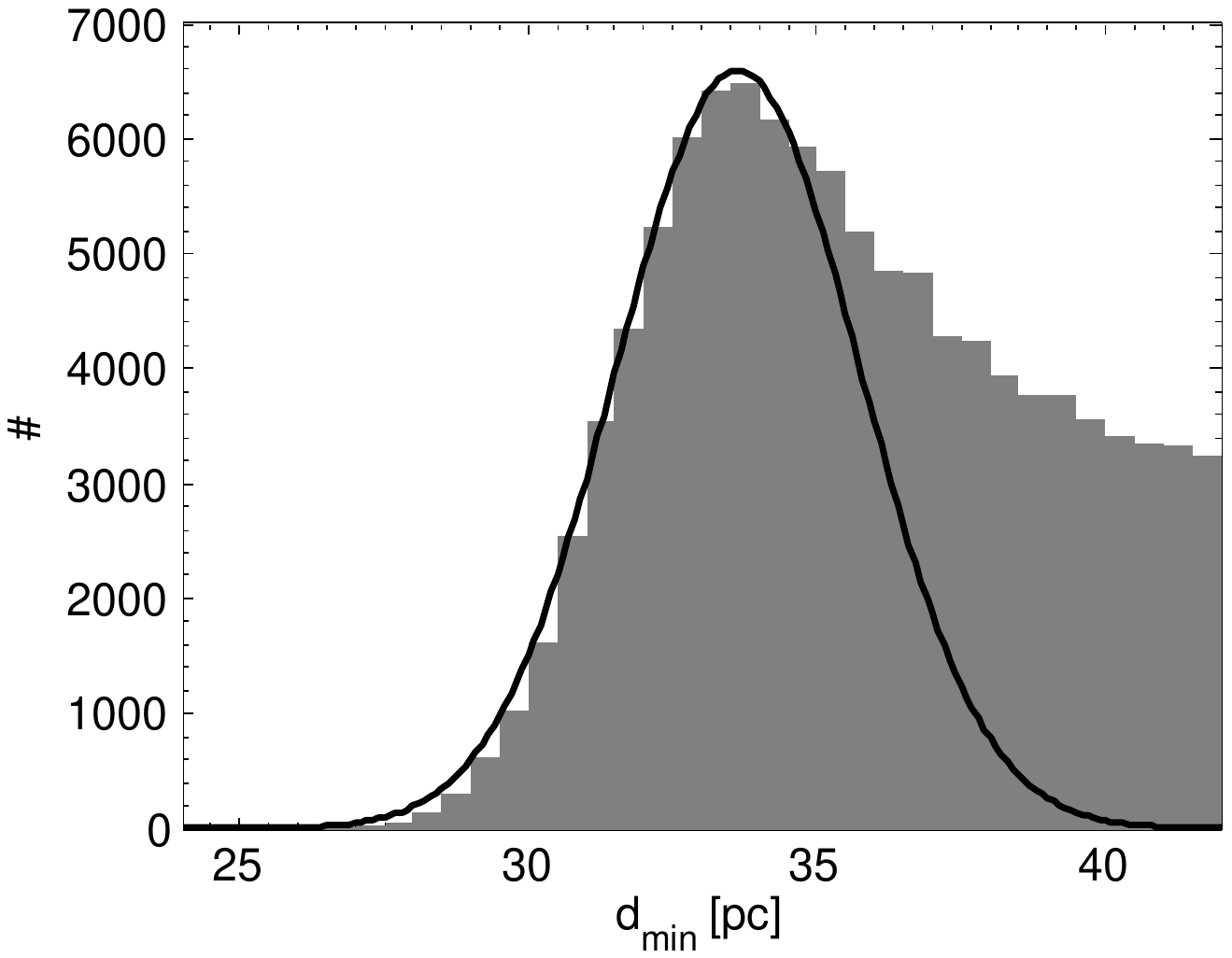}\label{subfig:1605_ExtRCrA_dmin}}\nolinebreak
\subfigure[AB Dor]{\includegraphics[width=0.25\textwidth, viewport= 85 265 480 590]{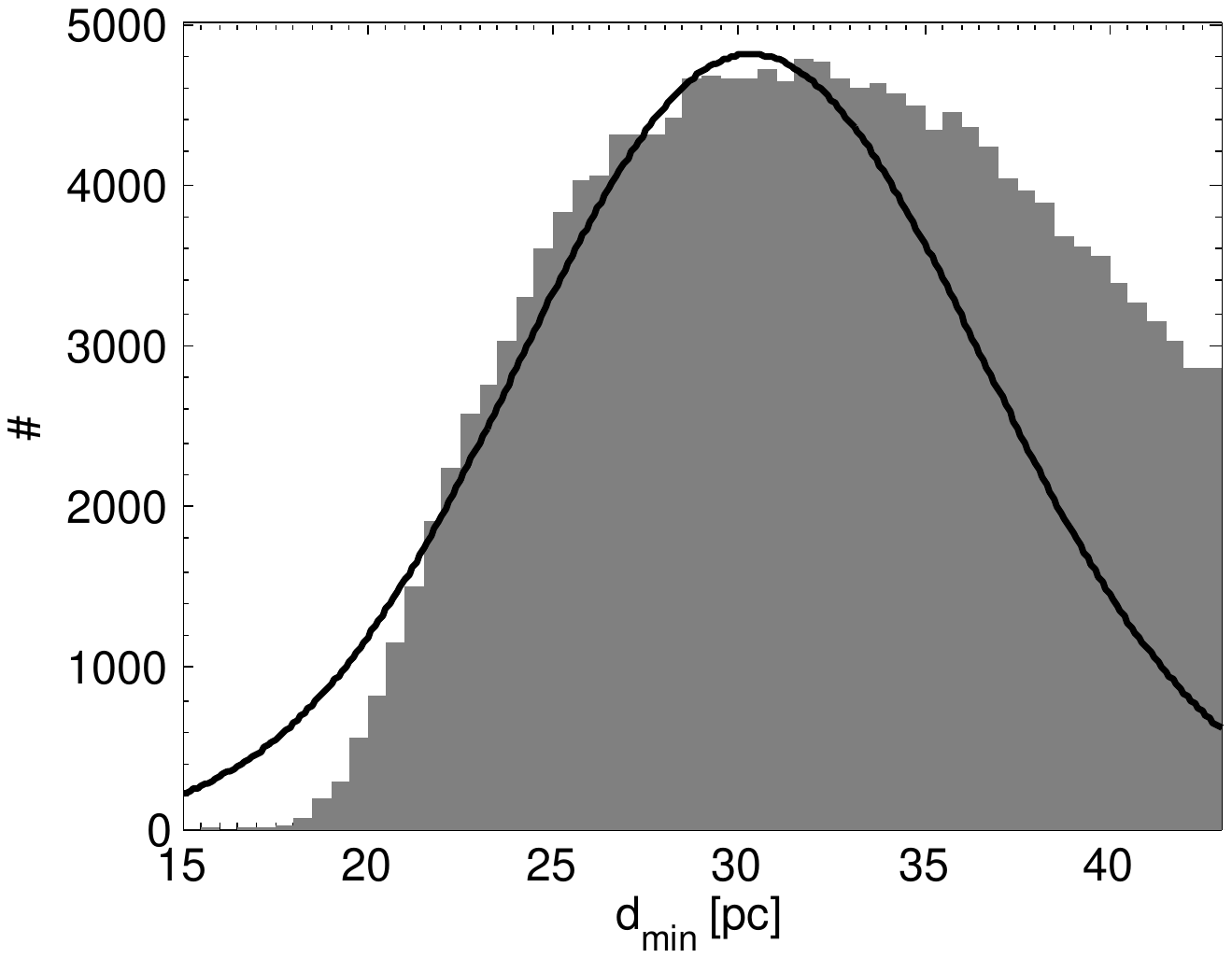}\label{subfig:1605_ABDor_dmin}}
\subfigure[US]{\includegraphics[width=0.25\textwidth, viewport= 85 265 480 590]{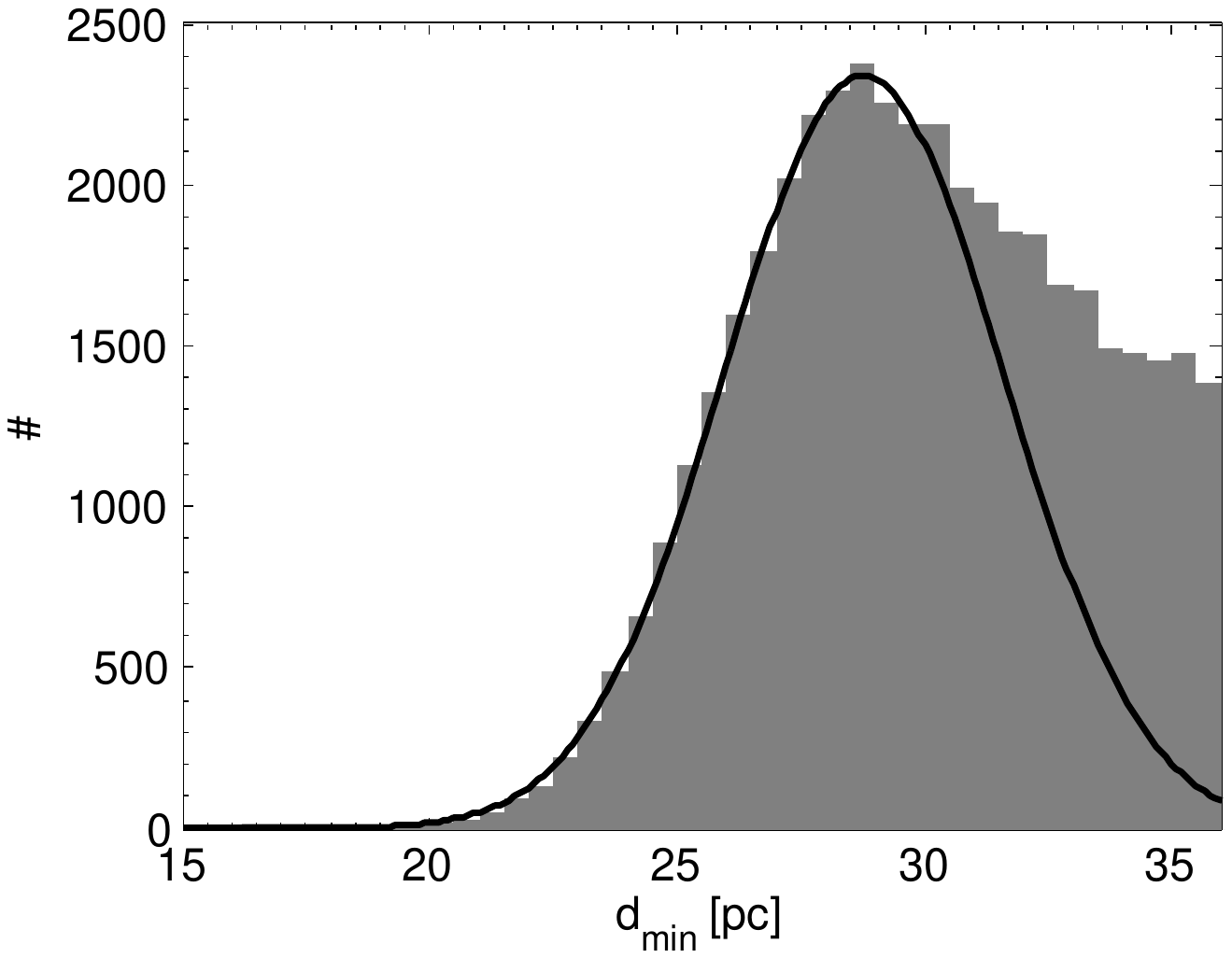}\label{subfig:1605_US_dmin}}\nolinebreak
\subfigure[UCL]{\includegraphics[width=0.25\textwidth, viewport= 85 265 480 590]{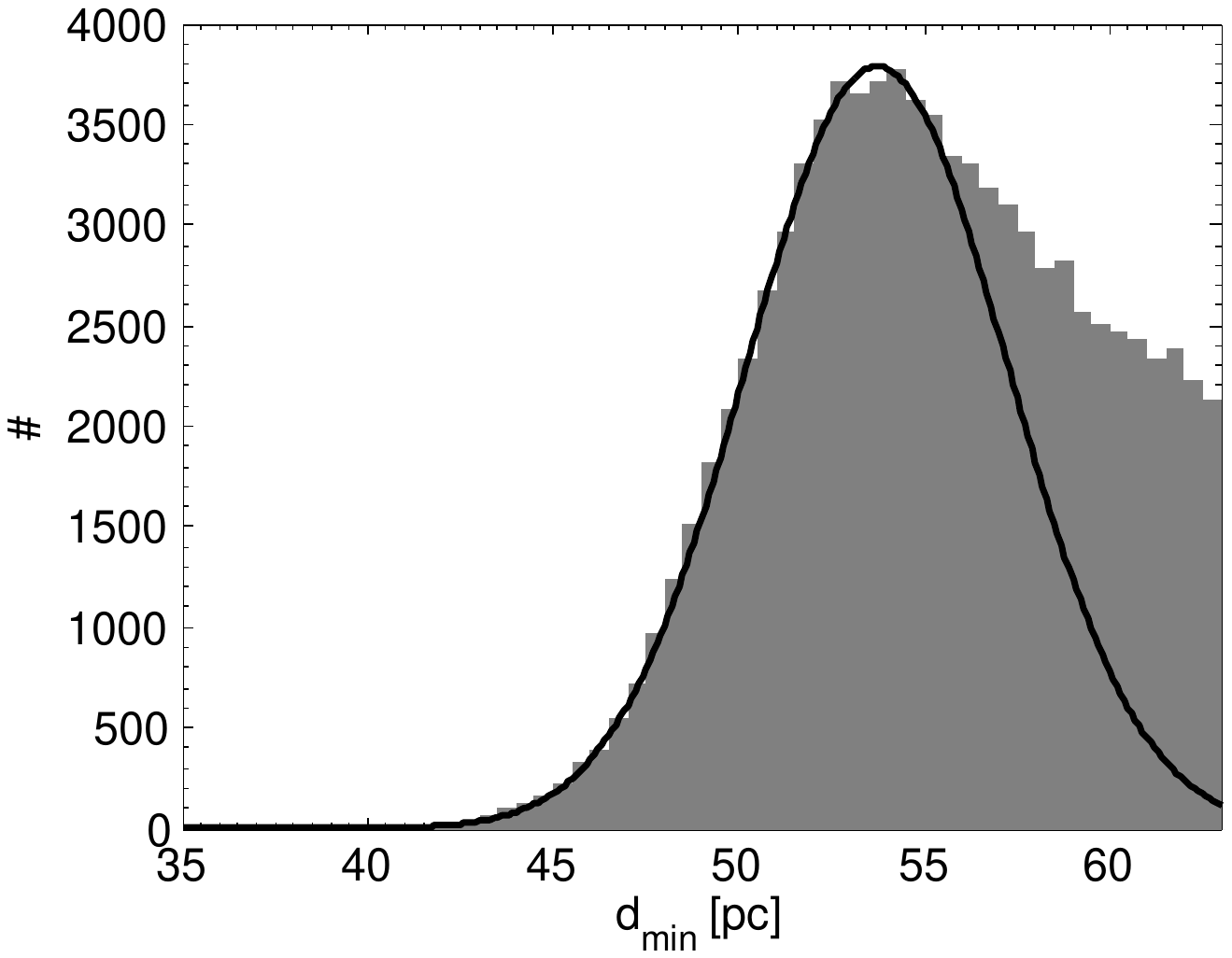}\label{subfig:1605_UCL_dmin}}\\
\caption{(a) Distribution of minimum separations $d_{min}$ between RX J1605.3+3249 and the centres of Tuc-Hor (a), $\beta$ Pic-Cap (b), Ext. R CrA (c), AB Dor (d), US (e) and UCL (f), respectively, within the defined time range $\tau$ (see \autoref{tab:detailed_scan_1605}, column 3). Shown as well are theoretical curves for three-dimensional Gaussian distributions (\autoref{eq:3DGauss_diff}) with $\mu~=\unit[9{.}5]{pc}$ and $\sigma~=\unit[1{.}0]{pc}$ for Tuc-Hor, $\mu~=\unit[8{.}5]{pc}$ and $\sigma~=\unit[0{.}4]{pc}$ for $\beta$ Pic-Cap, $\mu~=\unit[33{.}5]{pc}$ and $\sigma~=\unit[1{.}5]{pc}$ for Ext. R CrA, $\mu~=\unit[29{.}0]{pc}$ and $\sigma~=\unit[4{.}5]{pc}$ for AB Dor, $\mu~=\unit[28{.}5]{pc}$ and $\sigma~=\unit[2{.}0]{pc}$ for US and $\mu~=\unit[53{.}5]{pc}$ and $\sigma~=\unit[2{.}5]{pc}$ for UCL. Note that the curves are not fitted to the data (see also \autoref{fig:hist_1856_US}).}
\label{fig:hists_1605}
\end{figure}

\begin{figure}
\centering
\includegraphics[clip, width=0.3\textwidth, viewport= 57 0 663 537]{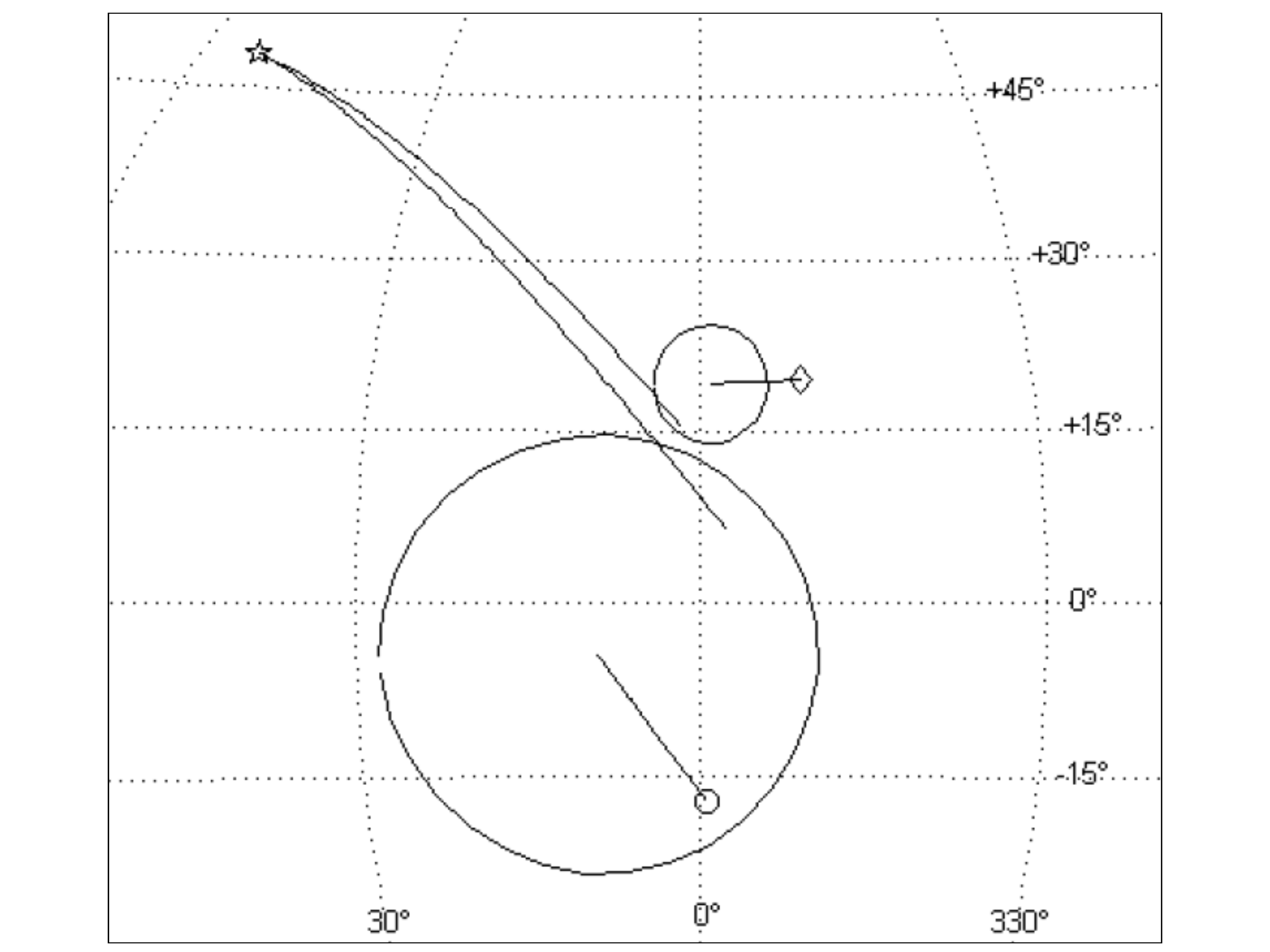}
\caption{Past trajectories for \rxjs{} and US and Ext. R CrA, respectively, projected on a Galactic coordinate system (for particular sets of input parameters consistent with \autoref{tab:detailed_scan_1605}). Present positions are marked with a star for the neutron star and a diamond for US and an open circle for Ext. R CrA. Large circles reflect association extensions (radii of $\unit[15]{pc}$ for US and $\unit[31]{pc}$ for Ext. R CrA).}
\label{fig:1605USExtRCrA_traj}
\end{figure}

Due to the small parallax needed for \rxjs{} to reach the Scorpius OB4 (Sco OB4) association ($\pi=\unit[1{.}1]{mas}$, column 7 of \autoref{tab:detailed_scan_1605}) we may exclude this association since this is even smaller than the lower limit of $\pi > \unit[2{.}4]{mas}$ (\autoref{tab:vierM7_Eigensch}).\\
The remaining four associations once again belong to the YLA (F08). We can explain the slopes of the histograms of minimum separations $d_{min}$ with three-dimensional Gaussians (\autoref{eq:3DGauss_diff}) that predict the place of the potential supernova within the boundaries of the associations (see \autoref{fig:hists_1605}). For the YLA, the simulations suggest a very small current distance of the neutron star to the Sun between $100$ and $\unit[250]{pc}$. From $n_H$ P07 derived a distance between $325$ and $\unit[390]{pc}$ (they obtained $n_H=\unit[2{.}0\cdot10^{20}]{cm^{-2}}$ without giving errors; however, for an $n_H$ of $\approx\unit[1\cdot10^{20}]{cm^{-2}}$ (\autoref{tab:vierM7_Eigensch}), the distance should at least be $\approx\unit[230]{pc}$, cf. P07 their Table 2). For that reason we find Tuc-Hor, $\beta$ Pic-Cap and AB Dor to be less likely parent associations of \rxjs{} since the neutron star would then need to have a current distance to the Sun smaller than $\unit[160]{pc}$.\footnote{Estimating probabilities, even rough values, is unreliable in the case of \rxjs{} because of the uncertain distance. Hence, an individual discussion is appropriate.} \\
In \autoref{tab:detailed_scan_1605} (bottom) we also give the results for two associations of the Scorpius-Centaurus (Sco-Cen) complex -- US and Upper Centaurus Lupus (UCL) -- for which slightly larger distances are required. Constraining the age to $\unit[1]{Myr}$ as a maximum and using current distances of $100$ to $\unit[500]{pc}$ as well as radial velocities in the range of $\unit[\pm700]{km/s}$ \citet{2005A&A...429..257M} proposed the origin of \rxjs{} in US. We consider the Sco-Cen complex also to be a likely birth site of \rxjs{}. The respective $d_{min}$ histograms are shown in \autorefs{subfig:1605_US_dmin} and \ref{subfig:1605_UCL_dmin}.\\
Hence, we conclude that \rxjs{} either originated from the extended Corona-Australis association $\approx\unit[0{.}5]{Myr}$ ago with a current distance of $\approx\unit[230]{pc}$ and radial velocity of $\approx\unit[320]{km/s}$ (space velocity $\approx\unit[360]{km/s}$) or was born in the Sco-Cen complex $\unit[28{.}5\pm2{.}0]{pc}$ from the centre of US about $\unit[0{.}5]{Myr}$ in the past (\autoref{fig:1605USExtRCrA_traj}). The first scenario is possible due to the existence of B-type stars (highest content within the YLA, F08), thus late O- or early B-type stars may have been born there. For the latter scenario the current distance to the Sun would be $\approx\unit[300]{pc}$ and the neutron star's radial velocity $\approx\unit[340]{km/s}$ (space velocity $\approx\unit[400]{km/s}$). Both cases would be in good agreement with a space velocity of $\approx\unit[400]{km/s}$ which reflects the mean velocity of the velocity distribution for pulsars from Ho05.\\
Accepting one of the two birth sites and taking the ages of the associations of $5$ to $\unit[10]{Myr}$ (US) and $10$ to $\unit[15]{Myr}$ (Ext. R CrA), we estimate the mass of the progenitor star to be $17$ (B0V) to $\unit[45]{M_{\odot}}$ (O5.5V) ($29$ to $\unit[45]{M_{\odot}}$ for T80, $18$ to $\unit[42]{M_{\odot}}$ for MM89, $17$ to $\unit[36]{M_{\odot}}$ for K97) for a birth in the Sco-Cen complex or $12$ (B1V) to $\unit[30]{M_{\odot}}$ (O7V) ($13$ to $\unit[30]{M_{\odot}}$ for T80, $13$ to $\unit[18]{M_{\odot}}$ for MM89, $12$ to $\unit[17]{M_{\odot}}$ for K97) for a birth in Ext. R CrA (spectral types from \citealt{Schmidt-Kaler1982}).\\
Such a nearby supernova (see \autoref{tab:detailed_scan_1605}, column 8) could also have contributed to the $^{10}$Be and $^{60}$Fe material found in the Earth's crust \citep[e.g.][]{1987Natur.326..273R,2004PhRvL..93q1103K}.


\subsection{RBS 1223 (RX J1308.8+2127)}

\begin{table}
\centering
\caption{Associations for which the smallest separation between RBS 1223 and the association centre found (min. $d_{min}$) after 2 million runs was within the radius of the association, as \autoref{tab:1856_smallestsep}.}\label{tab:1308_smallestsep}
\begin{tabular}{l d{2.2} d{2.2}}
\toprule
Association	&		\multicolumn{1}{c}{min. $d_{min}$ [pc]} & \multicolumn{1}{c}{$R_{Assoc}$ [pc]}\\\midrule
US					&		1,1																			& 15\\
Tuc-Hor			&		10,3				  													& 50\\
$\beta$ Pic-Cap & 0,3 																	& 57\\
HD 141569   &   0,1																			& 16\\
Ext. R CrA	&		0,3 																		& 31\\
AB Dor			&		7,9 																		& 43\\
Her-Lyr			&   9,7																			& 13\\
Sgr OB5			&   6,2 																		& 111\\
NGC 6530		&   1,3																			& 6\\
Sgr OB1			&		2,2																			& 104\\
Sgr OB7			&		3,0																			& 29\\
Sgr OB4			&		5,0																			& 27\\
Sgr OB6			&		1,8																			&	17\\
M 17				&		3,5																			& 10\\
Ser OB1			&		1,1																			& 36\\
NGC 6611		&		0,8																			& 7\\
Sct OB3			&   2,2																			& 17\\
Ser OB2			&		3,7																			& 31\\
NGC 6604		&		4,5																			& 8\\
Sct OB2			&		6,6																			& 31\\
NGC 6383		&   3,0																			& 5\\
M6					&   2,5																			& 4\\
\bottomrule
\end{tabular}
\end{table}

For RBS 1223 both distance limits are available ($\unit[76\ldots700]{pc}$, \citealt{2005A&A...441..597S,2007astph.712..342M}, hence $\pi = \unit[1{.}4\ldots13]{mas}$). Since the span between those limits is rather large, we adopted a parallax of $\pi = \unit[7\pm5]{mas}$ which is, as for \rxjs{}, a source of large uncertainty. After performing 2 million Monte-Carlo runs, we found 22 associations for which the smallest separation between RBS 1223 and the association centre found was within the association radius (\autoref{tab:1308_smallestsep}). Five of these 23 show well defined areas of higher probability in their $\tau$-$d_{min}$ contour plots (\autoref{tab:detailed_scan_1308}), whereas for more distant ($>\unit[1]{kpc}$) associations as well as Serpens OB1 (Ser OB1) and M6 (below Her-Lyr in \autoref{tab:1308_smallestsep}) only very few runs yield small separations $d_{min}$ between the neutron star and the association centre. The latter behaviour does not change even if we constrain input parallaxes to a range of $\unit[3\pm2]{mas}$. Furthermore, an origin in most of those associations implies a parallax smaller than the lower limit of $\unit[1{.}4]{mas}$. For those reasons we may exclude them as probable birth associations for RBS 1223.\\

\begin{table*}
\centering
\caption{Potential parent associations of RBS 1223, columns as in \autoref{tab:detailed_scan_1856}.}
\label{tab:detailed_scan_1308}
\begin{tabular}{c d{2} >{$}c<{$} >{$}c<{$} >{$}r<{$} >{$}c<{$} >{$}r<{$} >{$}c<{$} >{$}r<{$} >{$}r<{$} >{$}r<{$}}
\toprule
Association	&		\multicolumn{1}{c}{$d_{min}$}	&	\multicolumn{1}{c}{$\tau$}		&	\multicolumn{1}{c}{$v_r$}			& \multicolumn{1}{c}{$\mu_{\alpha}^*$} & \multicolumn{1}{c}{$\mu_{\delta}$} & \multicolumn{1}{c}{$v_{space}$} &	\multicolumn{1}{c}{$\pi$}						&	d_{\odot} 	& \multicolumn{1}{c}{$\alpha$}	& \multicolumn{1}{c}{$\delta$}\\ 
						&	\multicolumn{1}{c}{[pc]} & \multicolumn{1}{c}{[Myr]} & \multicolumn{1}{c}{[km/s]} & \multicolumn{1}{c}{[mas/yr]} & \multicolumn{1}{c}{[mas/yr]} & \multicolumn{1}{c}{[km/s]} & \multicolumn{1}{c}{[mas]} & \multicolumn{1}{c}{[pc]} & \multicolumn{1}{c}{[$^\circ$]} & \multicolumn{1}{c}{[$^\circ$]}\\\midrule
Tuc-Hor							&  23\ldots30			& 0{.}17\ldots0{.}31				& 463^{+94}_{+82}	& -206\pm19 & 84\pm18 & 473^{+100}_{-86} & 10{.}85^{+3{.}38}_{-2{.}16}& 37^{+4}_{-3}			& 339{.}68^{+14{.}32}_{-9{.}68} & -29{.}20^{+2{.}56}_{-3{.}33}\\
$\beta$ Pic-Cap				&  5\ldots10			& 0{.}13\ldots0{.}22					& 482^{+106}_{-95}& -206\pm19 & 82\pm19 & 490^{+110}_{-98} & 12{.}06^{+2{.}45}_{-2{.}25}& 19^{+2}_{-3}		& 316{.}23^{+11{.}81}_{-7{.}80} & -29{.}63^{+3{.}70}_{-3{.}90}\\
HD 141569				&  5\ldots21			&  0{.}27\ldots0{.}40					& 332^{+103}_{-85}& -209\pm19 & 85\pm20	& 393^{+130}_{-99} & 5{.}09^{+0{.}91}_{-0{.}98}& 84\ldots122		& 236{.}46^{+7{.}70}_{-9{.}09}	& 1{.}55^{+5{.}55}_{-4{.}74}\\
Ext. R CrA		& 	 16\ldots40		&  0{.}38\ldots0{.}58			& 345^{+114}_{-90}	& -209\pm20	& 85\pm18	& 391^{+137}_{-103} & 5{.}78^{+1{.}33}_{-1{.}21}& 80^{+22}_{-10}		&	294{.}82^{+7{.}24}_{-7{.}92}	& -22{.}94^{+4{.}33}_{-6{.}29}\\
AB Dor				&  15\ldots28			& 0{.}15\ldots0{.}24				& 437^{+99}_{-95}	& -206\pm20	& 83\pm19 & 446^{+104}_{-97}	& 11{.}56^{+2{.}24}_{-2{.}05}& 18^{+4}_{-3}		&	225\ldots358	& -28{.}86^{+11{.}77}_{-6{.}85}\\\midrule
US			& 42\ldots64				& 0{.}38\ldots0{.}55				& 351^{+100}_{-104} & -210\pm20 & 83\pm18 & 420^{+154}_{-118} & 4{.}63^{+0{.}84}_{-1{.}27}	& 109^{+36}_{-19}			& 254{.}09^{+2{.}11}_{-2{.}34}	& -6{.}29^{+3{.}23}_{-5{.}09}\\
\bottomrule
\end{tabular}
\end{table*}

However, calculating past trajectories of RBS 1223, \citet{2007astph.712..342M} proposed that a small fraction of orbits passes though the Scutum OB2 (Sct OB2) association. More recently, \citet{2009A&A...497..423M} reinvestigated this association considering that it consists of two groups, Sct OB2A at $\unit[510]{pc}$ and Sct OB2B at $\unit[1170]{pc}$ \citep{1990Ap&SS.163..275R}. We adopt these two distances and the proper motion and radial velocity for Sct OB2 from \citet{2001AstL...27...58D} and calculate again 2 million orbits for both, the neutron star and each subgroup, as well as their separations $d_{min}$ at each time step $\tau$. Due to the large uncertainty in parallax and radial velocity of RBS 1223, we do not find a maximum in their $\tau$-$d_{min}$ plots, but can derive flight time ranges from the $\tau$ histograms for which smaller separations ($d_{min}\leq\unit[50]{pc}$) occur. Considering the input parameters of the runs in the derived time ranges, we obtain the neutron star properties and potential supernova positions as given in \autoref{tab:detailed_scan_1308_SctOB2AB}.\\

\begin{table*}
\centering
\caption{Results for RBS 1223 and Sct OB2A and B
. Minimum separations have been constrained to $d_{min}\leq\unit[50]{pc}$. Column 2 gives the smallest separation $d_{min}$ found after 2 million runs, column 3 gives the maximum of the time ($\tau$) histogram along with its 68\% interval (note that this is not a $1\sigma$ error). Columns 4 to 10 indicate the current neutron star properties needed to yield small separations to the particular association as well as the position of the potential supernova (see also \autoref{tab:detailed_scan_1856}). Note that receding radial velocities are positive contrary to \citet{2009A&A...497..423M}.}
\label{tab:detailed_scan_1308_SctOB2AB}
\begin{tabular}{c d{2} >{$}c<{$} >{$}c<{$} >{$}r<{$} >{$}c<{$} >{$}r<{$} >{$}c<{$} >{$}r<{$} >{$}r<{$} >{$}r<{$}}
\toprule
Association	&		\multicolumn{1}{c}{min. $d_{min}$}	&	\multicolumn{1}{c}{$\tau$}		&	\multicolumn{1}{c}{$v_r$}			& \multicolumn{1}{c}{$\mu_{\alpha}^*$} & \multicolumn{1}{c}{$\mu_{\delta}$} & \multicolumn{1}{c}{$v_{space}$} &	\multicolumn{1}{c}{$\pi$}						&	d_{\odot} 	& \multicolumn{1}{c}{$\alpha$}	& \multicolumn{1}{c}{$\delta$}\\ 
						&	\multicolumn{1}{c}{[pc]} & \multicolumn{1}{c}{[Myr]} & \multicolumn{1}{c}{[km/s]} & \multicolumn{1}{c}{[mas/yr]} & \multicolumn{1}{c}{[mas/yr]} & \multicolumn{1}{c}{[km/s]} & \multicolumn{1}{c}{[mas]} & \multicolumn{1}{c}{[pc]} & \multicolumn{1}{c}{[$^\circ$]} & \multicolumn{1}{c}{[$^\circ$]}\\\midrule
Sct OB2A							&  1{.}6			& 1{.}07^{+0{.}56}_{-0{.}34}				& 75\ldots310	& -216\pm19 & 53\pm10 & 272$\ldots$597 & 2{.}70^{+0{.}96}_{-0{.}44}& 517^{+18}_{-19}			& 277{.}15^{2{.}47}_{2{.}60} & -12{.}04^{+2{.}13}_{-1{.}83} \\
Sct OB2B							&  3{.}2			& 1{.}56^{+0{.}80}_{-0{.}42}				& 285^{+107}_{-129}	& -221\pm20 & 46\pm7 & 731^{+313}_{-172} & 1{.}59^{+0{.}22}_{-0{.}38}& 1199^{+19}_{-19}			& 277{.}93^{+1{.}22}_{-1{.}41} & -10{.}12^{+1{.}93}_{-0{.}63}\\
\bottomrule
\end{tabular}
\end{table*}

\begin{figure}
\subfigure[Ext. R CrA]{\includegraphics[width=0.25\textwidth, viewport= 85 265 480 590]{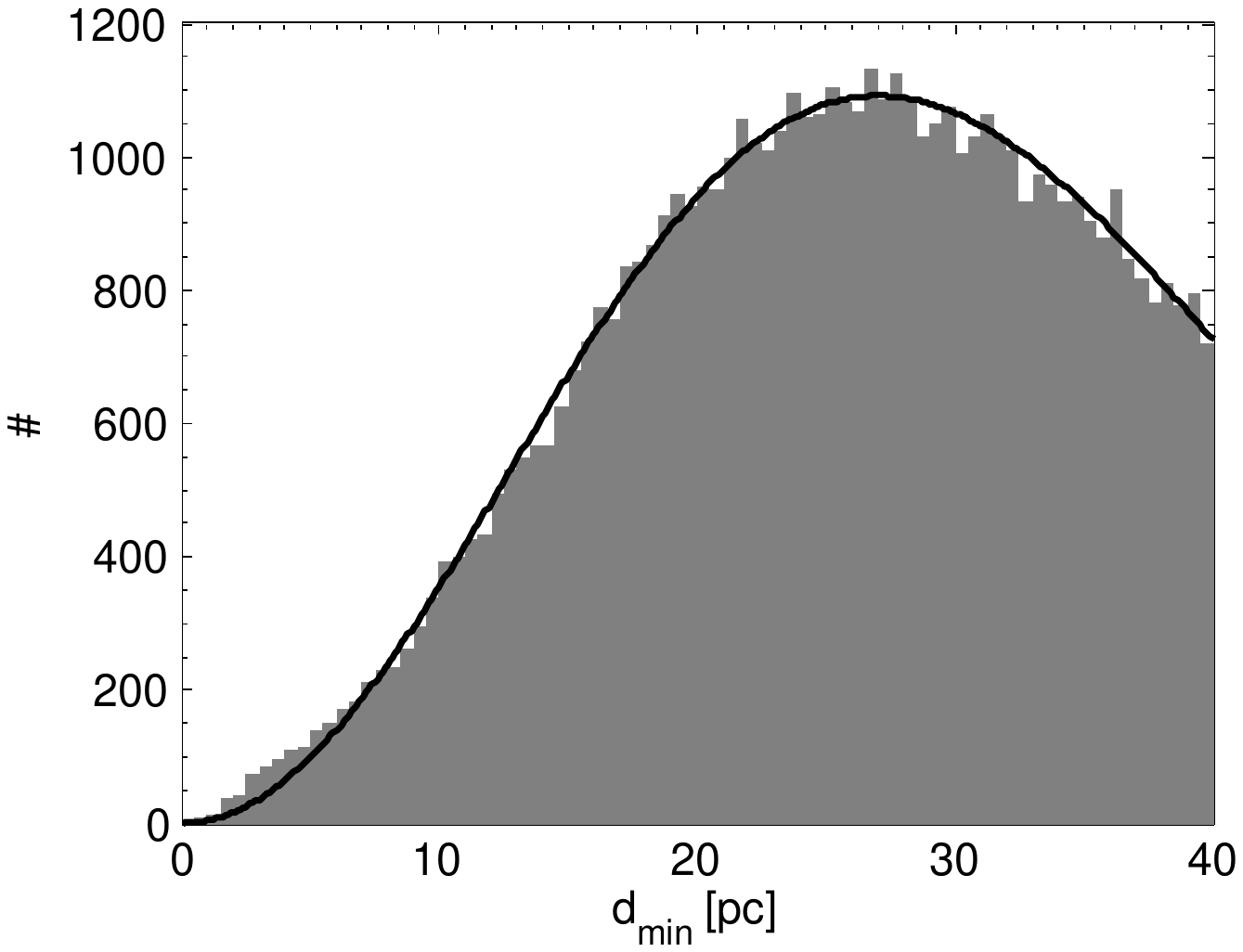}\label{subfig:hist_1308_ExtRCrA}}\nolinebreak
\subfigure[US]{\includegraphics[width=0.25\textwidth, viewport= 85 265 480 590]{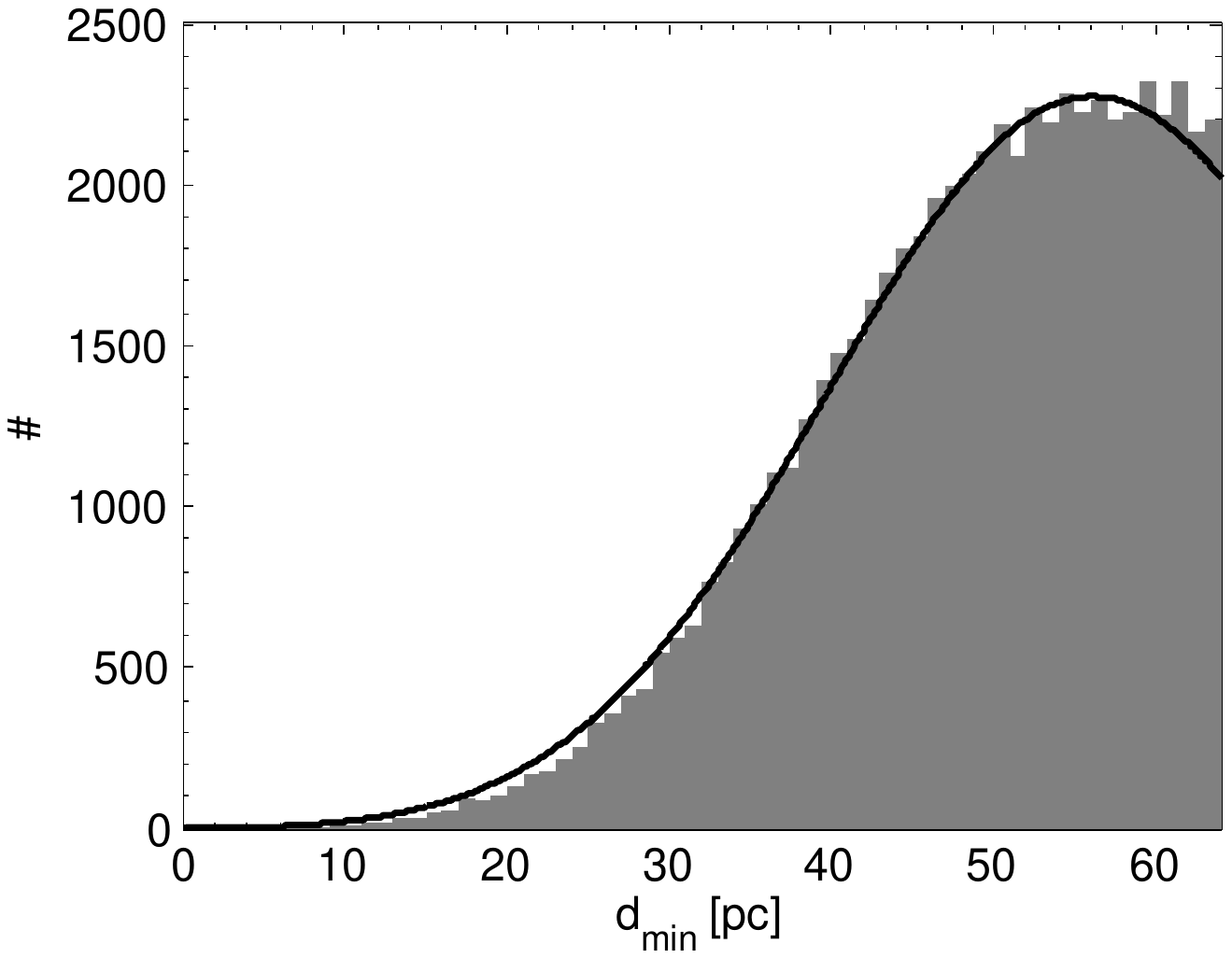}\label{subfig:1308_US_dmin}}\nolinebreak
\caption{Distribution of minimum separations $d_{min}$ between RBS 1223 and the centre of the extended Corona-Australis association (a) and US (b) within the defined time range $\tau$ (see \autoref{tab:detailed_scan_1308}, column 3). Shown as well is a theoretical curve for a three-dimensional Gaussian distributions (\autorefs{eq:3DGauss_diff} and \ref{eq:3DGaussmunull_diff}) with $\mu~=\unit[0]{pc}$ and $\sigma~=\unit[13{.}5]{pc}$ for Ext. R CrA and $\mu~=\unit[51]{pc}$ and $\sigma~=\unit[12]{pc}$ for US. Note that the curves are not fitted to the data but shall just give an explanation of the slopes of the histograms.}
\label{fig:hists_1308}
\end{figure}

\begin{figure}
\centering
\includegraphics[clip, width=0.3\textwidth, viewport= 164 0 565 535]{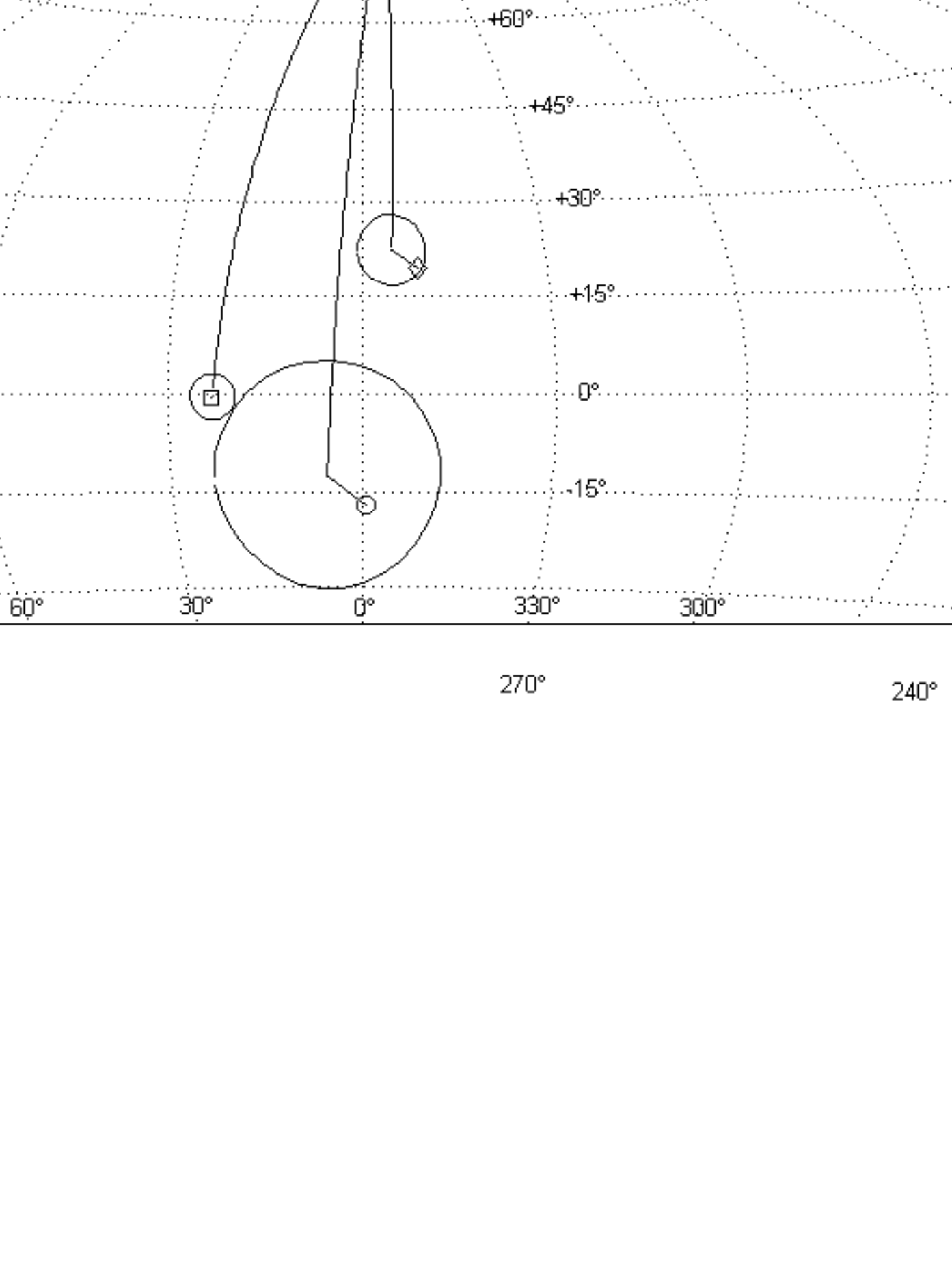}
\caption{Past trajectories for RBS 1223 and US, Ext. R CrA and Sct OB2, respectively, projected on a Galactic coordinate system (for a particular set of input parameters consistent with \autoref{tab:detailed_scan_1308}). Present positions are marked with a star for the neutron star, a diamond for US, an open circle for Ext. R CrA and a square for Sct OB2. Large circles reflect association extensions (radii of $\unit[15]{pc}$ for US, $\unit[31]{pc}$ for Ext. R CrA) and $\unit[31]{pc}$ for Sct OB2.}
\label{fig:1308USExtRCrASctOB2_traj}
\end{figure}

All associations for which we found a peak in their $\tau$-$d_{min}$ contour plot that was consistent with the association boundaries (\autoref{tab:detailed_scan_1308}, top) belong to the YLA (F08) and imply a rather small present distance of RBS 1223 to the Sun. Regarding the hydrogen column density of $n_H = \unit[1{.}8\pm0{.}2\cdot10^{20}]{cm^{-2}}$ \citep{2007Ap&SS.308..619S} which is comparable to other M7 members and even a bit higher (see \autoref{tab:vierM7_Eigensch}), the distance of RBS 1223 to the Sun is supposed to be at least $\unit[180]{pc}$ (the distance to the Sun of the probably closest member \rxja{} is $\approx\unit[180]{pc}$), corresponding to a parallax smaller than approximately $\unit[5{.}5]{mas}$. For that reason, we can exclude three of the YLA -- Tuc-Hor, $\beta$ Pic-Cap and AB Dor -- for which the current distance to RBS 1223 would be smaller than $\unit[100]{pc}$.\footnote{As for \rxjs{}, the distance of RBS 1223 is too poorly known to estimate probabilities for the potential parent associations. For that reason, we give an individual discussion.} From the $\tau$-$d_{min}$ contour plot for HD 141569, a potential supernova in HD 141569 would have occurred near the edge of the group our even outside (between $5$ and $\unit[21]{pc}$, the radius of HD 141569 is $\approx\unit[16]{pc}$) and thus is unlikely.\\
Indeed, the case Ext. R CrA is very promising, since the shape of the $d_{min}$ distribution is in good agreement with a three-dimensional Gaussian distribution with $\mu=0$ and $\sigma=13{.}5$ (see \autoref{subfig:hist_1308_ExtRCrA}, \autoref{eq:3DGaussmunull_diff}). Hence, a potential supernova which created RBS 1223 would have occurred near the centre of the association. Assuming this, RBS 1223 would now be $\approx\unit[0{.}5]{Myr}$ old. Its current radial velocity would be $\approx\unit[350]{km/s}$ (space velocity $\approx\unit[400]{km/s}$). Furthermore, it is very likely that Ext. R CrA experienced a supernova in the near past owing to the existence of B-type stars (highest content within the YLA, F08).\\
In \autoref{tab:detailed_scan_1308} (bottom) we also give the results for the US association which has also been suggested as a possible birth place for RBS 1223 by \citet{2009A&A...497..423M}. Although the $d_{min}$ histogram in this case does not support a birth in US (a supernova is predicted at a distance from the US centre of $\unit[51\pm12]{pc}$, see \autoref{subfig:1308_US_dmin}), it seems possible that the neutron star was born in the Sco-Cen complex (similar to \rxjs{}, see previous section) in the vicinity of US. In this case, the present distance of RBS 1223 to the Sun would be slightly larger as for Ext. R CrA ($\approx\unit[215]{pc}$) whereas age and radial velocity are similar to those for Ext. R CrA ($\tau\approx\unit[0{.}5]{Myr}$, $v_r\approx\unit[350]{km/s}$, space velocity $\approx\unit[420]{km/s}$). Adopting a near birth scenario either in the Sco-Cen complex or Ext. R CrA $\unit[0{.}5]{Myr}$ ago (\autoref{fig:1308USExtRCrASctOB2_traj}), we can estimate the mass of the progenitor of RBS 1223. Given association ages of $5$ to $\unit[10]{Myr}$ for US and $10$ to $\unit[15]{Myr}$ for Ext. R CrA, the mass would be $17$ (B0V) to $\unit[45]{M_{\odot}}$ (O5.5V) (US) or $12$ (B1V) to $\unit[30]{M_{\odot}}$ (O7V) (Ext. R CrA) (T80, MM89, K97; spectral types from \citealt{Schmidt-Kaler1982}).\\

However, since the distance to RBS 1223 is very uncertain and following P07 can only hardly get constrained through $n_H$, we eventually cannot completely exclude the associations in the lower part of \autoref{tab:1308_smallestsep}. Infact, in the case RBS 1223 P07 reached the limit of their modelling which was set at $\unit[1]{kpc}$. Thus, the neutron star might be located at somewhat larger distances which would be consistent with its origin lying in the Sct region. If so, we can again estimate the mass of the progenitor star adopting an age of RBS 1223 of $\unit[1]{Myr}$ (\autoref{fig:1308USExtRCrASctOB2_traj}) and Sct OB2 of $\unit[6]{Myr}$ \citep{1985A&A...143L...7S}. Models from T80, MM89 and K97 yield masses of $\unit[44]{M_{\odot}}$, $\unit[38]{M_{\odot}}$ and $\unit[32]{M_{\odot}}$ ($\approx$O6 on the main sequence).


\subsection{Remarks}\label{subsec:remarks}

It has to be mentioned that any association identified for which it is possible to trace our neutron stars back such that they reach the association boundary (although it might be only fulfilled for a very small fraction of runs) is a potential parent association. However, for a few associations, theoretical curves (\autorefs{eq:3DGauss_diff} and \ref{eq:3DGaussmunull_diff}) fit the $d_{min}$ histograms remarkably well. Hence, we conclude that they reflect the real separation at the encounter time (flight time).\\
Furthermore, it is important to bear in mind that the initially input distribution for radial velocities (Ho05) somewhat biases the outcoming radial velocities to median ranged values ($\approx\unit[\pm400]{km/s}$, depending on the transverse velocity). 
Due to this bias, the range of $\tau$ (flight time) might be underestimated. Nonetheless, our method allows us to also identify associations requiring either very small or high radial velocities. \\

Compared to previous investigations of the origin of M7 members as done by e.g. \citet{2002ApJ...576L.145W}, \citet{2003A&A...408..323M}, \citet{2007ApJ...660.1428K} and \citet{2009A&A...497..423M}, our calculations do not only cover certain radial velocity ranges, but the whole spectrum of possible radial velocities ranging from $-1,500$ to $\unit[+1,500]{km/s}$. To account for the velocity distribution for neutron stars we used the probability distribution of Ho05. Furthermore, we vary the distances for \rxjs{} and RBS 1223 in a wide range to account for their large uncertainties.\\
We compiled a list of 140 OB associations and clusters to cover most\footnote{However, some neutron stars may have been former massive runaway stars.} of the potential birth sites of neutron stars.\\

For estimating the supernova progenitor star mass from the difference between association age and kinematic age, we assumed that all stars formed about at the same time in that association. This assumption is justfied, because once an O or B star is formed, its strong wind and slightly later its supernova shock front will blow away all the remaining gas, so that star formation will cease in that region. This assumption holds for OB associations and is consistent with observations. It does not neccessary hold for associations without an O- or B-type star, like possibly TWA, which however may have had an early B-type star. 


\section{Conclusions}\label{sec:conclusions}

\begin{table*}
\centering
\caption{Summary of the results discussed in \autoref{sec:idparassoc}. Given here are those associations that we conclude to be the most probable birth places of the four M7 members. Columns 3 to 5 give the position of the potential supernova (SN) as equatorial coordinates and distance to the Sun, column 6 gives the approximate time in the past at which the SN may have occurred and columns 7 to 10 give constraints on the neutron star parameters as they would be needed to reach the respective association at the given time.}\label{tab:summary}
\footnotesize
\setlength\extrarowheight{2pt}
\begin{tabular}{l | l | >{$}c<{$} >{$}c<{$} >{$}c<{$} | c | >{$}c<{$} >{$}c<{$} o{3.2} o{3.2}}
\toprule
RX J & Association & \multicolumn{3}{c|}{Position of SN} & Time& \multicolumn{4}{c}{Present-Day Parameters}\\
							 &						 & \multicolumn{1}{c}{$\unit[\alpha]{[^\circ]}$} & \multicolumn{1}{c}{$\unit[\delta]{[^\circ]}$} & \multicolumn{1}{c|}{$\unit[d_{\odot}]{[pc]}$} & of SN & \multicolumn{1}{c}{$\unit[\pi]{[mas]}$} & \multicolumn{1}{c}{$\unit[v_r]{[km/s]}$} &  \multicolumn{1}{c}{$\unit[\mu_{\alpha}^*]{[mas/yr]}$} & \multicolumn{1}{c}{$\unit[\mu_{\delta}]{[mas/yr]}$}\\\midrule
1856.5-3754 & US						 & 243{.}21^{+0{.}58}_{-0{.}52} & -23{.}66^{+0{.}31}_{-0{.}29} & 140\ldots163 & $\unit[0{.}3]{Myr}$ & 5{.}41^{+0{.}33}_{-0{.}28} & 193^{+45}_{-32} & 326{.}7+0{.}8 & -59{.}1+0{.}7\\\hline
0720.4-3125 & TWA	       & 187{.}10^{+3{.}44}_{-2{.}91} & -37{.}98^{+1{.}53}_{-1{.}38} & 48\ldots67 & $\unit[0{.}4]{Myr}$ & 4{.}06^{+0{.}47}_{-0{.}49} & 502^{+111}_{-88} & -93{.}9+2{.}2 & 52{.}8+2{.}3\\
							 & Tr 10	       & 135\ldots140 & -40{.}19^{+0{.}58}_{-0{.}59} & 335\ldots375 & $\unit[0{.}5]{Myr}$ & 41{.}93^{+0{.}24}_{-0{.}20} & 290^{+143}_{-110} & -93{.}8+2{.}2	& 53{.}1+2{.}2\\\hline
1605.3+3249  & Sco-Cen				 & 255{.}09^{+0{.}57}_{-0{.}66} & -18{.}44^{+1{.}00}_{-1{.}06} & 129\ldots171 & $\unit[0{.}5]{Myr}$ & 3{.}38^{+0{.}47}_{-0{.}41} & 343^{+86}_{-80} & -43{.}7+1{.}6 & 148{.}8+2{.}6\\
								& Ext. R CrA& 261{.}99^{+1{.}12}_{-0{.}99} & -39{.}84^{+2{.}10}_{-2{.}00} & 70\ldots115 & $\unit[0{.}5]{Myr}$ & 4{.}41^{+0{.}90}_{-0{.}85} & 321^{+109}_{-58} & -43{.}7+1{.}7 & 148{.}8+2{.}6\\\hline
RBS 1223			  & Sco-Cen		 & 254{.}09^{+2{.}11}_{-2{.}34} & -6{.}29^{+3{.}23}_{-5{.}09} & 109^{+36}_{-19} & $\unit[0{.}5]{Myr}$ & 4{.}63^{+0{.}84}_{-1{.}27} & 351^{+100}_{-104} & -210+20 & 83+18\\
							  & Ext. R CrA		 & 294{.}82^{+7{.}24}_{-7{.}92} & -22{.}94^{+4{.}33}_{-6{.}29} & 80^{+22}_{-10} & $\unit[0{.}5]{Myr}$ & 5{.}78^{+1{.}33}_{-1{.}21} & 345^{+114}_{-90} & -209+20 & 85+18\\
							  & Sct OB2A		 & 277{.}15^{+2{.}47}_{-2{.}60} & -12{.}04^{+2{.}13}_{-1{.}83} & 517^{+18}_{-19} & $\unit[1]{Myr}$ & 2{.}70^{+0{.}96}_{-0{.}44} & 75\ldots310 & -216+19 & 53+10\\\bottomrule
\end{tabular}
\end{table*}

We reviewed the prominent neutron star/ runaway case PSR B1929+10/ $\zeta$ Oph (H01) and reinvestigated it with more recent data for the pulsar. Although we did not put constraints on the radial velocity, we could confirm previous results from \citet{2008AstL...34..686B} and eventually derive a radial velocity of $\approx\unit[250]{km/s}$ of the pulsar which is needed to find the two objects at nearly the same position in US at the same time in the past. However, we should bear in mind that we had to increase the errors of the parallax and proper motion components. If they are truly as small as published by \citet{2004ApJ...604..339C} the scenario is less likely.\\

We investigated the motion of four members of the M7 and for each neutron star simultaneously the motion of 140 OB associations and clusters to confine potential parent associations of the neutron stars. A summary of our results is given in \autoref{tab:summary} (for individual discussions please see previous sections).\\
We find that \rxja{} most probably originated from US about $\unit[0{.}3]{Myr}$ ago. \rxjn{} was very likely born in TWA about $\unit[0{.}4]{Myr}$ ago or may come from Tr 10 where the supernova then would have occured $\unit[0{.}5]{Myr}$ in the past. \rxjs{} and RBS 1223 also were seemingly born from a close young association such as the Scorpius-Centraurus complex or Ext. R CrA. For RBS 1223 also a birth in Scutum OB2 is possible.\\

Thus, we find that the most probable birth places were located close to the Sun. The supernova distances are consistent with the Local Bubble and hence may have contributed to its formation or reheating (\citealt{2002A&A...390..299B}; F08). There is no doubt that the Sco OB2 association to which US belongs experienced some supernovae in the past (\citealt{2001ApJ...560L..83M,2002A&A...390..299B}; F08).  There have been other investigations showing that many neutron stars may have been born within the Gould Belt \citep{2003A&A...406..111P}, a torus-like structure around the Sun with a radius of $\approx\unit[600]{pc}$ \citep{2000A&A...359...82T}, to which most of the associations just mentioned also belong.\\
Some of the four investigated M7 neutron stars may have formed within $\approx\unit[250]{pc}$ within the last $\unit[0{.}5]{Myr}$, so that they could have contributed to the $^{10}$Be and $^{60}$Fe found in the Earth's crust.

\begin{figure}
\includegraphics[width=0.45\textwidth, viewport= 85 265 480 590]{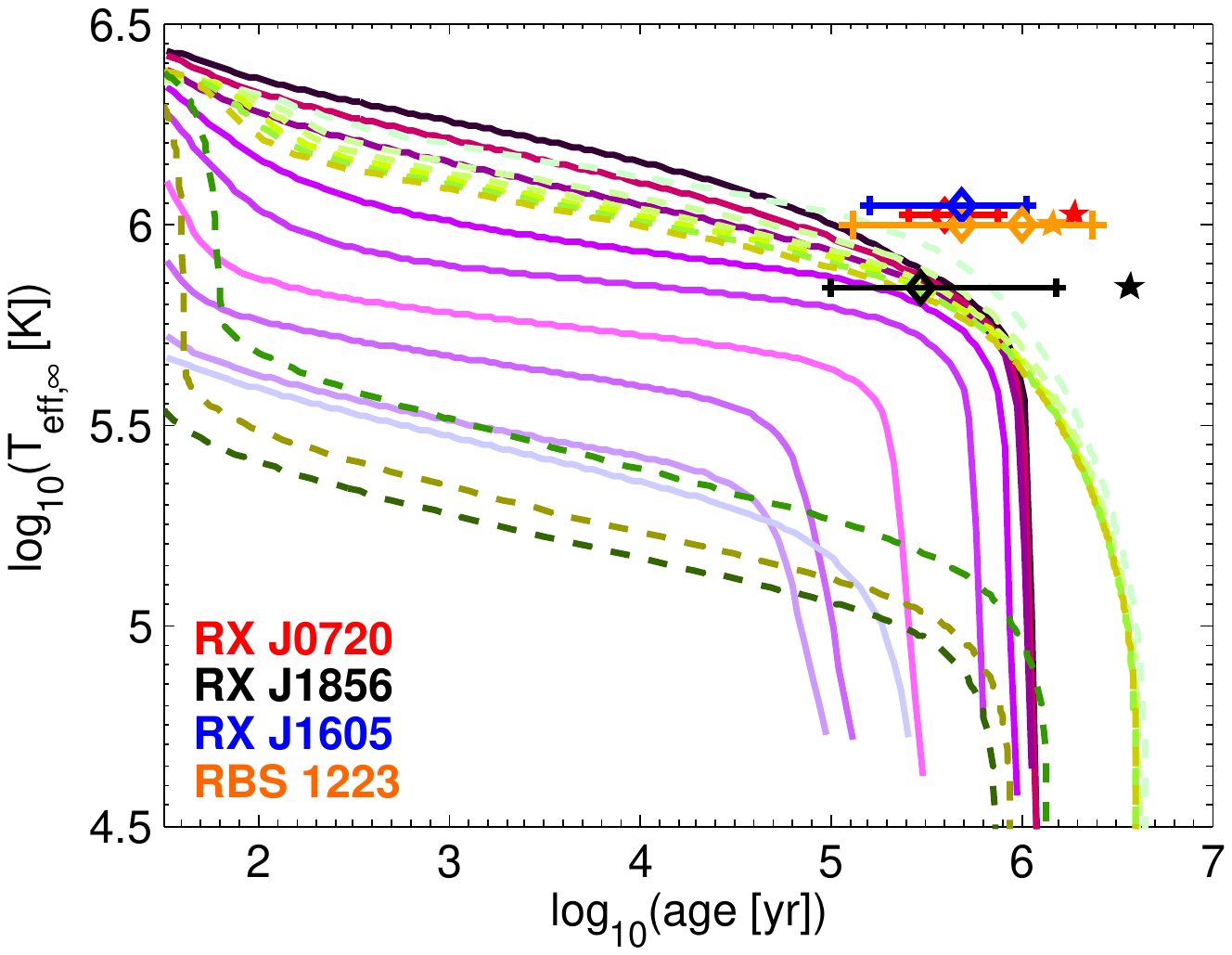}
\caption{The four M7 members inserted into a cooling diagram. Filled stars mark the characteristic spin-down age (see \autoref{tab:agemass}) whereas horizontal lines characterise an area of the kinematic age (lower and upper values from associations in tables of \autoref{sec:idparassoc}). Open diamonds show the kinematic age for the associations summarised in \autoref{tab:summary}. Effective temperatures can be found in \autoref{tab:agemass}.\newline
The purple set of cooling curves was adopted from \citet{2006PhRvC..74b5803P} (solid lines, for masses of $1{.}05$, $1{.}13$, $1{.}22$, $1{.}28$, $1{.}35$, $1{.}45$, $1{.}55$, $1{.}65$ and $\unit[1{.}75]{M_{\odot}}$ from top to bottom; Model from \citealt{2005PhRvC..71d5801G}), the green set has been kindly provided by A. D. Kaminker (dashed lines, includes superconductive protons and neutrons, for masses from $1{.}1$ to $\unit[1{.}9]{M_{\odot}}$ from top to bottom; see also \citealt{2005MNRAS.363..555G}).}
\label{fig:cooling}
\end{figure}

Adopting a birth scenario of each neutron star, we can derive estimations of the mass of the progenitor star assuming contemporary star formation in the association as well as the kinematic neutron star age which we here always find to be lower than the characteristic spin-down age (\autoref{tab:agemass}) that is an upper limit to the true age. From cooling models we know that, given the surface temperature, for the neutron stars investigated here, the characteristic age is too high. This is also plausible since those objects are still very young. Thus, our kinematic ages better fit with cooling models (see \autoref{fig:cooling}).\\
Since \rxja{} is the coolest of the four M7 members investigated (\rxjn{}, \rxjs{} and RBS 1223 show similar temperatures), it should be the oldest. However, it seems, that \rxja{} is slightly younger than the other three (or has a similar kinematic age if we adopt the parallax of \citealt{2000IAUS..195..437W}). This might be owing to different parameters at birth or different cooling. Also the kinematic age of \rxja{} is of about a factor of $13$ smaller than its characteristic age which is a large difference. This may indicate further doubts on the characteristic age. \citet{2003ApJ...592L..75M} propose that \rxja{} is an isolated magnetar that has spun down by the propeller effect. In this case, dipole braking, which is assumed to calculate characteristic ages, is no longer valid. Moreover, \rxja{} does only show a very small pulse fraction making it difficult to derive $\dot{P}$ \citep{2008ApJ...673L.163V}.\\
In addition, effective temperatures may be influenced by hot spots and do not reflect the real surface temperature. Infact, this is the case for \rxjn{} \citep{2006A&A...451L..17H,2009A&A...498..811H} and RBS 1223 \citep{2005A&A...441..597S,2007Ap&SS.308..181H} which appear hotter, thus must be actually shifted somewhat to smaller temperatures (even more consistent with cooling curves).\\

\stepcounter{footnote}
\begin{table}
\centering
\caption{Kinematic ages $\tau_{kin}$ of the four M7 members compared with their characteristic spin-down ages $\tau_{char}$ (taken from the ATNF pulsar database${}^{\thefootnote}$, \citealt{2005AJ....129.1993M}). The fifth column gives the effective temperature as listed in Table 1 of \citet{2005fysx.conf...39H} (see references therein). Column 6 gives the median masses (for TWA a mass consistent with its mass function) of the progenitor stars predicted using evolutionary models from T80, MM89 and K97 (see text) given the difference between the ages of the associations and the time since the predicted supernovae.
}\label{tab:agemass}
\footnotesize
\setlength\extrarowheight{2pt}
\begin{tabular}{l l d{2.1} d{1.1} d{4.0} d{3.0} }
\toprule
RX J	&	Association	& \multicolumn{1}{c}{$\tau_{kin}$} & \multicolumn{1}{c}{$\tau_{char}$}  &	\multicolumn{1}{c}{$T_{eff,\infty}$} & \multicolumn{1}{c}{$M_{prog}$} \\
							&							&	\multicolumn{1}{c}{[Myr]}				 & \multicolumn{1}{c}{[Myr]}					& \multicolumn{1}{c}{[$\unit[10^3]{K}$]} & \multicolumn{1}{c}{[$\mathrm{M_{\odot}}$]} \\\midrule
1856.5-3754 & US 					&  \approx0,3		& 3,8  				& 696		  	&  \approx29  \\\hline
0720.4-3125 & TWA   			&  \approx0,4		& 1,9	 				& 1044  		&  \approx10  \\
						& Tr 10 			&  \approx0,5		& 	 					&    				&  \approx10  \\\hline
1605.3+3249 & Sco-Cen			&  \approx0,5		& \multicolumn{1}{c}{--}	&  1113   & \approx29 \\
						& Ext. R CrA	&  \approx0,5		& 						&   			  & \approx15   \\\hline
RBS 1223		&	Sco-Cen			&  \approx0,5		& 1,5  				& 998    		&  \approx29  \\
						& Ext. R CrA  &  \approx0,5		& 						&						&  \approx15  \\
						& Sct OB2			&  \multicolumn{1}{c}{$\approx1$}	&		&		&  \approx38  \\
\bottomrule
\end{tabular}
\end{table}
\footnotetext{\href{http://www.atnf.csiro.au/research/pulsar/psrcat/}{http://www.atnf.csiro.au/research/pulsar/psrcat/}}

\section*{Acknowledgments}


We thank Thomas Eisenbeiss for data on the Hercules-Lyrae assocation and Sergei B. Popov and Alexandr Loktin for their assistance with the preparation of our sample of OB associations and clusters. We also would like to thank Valeri V. Hambaryan for discussions and data on RBS 1223 as well as Dieter Breitschwerdt and Burkhard Fuchs for discussions on the Local Bubble. Furthermore, we thank A. D. Kaminker, Hovic Gregorian and David Blaschke for providing us their cooling curves. We also would like to thank Jos de Bruijne for discussions on the statistics and Fred Walter for carefully reading the manuscript and helping to improve the paper.\\
RN acknowledges general support from the German National Science Foundation (Deutsche Forschungsgemeinschaft, DFG) in grants NE 515/23-1 and SFB/TR-7, MMH and NT acknowledge partial support from DFG in the SFB/TR-7 Gravitational Wave Astronomy. The work of MMH and NT has been supported by CompStar, a research networking  programme of the European Science Foundation (ESF). GM acknowledges support from the EU in the FP6 MC ToK project
MTKD-CT-2006-042514. Our work has made use of the ATNF Pulsar Catalogue, version of February 2009 \citep{2005AJ....129.1993M}. \\
We would like to thank Arnold Benz, who has motivated this study with one sentence in his new book ``Das geschenkte Universum''.

\bibliographystyle{mn2e}
\bibliography{bib_M7paper}

\appendix

\section{Sample of OB associations and clusters}\label{appsec:Assoclist}

For our investigation we selected OB associations and clusters within $\unit[3]{kpc}$ from the Sun (\autoref{tab:Assoc_list}; for detailes on the selection criteria, please see \autoref{sec:assocsample}).

\begin{table*}
\caption{The sample of OB associations and clusters. The table gives the consecutive number, the designation, the position on the sky (J2000 galactic longitude $l$ and latitude $b$), the parallax $\pi$, heliocentric velocity components $U$, $V$ and $W$ as well as estimated ages and spatial extensions either as found in the literature or obtained from angular extensions and distances. Besides Sco OB2 and YLA, $U$, $V$ and $W$ were computed as in \citet{1987AJ.....93..864J} (errors include proper motion and radial velocity uncertainties only). Note that ages are often uncertain to a factor up to two. Many OB associations contain several clusters of different ages (many of them listed here as well). \newline The last column lists the references used for each group: A06 -- \citet{2006A&A...454..775A}, B08 -- \citet{2008AstL...34..686B}, BaL95 -- \citet{1995MNRAS.276..627B}, BH89 -- \citet{1989AJ.....98.1598B} (according to \citet{2001AstL...27...58D} distances taken from this publication have been reduced by 20\%), Bh07 -- \citet{2007BASI...35..383B}, Bl56 -- \citet{1956ApJ...123..408B}, Bl91 -- \citet{1991psfe.conf..125B}, BL04 -- \citet{2004A&AT...23..103B}, Br99 -- \citet{1999osps.conf..411B}, BYN06 -- \citet{2006A&A...459..511B}, C06 -- \citet{2006MNRAS.365..867C}, CH08 -- \citet{2008hsf1.book..899C}, Cl74 -- \citet{1974A&A....37..229C}, Co02 -- \citet{2002AJ....124.1585C}, D01 -- \citet{2001AstL...27...58D}, dG89 -- \citet{1989A&A...216...44D}, Di02 -- \citet{2002A&A...388..168D}, dlR06 -- \citet{2006AJ....131.2609D}, dZ99 -- \citet{1999AJ....117..354D}, E07 -- \citet{2007DiplA...Thomas}, F08 -- \citet{2008A&A...480..735F}, Fu04 -- \citet{2004AN....325....3F}, G94 -- \citet{1994ApJ...436..705G}, GB08 -- \citet{2008A&A...490.1071G}, GH08 -- \citet{2008hsf2.book....1G}, Gr71 -- \citet{1971AJ.....76.1079G}, GS92 -- \citet{1992A&AS...94..211G}, H01 -- \citet{2001A&A...365...49H}, Ha72 -- \citet{1972A&A....17..413H}, HR76 -- \citet{1976AJ.....81..840H}, Hu78 -- \citet{1978ApJS...38..309H}, J05 -- \citet{2005ApJ...619..945J}, K05 -- \citet{2005A&A...440..403K}, K07 -- \citet{2007AN....328..889K}, KG94 -- \citet{1994MNRAS.269..289K}, KM07 -- \citet{2007ApJ...667L.155K}, L06 -- \citet{2006ApJ...643.1160L}, LB01 -- \citet{2001AstL...27..386L}, LB03 -- \citet{2003ARep...47....6L}, Lo06 -- \citet{2006A&A...450..147L}, Loz86 -- \citet{1986Ap&SS.121..357L}, LS04 -- \citet{2004ApJ...609..917L}, Lu08 -- \citet{2008hsf2.book..169L}, M02 -- \citet{2002A&A...381..446M}, Ma07 -- \citet{2007ApJS..169..105M}, Mae87 -- \citet{1987A&A...178..159M}, Mas95 -- \citet{1995ApJ...454..151M}, Mas01 -- \citet{2001AJ....121.1050M}, MC68 -- \citet{1968ApJS...16..275M}, ME95 -- \citet{1995AstL...21...10M}, Me07 -- \citet{2007A&A...474..515M}, Mer08 -- \citet{2008A&A...485..303M}, Meri04 -- \citet{2004A&A...419..301M}, MP03 -- \citet[][\textit{\textsc{Webda}} database, http://www.univie.ac.at/webda]{2003A&A...410..511M}, MV73 -- \citet{1973A&AS...10..135M}, Mo53 -- \citet{1953ApJ...118..318M}, N00 -- \citet{2000A&AS..146..323N}, P91 -- \citet{1991A&AS...90..195P}, Pr02 -- \citet{2002AJ....124..404P}, S06 -- \citet{2006MNRAS.367..763S}, Sa03 -- \citet{2003A&A...404..913S}, SM85 -- \citet{1985A&A...143L...7S}, SN00 -- \citet{2000A&A...361..581S}, So04 -- \citet{2004MNRAS.351.1277S}, Sod98 -- \citet{1998ApJ...498..385S}, T76 -- \citet{1976ApJ...210...65T}, T79 -- \citet{1979A&A....76..350T}, T80a -- \citet{1980ApJ...235..146T}, T80b -- \citet{1980AJ.....85.1193T}, E07 -- \citet{2007DiplA...Thomas}, Te99 -- \citet{1999A&A...341L..79T}, U01 -- \citet{2001A&A...371..675U}, W07 -- \citet{2007AJ....133.1092W}, We63 -- \citet{1963MNRAS.127...71W}, Web99 -- \citet{1999ApJ...512L..63W}, Wh97 -- \citet{1997A&A...327.1194W}, Wo08 -- \citet{2008hsf2.book..388W}, Z01a -- \citet{2001ApJ...562L..87Z}, Z01b -- \citet{2001ApJ...559..388Z}.\label{tab:Assoc_list}}
\includegraphics*[viewport = 34 57 566 505]{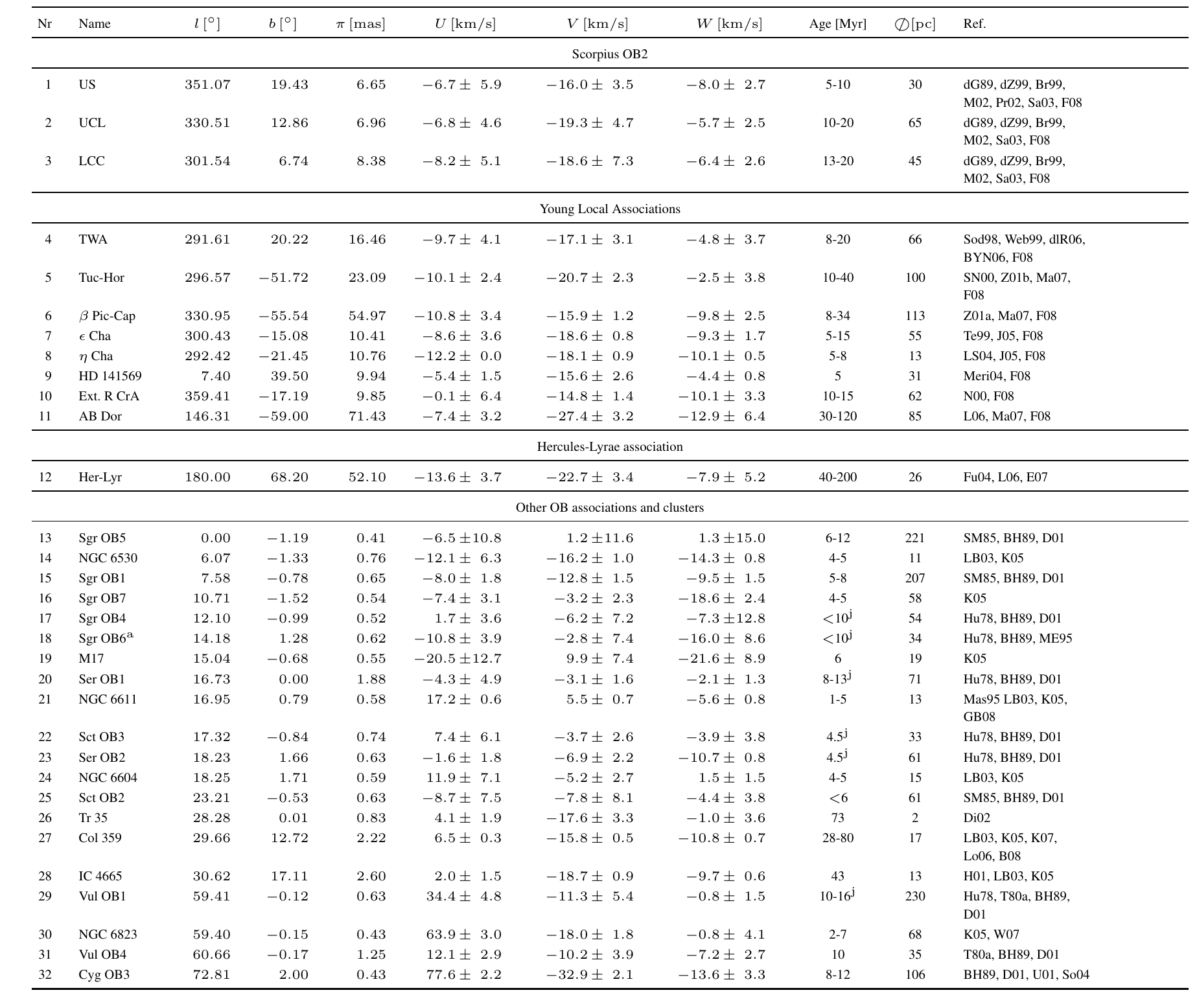}
\end{table*}
\setcounter{table}{-1}
\begin{table*}
\includegraphics[viewport = 41 51 566 744]{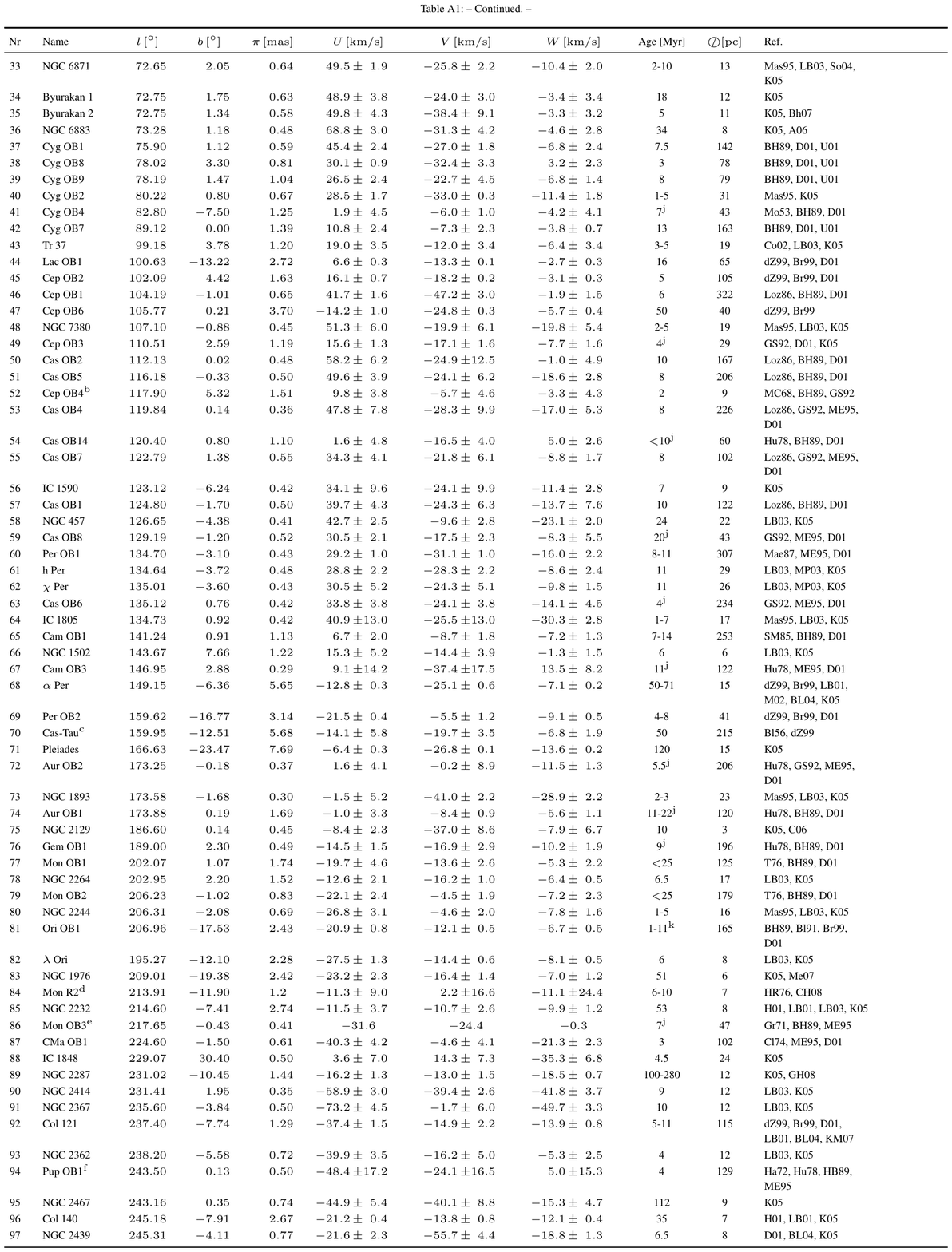}
\end{table*}
\setcounter{table}{-1}
\begin{table*}
\includegraphics[viewport = 35 178 566 744]{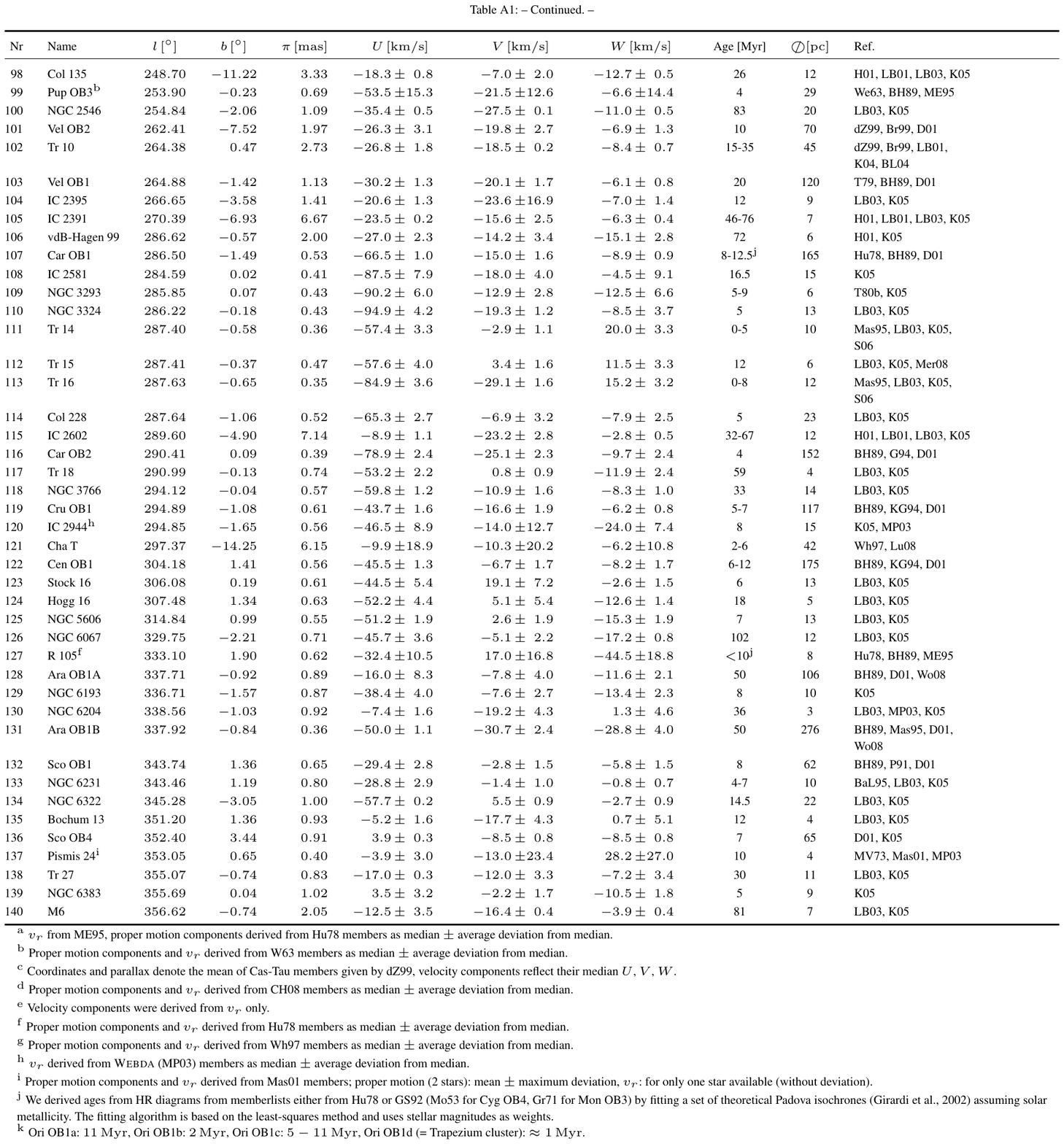}
\end{table*}

\newpage
\normalsize
\section{Derivation of present-day neutron star parameters and supernova position -- example \rxja{} coming from US}\label{appsec:colval}

\begin{figure}
\includegraphics[width=0.45\textwidth]{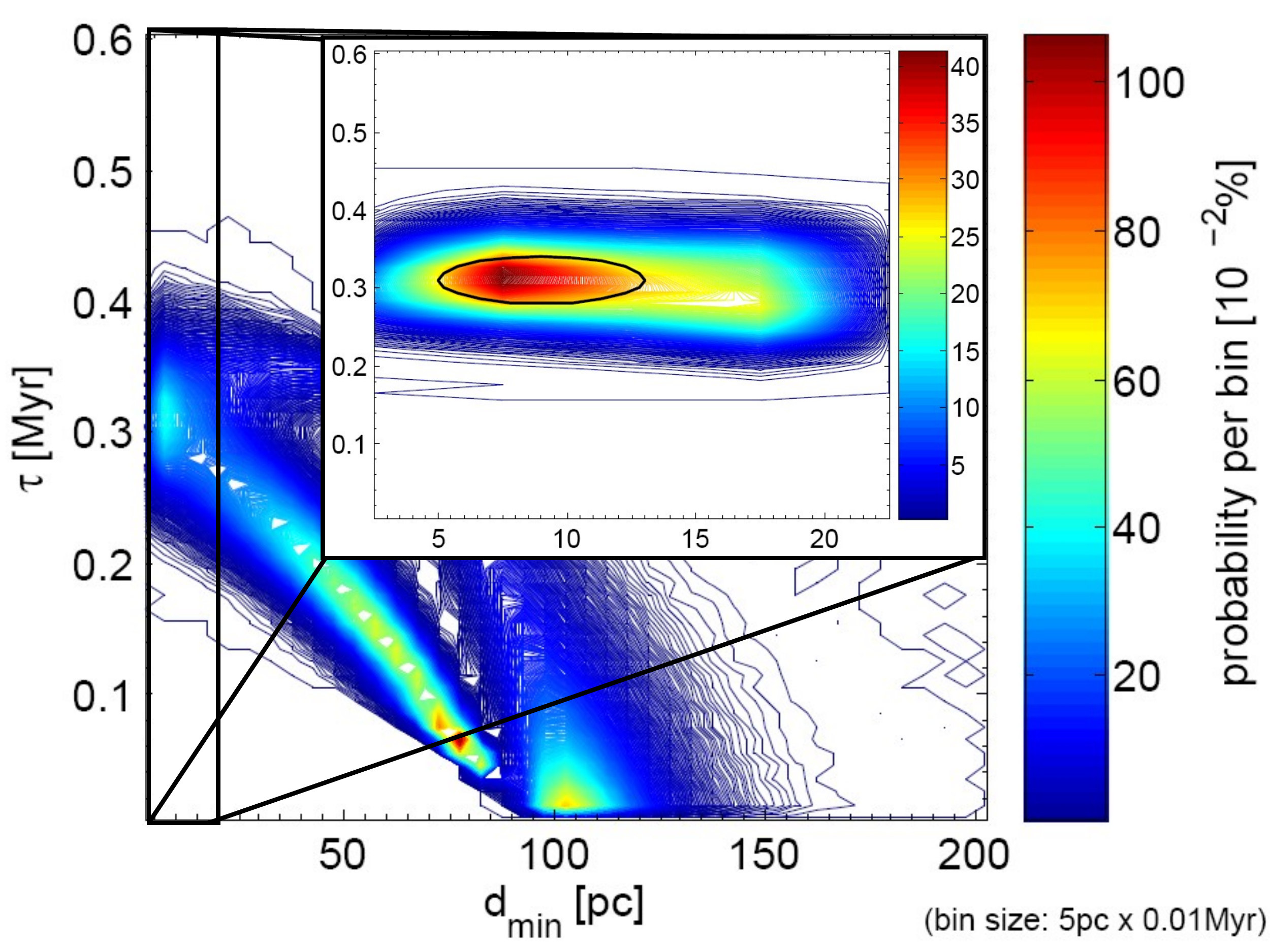}
\caption{Distribution of the number of runs in the $\tau$-$d_{min}$ space (excluding runs yielding $\tau~=~0$) for \rxja{} and US. The ellipse marks the region for determination of the parameters given in \autoref{tab:detailed_scan_1856} and \autoref{fig:hists_US1856}.}
\label{fig:KonturplotUS1856}
\end{figure}

\begin{figure*}
\centering
\subfigure[Parallax $\pi$.]{\includegraphics[width=0.25\textwidth, viewport= 85 265 480 590]{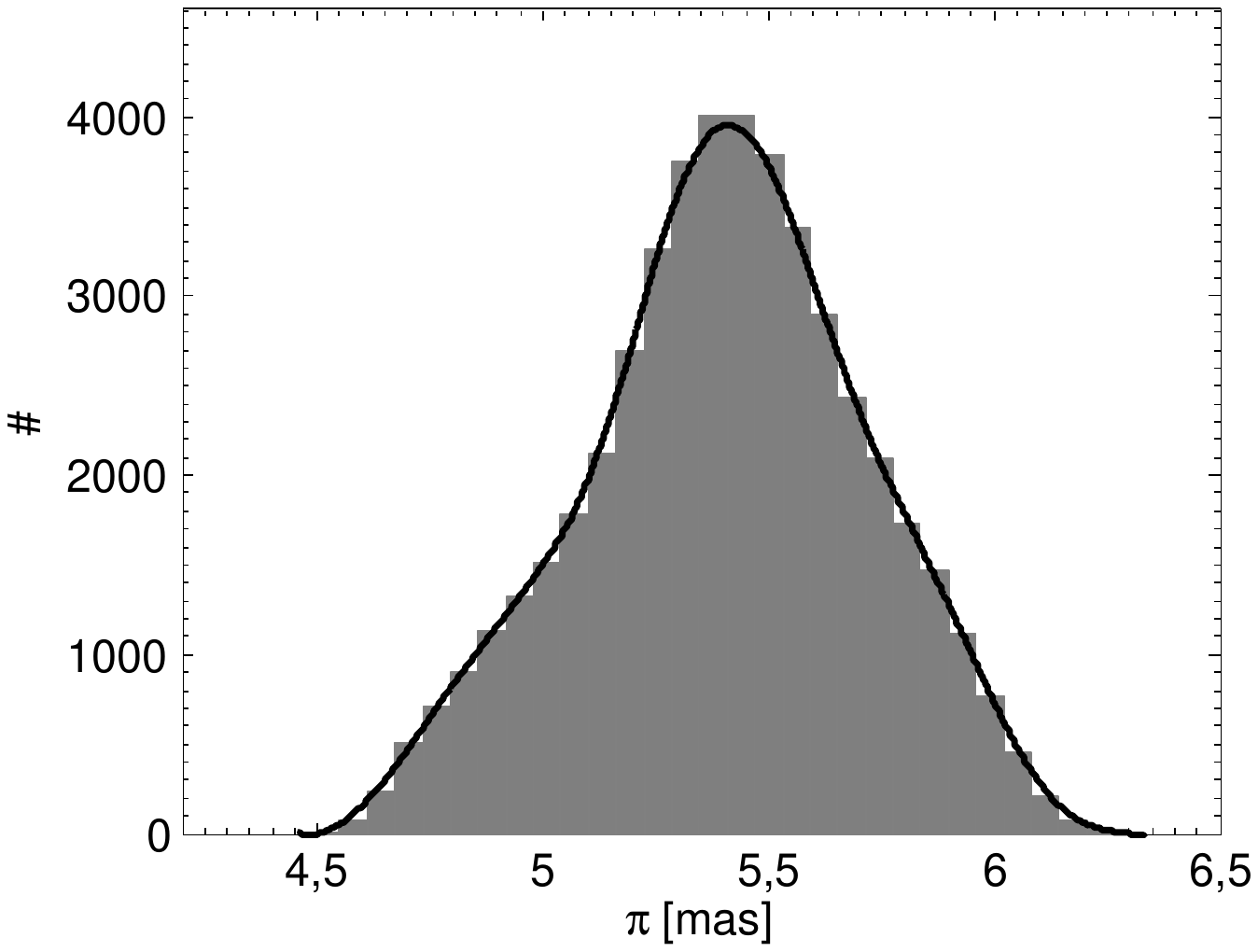}\label{subfig:hist_plx_1856US}}\hspace{1em}\nolinebreak
\subfigure[Radial velocity $v_r$.]{\includegraphics[width=0.25\textwidth, viewport= 85 265 480 590]{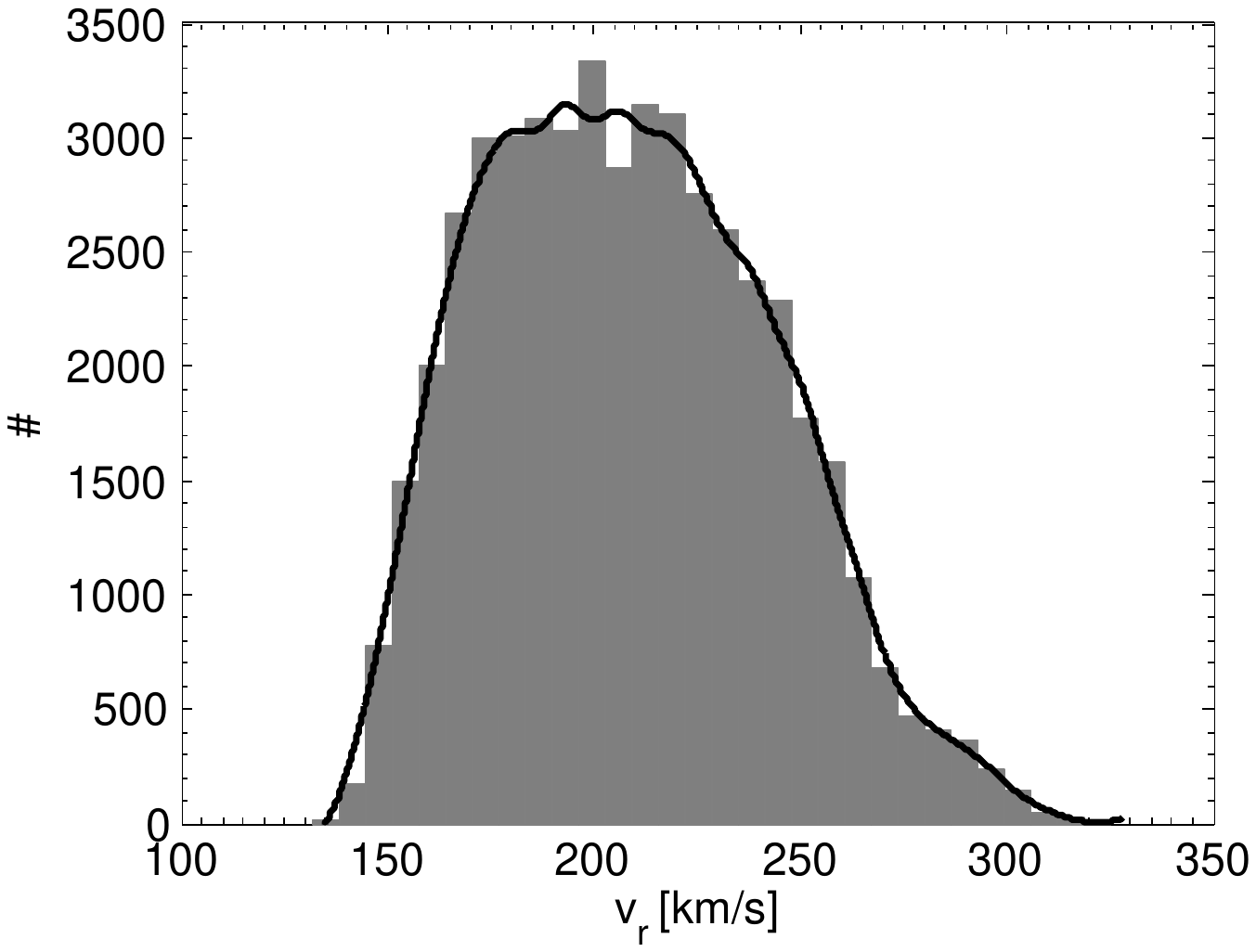}\label{subfig:hist_RV_1856US}}\nolinebreak
\subfigure[Proper motion in right ascension $\mu_{\alpha}^*$.]{\includegraphics[width=0.25\textwidth, viewport= 85 265 480 590]{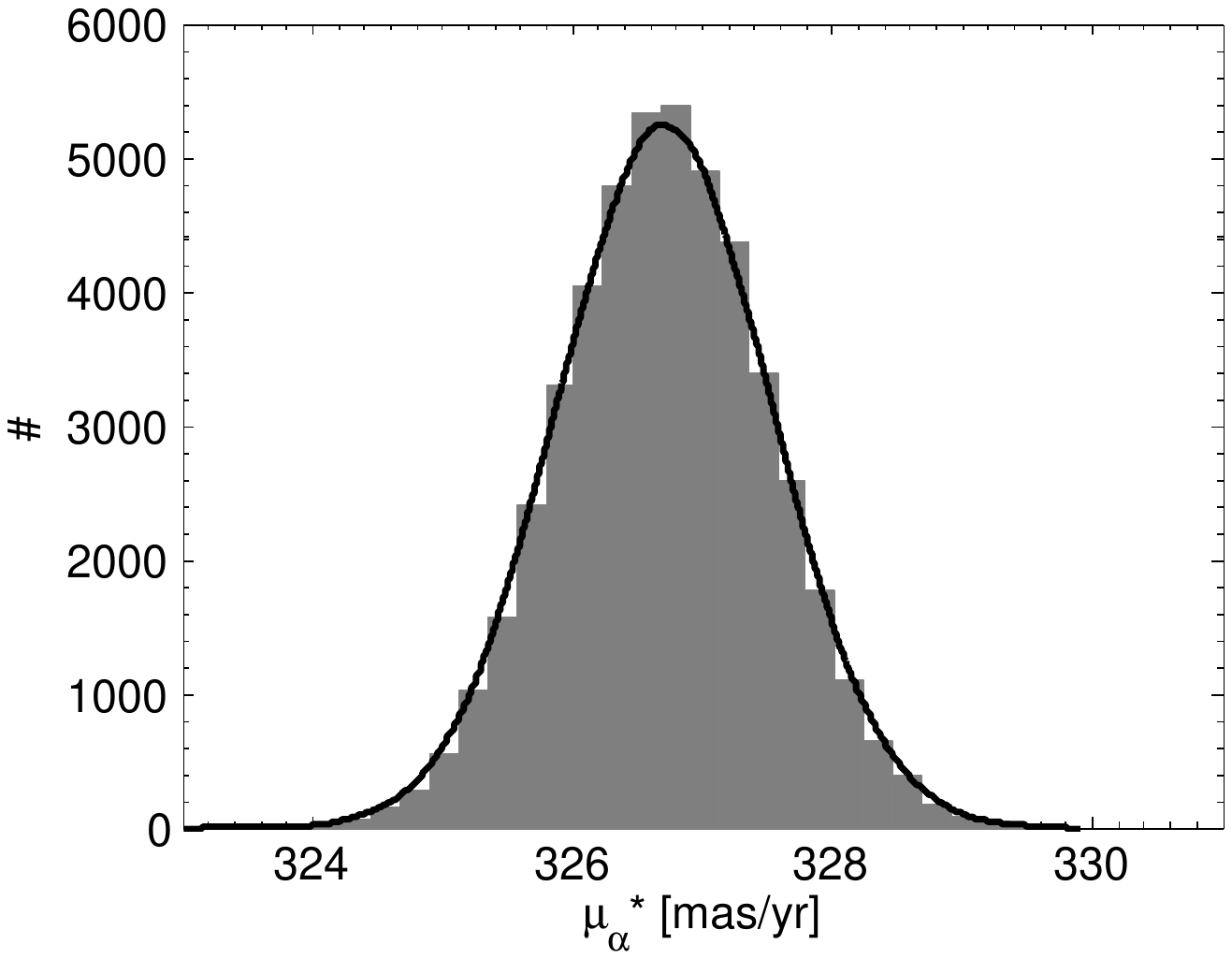}\label{subfig:hist_muRA_1856US}}\hspace{1em}\nolinebreak
\subfigure[Proper motion in declination $\mu_{\delta}$.]{\includegraphics[width=0.25\textwidth, viewport= 85 265 480 590]{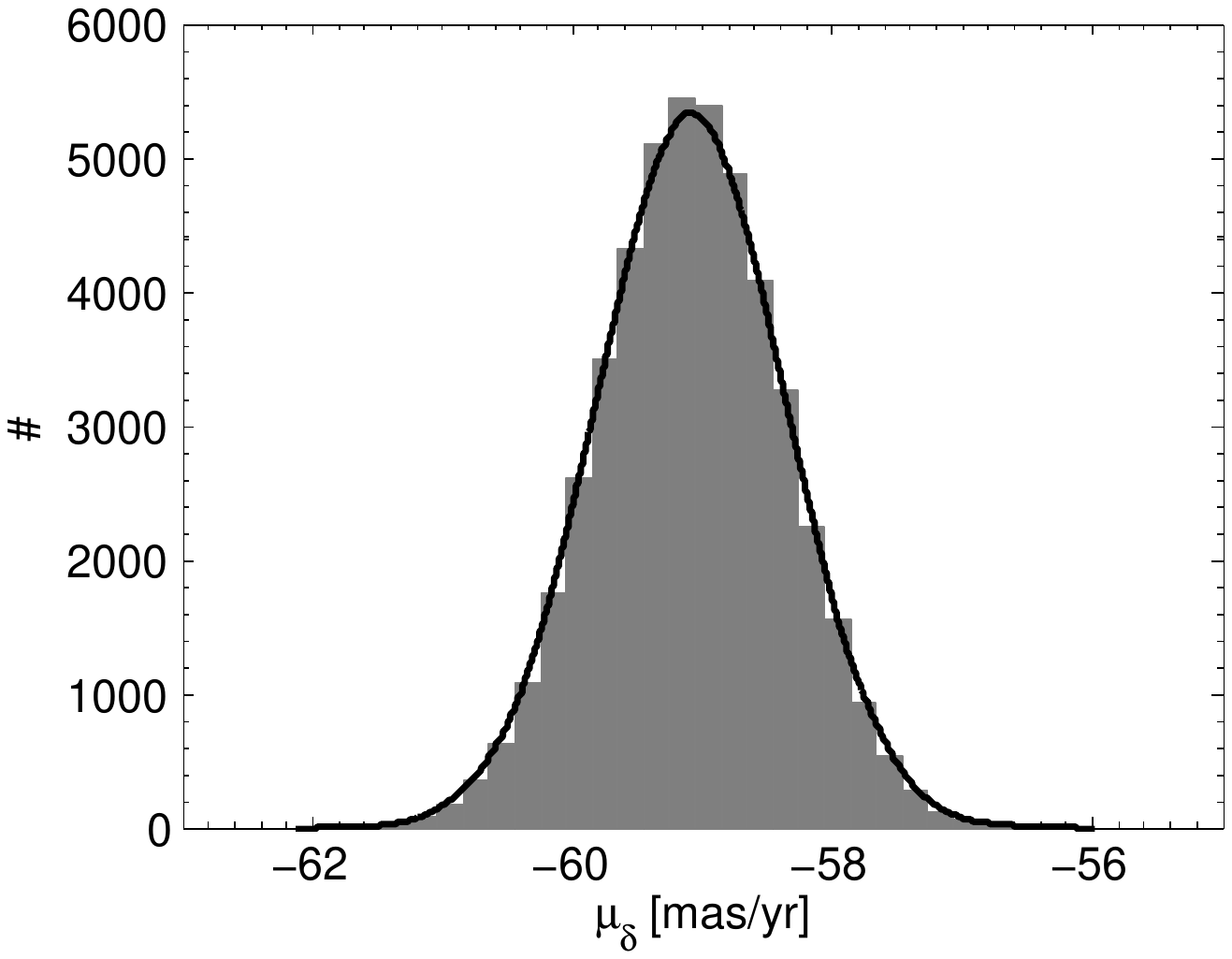}\label{subfig:hist_muDEC_1856US}}\\
\subfigure[Right ascension $\alpha$.]{\includegraphics[width=0.25\textwidth, viewport= 85 265 480 590]{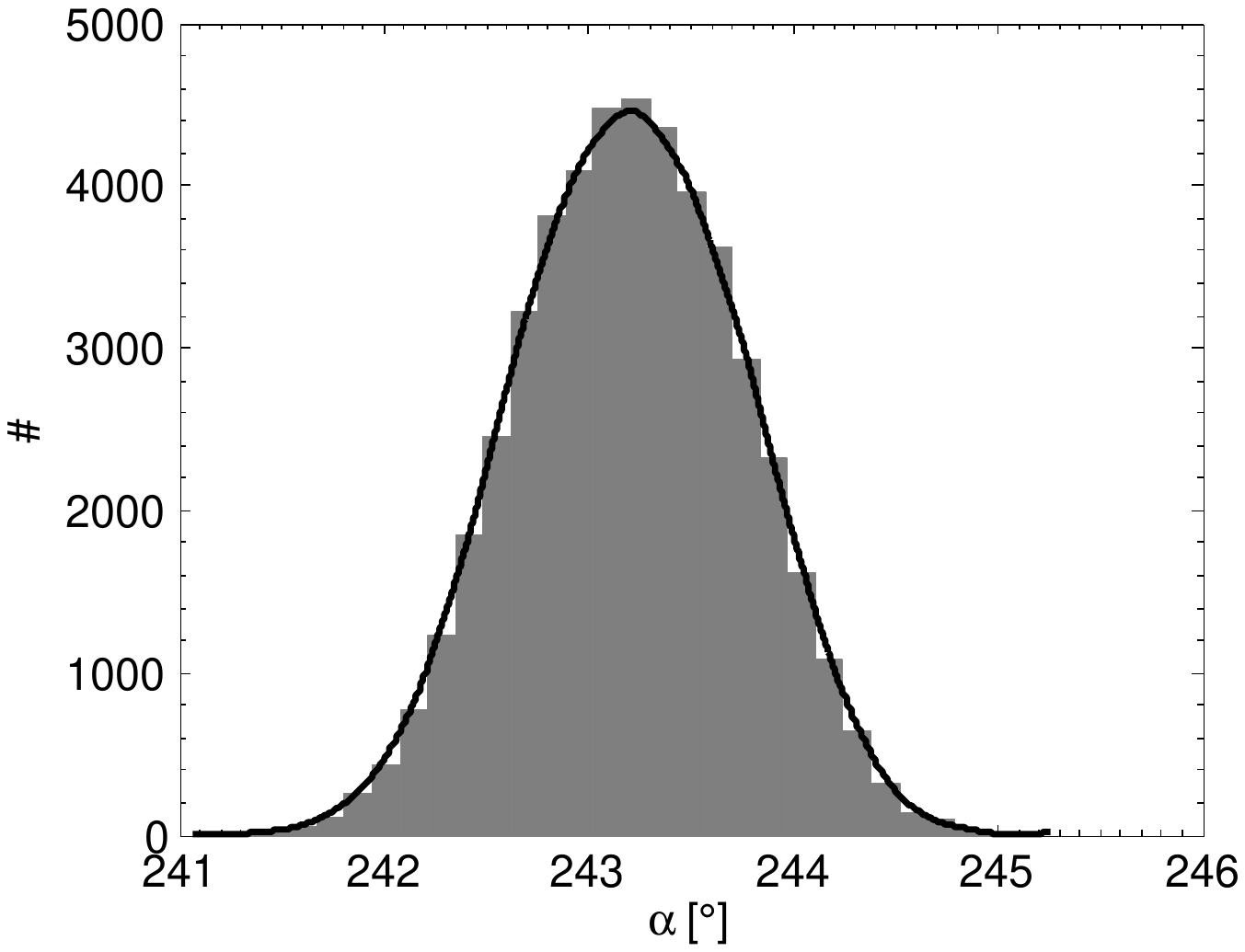}\label{subfig:hist_RA_1856US}}\hspace{1em}\nolinebreak
\subfigure[Declination $\delta$.]{\includegraphics[width=0.25\textwidth, viewport= 85 265 480 590]{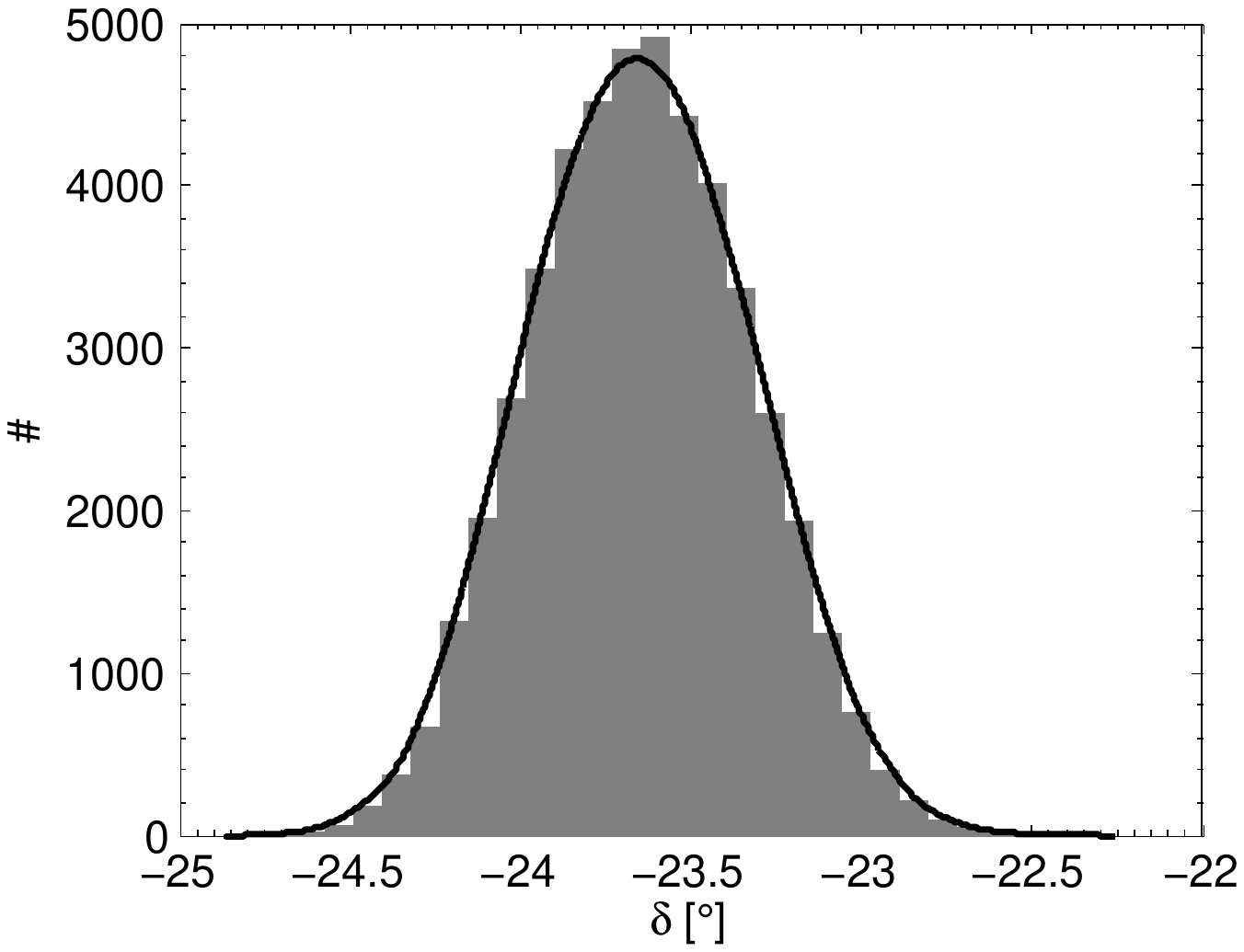}\label{subfig:hist_DEC_1856US}}\nolinebreak
\subfigure[Distance to Sun $d_{\odot}$.]{\includegraphics[width=0.25\textwidth, viewport= 85 265 480 590]{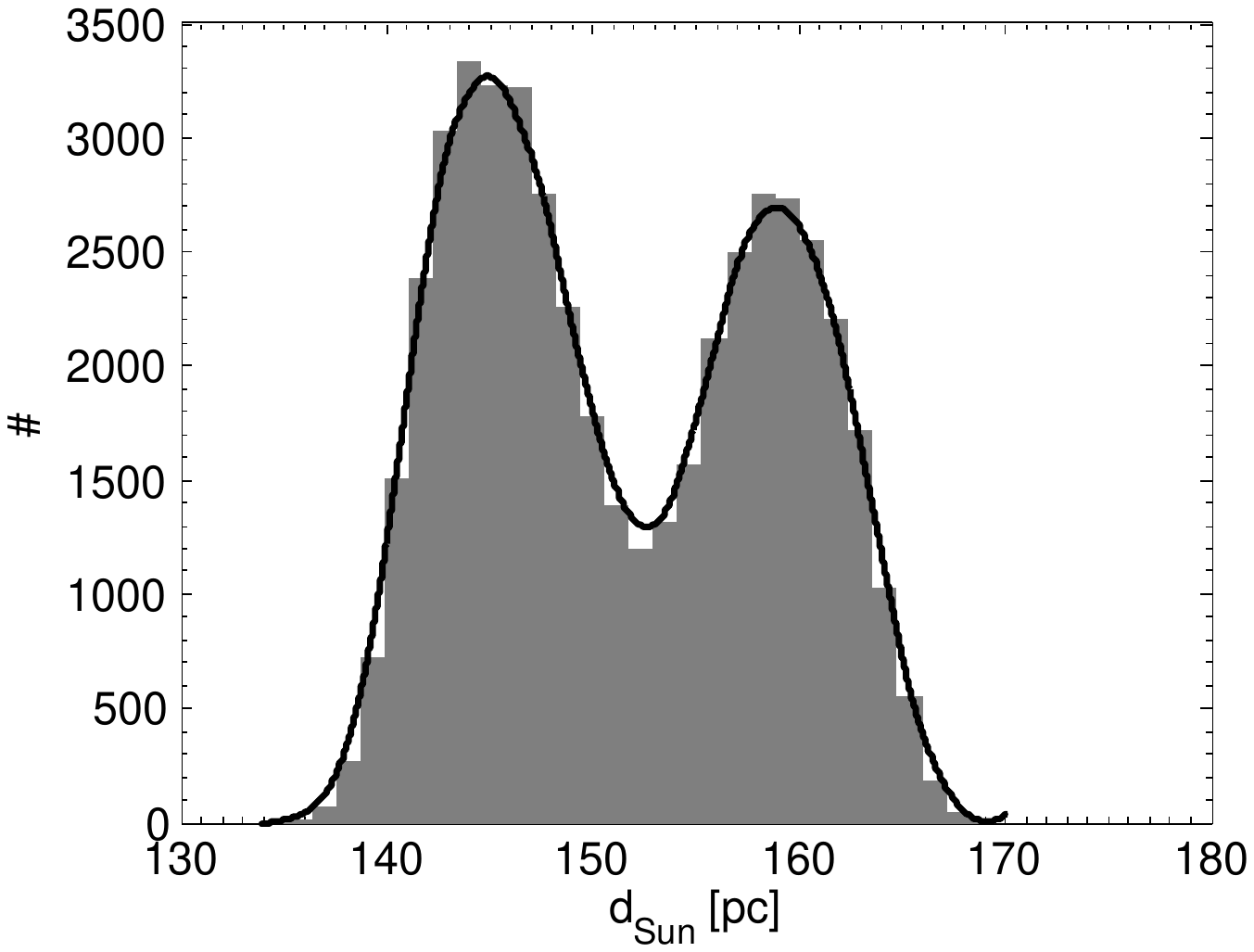}\label{subfig:hist_dErde_1856US}}
\caption{(a)-(d): Distributions of present-day parameters for \rxja{} supporting that it was created in a supernova in US about $\unit[0{.}3]{Myr}$ ago. (e)-(g): Position of the potential supernova. Lines are interpolations to easier see the shape of the histogram and determine the confidence intervals of the parameters.}
\label{fig:hists_US1856}
\end{figure*}

In \autorefs{tab:detailed_scan_1856}, \ref{tab:detailed_scan_0720}, \ref{tab:detailed_scan_1605} and \ref{tab:detailed_scan_1308} we list the current proper motions ($\mu^*_\alpha$, $\mu_\delta$; columns 5 and 6), radial velocities ($v_r$; column 4) and parallaxes ($\pi$; column 8) which the neutron stars would have if they originated from the particular associations as well as the positions of the predicted supernova (distance to the Sun and equatorial coordinates; columns 9 to 11). Here we describe how those values are obtained.\\
We first define an area within the $\tau$-$d_{min}$ contour plot such that its boundaries reflect a 68\% decline from the peak (columns 2 and 3 of the respective tables). Parameter values in columns 4 to 11 were derived by selecting the input parameters (parallax, proper motion, radial velocity) which correspond to this defined region. The distance to the Sun as well as equatorial coordinates at the time of the supernova were then calculated for each parameter set.\\
From the histogram of each parameter we obtain its value and error by drawing an interpolation curve to better characterise the shape of the distribution. The ``mean'' of the parameter is then given by the maximum of the curve. The error intervals include about 68\% of the histogram area (note that these are not $1\sigma$ errors). In the case of multiple peaks in the histogram, the interval between the two values which reflect a 68\% decrease of the maximum is given.\\
\Autoref{fig:KonturplotUS1856} gives an example of a $\tau$-$d_{min}$ contour diagram indicating the area described above. The plot shows the distribution of minimum separations between \rxja{} and the US centre in the $\tau$-$d_{min}$ space. Histograms corresponding to the entries for US in columns 4 to 6 and 8 to 11 of \autoref{tab:detailed_scan_1856} are shown in \autoref{fig:hists_US1856}.\\

The probability of an association being the birth place of a neutron star is roughly estimated from the probability of the parallax value needed for the scenario (Gaussian distribution according to the literature value) and the probability of the necessary space velocity (adopting the distribution by Ho05). Furthermore, we are taking into account the distance of the potential supernova from the centre of the association since due to a decrease of the star density with increasing distance from the centre of an association, supernovae occur more likely near the centre. To account for that we make a rough estimate of the probability of a potential supernova position by taking a Gaussian distribution with $\sigma$ being half the radius of the association (weighted according to the number of member stars). This is justified since radii of associations are derived from star density profiles in most cases. The three probabilities are then multiplied.

\bsp

\label{lastpage}

\end{document}